# First M87 Event Horizon Telescope Results. VI.
# The Shadow and Mass of the Central Black Hole

The Event Horizon Telescope Collaboration
(See the end matter for the full list of authors.)

## Abstract

We present measurements of the properties of the central radio source in M87 using Event Horizon Telescope data obtained during the 2017 campaign. We develop and fit geometric crescent models (asymmetric rings with interior brightness depressions) using two independent sampling algorithms that consider distinct representations of the visibility data. We show that the crescent family of models is statistically preferred over other comparably complex geometric models that we explore. We calibrate the geometric model parameters using general relativistic magnetohydrodynamic (GRMHD) models of the emission region and estimate physical properties of the source. We further fit images generated from GRMHD models directly to the data. We compare the derived emission region and black hole parameters from these analyses with those recovered from reconstructed images. There is a remarkable consistency among all methods and data sets. We find that >50% of the total flux at arcsecond scales comes from near the horizon, and that the emission is dramatically suppressed interior to this region by a factor >10, providing direct evidence of the predicted shadow of a black hole. Across all methods, we measure a crescent diameter of $42 \pm 3$ $\mu$as and constrain its fractional width to be <0.5. Associating the crescent feature with the emission surrounding the black hole shadow, we infer an angular gravitational radius of $GM/Dc^2 = 3.8 \pm 0.4$ $\mu$as. Folding in a distance measurement of $16.8^{+0.8}_{-0.7}$ Mpc gives a black hole mass of $M = 6.5 \pm 0.2|_{\rm stat} \pm 0.7|_{\rm sys} \times 10^9 M_\odot$. This measurement from lensed emission near the event horizon is consistent with the presence of a central Kerr black hole, as predicted by the general theory of relativity.

## 1. Introduction

Einstein's general theory of relativity not only predicts the existence of black holes, but also provides a means to directly observe them. Photons can escape from near the event horizon via an unstable circular orbit (von Laue [1921](); Bardeen [1973]()), whose observational manifestation would be a bright ring of emission surrounding a dark interior black hole "shadow" (Luminet [1979](); Falcke et al. [2000]()). The diameter of the shadow for a black hole of mass $M$ as seen by a distant observer is predicted to be $\simeq 9.6$–$10.4$ $GM/c^2$, which is larger than twice the coordinate radius of the event horizon due to light-bending effects (Takahashi [2004](); Johannsen & Psaltis [2010]()). The range results from different values of black hole spin and observer inclination angle. The black hole shadow can only be seen if (i) there are a sufficient number of emitted photons to illuminate the black hole, (ii) the emission comes from close enough to the black hole to be gravitationally lensed around it, and (iii) the surrounding plasma is sufficiently transparent at the observed wavelength. For nearby low-luminosity black holes accreting via a radiatively inefficient flow, these conditions can be met at millimeter wavelengths (e.g., Özel et al. [2000](); Ho [2008](); Yuan & Narayan [2014]()).

The nearby massive elliptical galaxy M87 provides an ideal laboratory to search for such a black hole shadow. It is relatively nearby ($D \simeq 16.8$ Mpc) and has long been known to host a bright, compact radio source at its center (Cohen et al. [1969]()). Starting with Young et al. ([1978]()) and Sargent et al. ([1978]()), several attempts have been made to "weigh" the supermassive black hole (SMBH) hypothesized to power the radio source. Recent stellar-dynamics observations by Gebhardt et al. ([2011]()) found $M = (6.6 \pm 0.4) \times 10^9 M_\odot$, while the latest gas dynamics observations by Walsh et al. ([2013]()) yielded a more modest $M = (3.5^{+0.9}_{-0.7}) \times 10^9 M_\odot$. Both values assumed a distance $D = 17.9$ Mpc. Strictly interpreted within the predictions of the general theory of relativity, these measurements make a strong case that M87 does harbor some sort of compact massive dark object at its center, but they have insufficient resolution to formally demonstrate that it is indeed an SMBH. For M87 the expected shadow angular diameter is $\simeq 20$ or $\simeq 38\mu$ as, which is now accessible using global very long baseline interferometry (VLBI) at millimeter wavelengths (EHT Collaboration et al. [2019a](), hereafter Paper II).

Accreting black holes are powered by matter flowing in via an accretion disk that in many cases launches a powerful jet of magnetized, relativistic plasma (e.g., Blandford & Znajek [1977](); Blandford & Payne [1982]()). M87 exhibits the characteristic flat-to-inverted radio/millimeter (mm) synchrotron spectrum considered to be a hallmark of the compact innermost jet core in low-luminosity active galactic nuclei (AGNs; e.g., Blandford & Königl [1979](); Ho [1999]()). In this picture, the jet photosphere moves inward with increasing frequency up to the spectral break, at which point the entire jet becomes optically thin. The average broadband spectrum of M87 (Reynolds et al. [1996](); Di Matteo et al. [2003](); Prieto et al. [2016]()) indicates that the mm-band should straddle this transition. While images at longer wavelengths reveal extended jet structure (e.g., Asada & Nakamura [2012](); Hada et al. [2016](); Mertens et al. [2016](); Kim et al. [2018](); Walker et al. [2018]()), both the observed core shift (Hada et al. [2011]()) and compact size of $\simeq 40$ $\mu$as from past mm-VLBI (Doeleman et al. [2012]();

Akiyama et al. 2015) are consistent with an origin of the mm-band emission near the event horizon of the central black hole.

Although the photon ring and shadow predictions are clear, the image morphology will depend on the physical origin of the surrounding emission and spacetime of the black hole. If the observed synchrotron radiation at 1.3 mm originates far from the black hole, the forward jet will dominate the observed emission and the lensed emission and shadow feature should be weak (Broderick & Loeb 2009). If instead the emission comes from near the event horizon, either the counter-jet or the accretion flow can produce a compact ring- or crescent-like image surrounding the shadow (Dexter et al. 2012). This type of image is now known to be commonly produced in radiative models of M87 based on general relativistic magnetohydrodynamic (GRMHD) simulations (Dexter et al. 2012; Mościbrodzka et al. 2016; Ryan et al. 2018; Chael et al. 2019b; see EHT Collaboration et al. 2019d, hereafter Paper V).

The outline of the shadow is expected to be nearly circular, if the central object in M87 is a black hole described by the Kerr metric (Bardeen 1973; Takahashi 2004). Violations of the no-hair theorem generically change the shadow shape and size (e.g., Bambi & Freese 2009; Johannsen & Psaltis 2010; Falcke & Markoff 2013; Broderick et al. 2014; Cunha & Herdeiro 2018). Therefore, detecting a shadow and extracting its characteristic properties, such as size and degree of asymmetry, offers a chance to constrain the spacetime metric.

The high resolution necessary to resolve horizon scales for M87 has been achieved for the first time by the Event Horizon Telescope (EHT) in April 2017, with an array that spanned eight stations in six sites across the globe (Section II). The EHT observed M87 on four days (April 5, 6, 10, and 11) at 1.3 mm (EHT Collaboration et al. 2019b, hereafter Paper III), and imaging techniques applied to this data set reveal the presence of an asymmetric ring structure (EHT Collaboration et al. 2019c, hereafter Paper IV). A large library of model images generated from GRMHD simulations generically finds such features to arise from emission produced near the black hole (Paper V) which is strongly lensed around the shadow.

In this Letter we use three different methods to measure properties of the M87 230 GHz emission region using EHT 2017 observations. In Section 2, we describe the EHT data set used. We present in Section 3 a pedagogical description showing how within compact ring models the emission diameter and central flux depression (shadow) can be inferred directly from salient features of the visibility data. In Section 4 we describe the three analysis codes used to infer parameters from the data, and in Section 5 and Section 6 we fit geometric and GRMHD-based models. In Section 7 we extract properties of the reconstructed images from Paper IV.

We show that asymmetric ring ("crescent") geometric source models with a substantial central brightness depression provide a better statistical description of the data than other comparably complex models (e.g., double Gaussians). We use Bayesian inference techniques to constrain the size and width of this crescent feature on the sky, as well as the brightness contrast of the depression at its center compared to the rim. We show that all measurements support a source structure dominated by lensed emission surrounding the black hole shadow.

To extract the physical scale of the black hole at the distance of M87, $GM/Dc^2$, from the observed ring structure in geometric models and image reconstructions, we do not simply assume that the measured emission diameter is that of the photon ring itself.

We instead directly calibrate to the emission diameter found in model images from GRMHD simulations. The structure and extent of the emission preferentially from outside the photon ring leads to a $\lesssim 10\%$ offset between the measured emission diameter in the model images and the size of the photon ring. The scatter over a large number of images, which constitutes a systematic uncertainty, is found to be of the same magnitude.

We use independent calibration factors obtained for the geometric models and reconstructed images (Paper IV), providing two estimates of $GM/Dc^2$. We also fit the library of GRMHD images described in Paper V directly to the EHT data, which provides a third. All three methods are found to be in remarkable agreement. We consider prior dynamical measurements of $M/D$ and $D$ for M87 in Section 8. In Sections 9 and 10, we discuss the evidence for the detection of lensed emission surrounding the shadow of a black hole in EHT 2017 data. We use the prior distance information to convert the physical scale to a black hole mass and show that our result is consistent with prior stellar, but not gas, dynamical measurements. We further discuss the implications for the presence of an event horizon in the central object of M87. Further technical detail supporting the analyses presented here have been included as Appendices.

## 2. Observations and Data

Operating as an interferometer, the EHT measures complex visibilities on a variety of baselines $b_{ij}$ between stations $i$ and $j$. A complex visibility is a Fourier component of the source brightness distribution $I(x, y)$,

$$\mathcal{V}(u, v) = \iint e^{-2\pi i(ux+vy)} I(x, y) dx dy, \qquad (1)$$

where $(x, y)$ are angular coordinates on the sky, and $(u, v)$ are projected baseline coordinates measured in units of wavelengths (Thompson et al. 2017; hereafter TMS).

The 2017 EHT observations of M87 and their subsequent correlation, calibration, and validation are described in detail in Paper III. On each of the four days—April 5, 6, 10, and 11—the EHT observed M87 in two 2 GHz frequency bands centered on 227.1 GHz (low-band; LO) and 229.1 GHz (high-band; HI); the baseline coverage for April 11 is shown in Figure 1. For the modeling results presented in this Letter we analyze all four observing days and both bands. We use Stokes $I$ visibility data reduced via the EHT-Haystack Observatory Processing System (HOPS) pipeline (Blackburn et al. 2019), coherently averaged in time by scan. Scan averaging decreases the data volume with negligible coherence losses (see Paper III) and it further serves to increase the signal-to-noise ratio (S/N) of the data. Fewer timestamps correspond to fewer gain terms (see Section 2.1) and higher S/N improves the validity of our Gaussian likelihood functions (see Section 4.1).

### 2.1. Data Products

Visibility measurements are affected by a combination of thermal noise and systematic errors. The thermal noise $\epsilon_{ij}$ is distributed as a zero-mean complex Gaussian random variable with variance determined by the radiometer equation (TMS), while the dominant systematic noise components are associated with station-based complex gains $g_i$. The measured visibilities can thus be expressed as

$$V_{ij} = g_i g_j^* \mathcal{V}_{ij} + \epsilon_{ij} = |V_{ij}| e^{i\phi_{ij}}, \qquad (2)$$

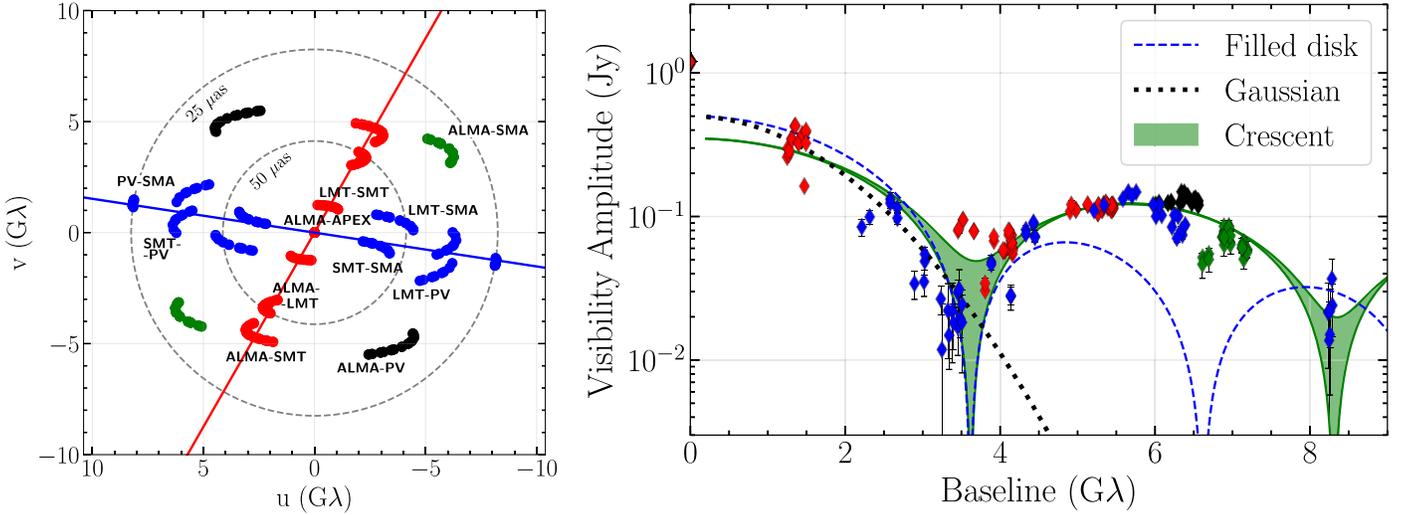

**Figure 1.** $(u, v)$-coverage (left panel) and visibility amplitudes (right panel) of M87 for the high-band April 11 data. The $(u, v)$-coverage has two primary orientations, east–west in blue and north–south in red, with two diagonal fillers at large baselines in green and black. Note that the Large Millimeter Telescope (LMT) and the Submillimeter Telescope (SMT) participate in both orientations, and that the LMT amplitudes are subject to significant gain errors. There is evidence for substantial depressions in the visibility amplitudes at ∼3.4 Gλ and ∼8.3 Gλ. The various lines in the right panel show the expected behavior of (dotted line) a Gaussian, (dashed line) a filled disk, and (green area) a crescent shape along different orientations. The image of M87 does not appear to be consistent with a filled disk or a Gaussian.

where $|V_{ij}|$ and $\phi_{ij} = \arg(V_{ij})$ are the measured visibility amplitude and phase. Measured visibility amplitudes are biased upward by thermal noise, so we use $A_{ij}$ to denote debiased visibility amplitude measurements (see TMS).

All noise sources in Equation (2) are functions of time and frequency, but the gain phase variations are particularly important for EHT data. Characteristic atmospheric timescales at 230 GHz are on the order of seconds, rendering visibility phase calibration unfeasible (Paper II; Paper III). We instead recover source phase information via the construction of closure phases $\psi_{\rm C}$, given by the argument of a product of visibilities around a triangle of baselines,

$$\psi_{{\rm C},ijk} = \arg(V_{ij} V_{jk} V_{ki}) = \phi_{ij} + \phi_{jk} + \phi_{ki}. \quad (3)$$

Because each gain term in the triple product gets multiplied by its complex conjugate, closure phases are immune to gain phase corruptions (Rogers et al. 1974).

Visibility amplitudes suffer less severely than visibility phases from station gain noise but the use of gain-free amplitude quantities can still aid modeling efforts. Closure amplitudes $A_{\rm C}$ are constructed from four visibilities on a quadrangle of baselines,

$$A_{{\rm C},ijk\ell} = \frac{A_{ij} A_{k\ell}}{A_{ik} A_{j\ell}}. \quad (4)$$

The appearance of each station in both the numerator and denominator of this expression causes the station gain amplitudes to cancel out. Because the closure amplitude is constructed from products and ratios of visibility amplitudes, it is often convenient to work instead with the logarithm of the closure amplitude,

$$\ln A_{{\rm C},ijk\ell} = \ln A_{ij} + \ln A_{k\ell} - \ln A_{ik} - \ln A_{j\ell}. \quad (5)$$

### 2.2. Data Selection and Preparation

From the scan-averaged visibility data, we increase the uncertainty associated with the debiased visibility amplitudes,

$A$, by adding a 1% systematic uncertainty component in quadrature to the thermal noise (Paper III; Paper IV); we refer to this increased uncertainty as the "observational error." These debiased visibility amplitudes are then used to construct a set of logarithmic closure amplitudes, $\ln A_{\rm C}$, per Equation (5). Closure amplitude measurements are generally not independent because a pair of quadrangles may have up to two baselines in common, and a choice must be made regarding which minimal (or "non-redundant") subset of closure amplitudes to use. We select the elements of our minimal set by starting with a maximal (i.e., redundant) set, from which we systematically remove the lowest-S/N quadrangles until the size of the reduced set is equal to the rank of the covariance matrix of the full set (see L. Blackburn et al. 2019, in preparation). This construction procedure serves to maximize the final S/N of the resulting closure amplitudes.

We construct closure phases, $\psi_{\rm C}$, from the visibilities using Equation (3), after first removing visibilities on the short intra-site baselines (James Clerk Maxwell Telescope–Submillimeter Array (JCMT–SMA), Atacama Large Millimeter/submillimeter Array–Atacama Pathfinder Experiment (ALMA–APEX)) which produce only "trivial" closure phases ≃0° (Paper III; Paper IV). As with closure amplitudes, closure phase measurements are in general not independent of one another because a pair of triangles may share a baseline. However, a suitable choice of non-redundant closure phase subset can minimize the covariance between measurements. We select our subset such that the highest-S/N baselines are the most frequently shared across triangles. Such a subset can be obtained by selecting one station to be the reference and then choosing all triangles containing that station (TMS). Because ALMA is so much more sensitive than the other stations in the EHT 2017 array, this construction procedure using ALMA as a reference station ensures the near-diagonality of the closure phase covariance matrix (see Section 4.1).

We list the number of data products in each class, along with the number of station gains, in Table 1 for each observing day and band. Information about accessing SR1 data and the

**Table 1**
The Number of Data Product and Gain Terms

| Day | Band | Including Intra-site | | Excluding Intra-site | | $N_{\psi C}$ | $N_{\ln A_C}$ |
|-----|------|------|------|------|------|------|------|
| | | $N_A$ | $N_g$ | $N_A$ | $N_g$ | | |
| April 5 | HI | 168 | 89 | 152 | 88 | 81 | 78 |
| | LO | 168 | 89 | 152 | 88 | 81 | 77 |
| April 6 | HI | 284 | 134 | 250 | 133 | 141 | 150 |
| | LO | 274 | 125 | 242 | 125 | 141 | 149 |
| April 10 | HI | 96 | 40 | 86 | 40 | 53 | 56 |
| | LO | 91 | 43 | 82 | 42 | 47 | 48 |
| April 11 | HI | 223 | 106 | 194 | 100 | 110 | 117 |
| | LO | 216 | 103 | 189 | 98 | 107 | 113 |

**Note.** $N_A$ is the number of visibility amplitudes, $N_g$ is the number of gain terms, $N_{\psi C}$ is the number of closure phases, and $N_{\ln A_C}$ is the number of logarithmic closure amplitudes. We show counts for the visibility amplitudes both with and without the inclusion of short intra-site baselines (ALMA–APEX and JCMT–SMA). The visibility amplitudes including intra-site baselines are used in Section 5, while those without are used in Section 6. The closure phase count always excludes triangles containing intra-site baselines, while the logarithmic closure amplitude count always includes quadrangles containing intra-site baselines. Both closure phase and logarithmic closure amplitude counts are for minimal (non-redundant) sets.

software used for analysis can be found on the Event Horizon Telescope website's data portal.[107]

### 3. Descriptive Features of the Visibility Data

In Paper IV, different image reconstruction methods all obtained similar looking images of M87 from the 2017 EHT observations, namely, a nearly circular ring with a dark center and azimuthally varying intensity. In this Letter, we consider a range of source models and calculate the corresponding visibilities as a function of the model parameters. We then employ statistical tools to select between models and to estimate model parameters by comparing the model visibilities to the observed ones. Because imaging and visibility domain analysis rely on the entire complex visibility data set and/or closure quantities, it is useful as a pedagogical guide to show first how some of the simple image characteristics are imprinted on the visibility data. We emphasize that a complete and accurate description of the source requires imaging analysis and visibility modeling, which we perform in Paper IV and in later sections of this Letter.

Here we show that the data are consistent with the presence of a ring structure with a characteristic emission diameter of ~45 μas. These aspects also match the predictions for an image dominated by lensed emission near the photon ring surrounding the black hole shadow of M87 (Paper V).

The 2017 EHT observations of M87 have good $(u, v)$-coverage, primarily along an east–west (blue) and a north–south (red) orientation, with additional diagonal long baselines (green and black; see Figure 1 and also Paper III). The right panel of this figure shows the visibility amplitudes observed on April 11 color coded by the orientation of the baselines.

There is evidence for a minimum of the visibility amplitudes at baseline lengths of ~3.4 Gλ, followed by a second peak around ~6 Gλ. Such minima are often associated with edges or gaps in the image domain.

Further, the visibility amplitudes are similar in the north–south and east–west directions, suggestive of a similar characteristic image size and shape in both directions (Figure 1). This is naturally accomplished if the image possesses a large degree of azimuthal symmetry, such as in a ring or disk. Differences in the visibilities as a function of baseline length for different orientations do exist, however, particularly in the depth of the first null, indicating that the source is not perfectly symmetric.

Next, we consider the presence of the central flux depression. For the case of a uniform disk model, the second visibility amplitude minimum occurs at 1.8 times the location of the first minimum, i.e., at ~6.3–7 Gλ, which is not seen in the visibility amplitudes. For a ring or annular model, however, the second minimum moves to longer baseline lengths, consistent with what is seen in the visibility amplitudes.

Indeed, the Fourier transform of an infinitesimally thin ring structure shows the first minimum in visibility amplitude at a baseline length $b_1$ for which the zeroth-order Bessel function is zero (see TMS). This allows us to estimate the source size for a ring model as

$$d_0 \simeq 45 \left( \frac{b_1}{3.5 \text{ G}\lambda} \right)^{-1} \mu\text{as}. \tag{6}$$

In subsequent sections, we quantify our characterization of this model through fitting in the visibility and image domains.

### 4. Model Fitting to Interferometric Data

We utilize three independent algorithms for parameter space exploration to quantify the size, shape, and orientation of this asymmetric ring structure. We fit both geometric and GRMHD models to the 2017 EHT interferometric data. In this section we outline the modeling framework used to extract parameter values from the M87 data. We first detail the construction of the corresponding likelihood functions in Section 4.1, and we then describe in Section 4.2 the three different codes we have used to estimate model parameters.

#### 4.1. Likelihood Construction

Our quantitative modeling approach seeks to estimate the posterior distribution $P(\Theta|\boldsymbol{D})$ of some parameters $\Theta$ within the context of a model and conditioned on some data $\boldsymbol{D}$,

$$P(\Theta|\boldsymbol{D}) = \frac{P(\boldsymbol{D}|\Theta)P(\Theta)}{P(\boldsymbol{D})} \equiv \frac{\mathcal{L}(\Theta)\pi(\Theta)}{\mathcal{Z}}. \tag{7}$$

Here, $\mathcal{L}(\Theta) \equiv P(\boldsymbol{D}|\Theta)$ is the likelihood of the data given the model parameters, $\pi(\Theta) \equiv P(\Theta)$ is the prior probability of the model parameters, and

$$\mathcal{Z} \equiv P(\boldsymbol{D}) = \int \mathcal{L}(\Theta)\pi(\Theta)d\Theta \tag{8}$$

is the Bayesian evidence. In this section we define our likelihood functions $\mathcal{L}(\Theta)$, which we note differ in detail from those adopted for the regularized maximum likelihood (RML) imaging procedures presented in Paper IV.

For each scan, the measured visibility amplitude, $A$, corresponds to the magnitude of a random variable distributed according to a symmetric bivariate normal distribution (TMS). This magnitude follows a Rice distribution, which in the high-S/N limit reduces to a Gaussian near the mode,

$$\mathcal{L}_{A,ij} = \frac{1}{\sqrt{2\pi\sigma_{ij}^2}} \exp\left[-\frac{(A_{ij} - |g_i||g_j|\hat{A}_{ij})^2}{2\sigma_{ij}^2}\right]. \tag{9}$$

Here, $\sigma_{ij}^2$ is the variance of the visibility measurement, $\hat{A}_{ij}$ is the model visibility amplitude of the source, and $|g_i|$ and $|g_j|$ are the gain amplitudes for stations $i$ and $j$. Both the mean and standard deviation of Equation (9) are biased with respect to the true visibility amplitude distribution, but for $A/\sigma \gtrsim 2.0$ these biases are below 10%; at least 94% of our visibility amplitude data for any day and band meet this criterion.

For scan-averaged EHT 2017 data, the gain amplitudes constitute on the order of 100 additional nuisance parameters per data set (see Table 1). These numerous additional parameters may be efficiently addressed by directly marginalizing the likelihood in Equation (9), a procedure detailed in Appendix A and A. E. Broderick et al. (2019, in preparation). Once the gain amplitudes have been reconstructed, the joint likelihood function for all visibility amplitude measurements within a data set is then given by the product over the individual likelihoods,

$$\mathcal{L}_A = \prod \mathcal{L}_{A,ij}, \tag{10}$$

where this product is taken over all baselines and scans.

The logarithm of the visibility amplitudes also follows a Gaussian distribution in the high-S/N limit, with an effective logarithmic uncertainty of

$$\sigma_{\ln A} = \frac{\sigma}{A}. \tag{11}$$

The Gaussianity of the logarithmic visibility amplitudes implies that the logarithm of the closure amplitudes will similarly be Gaussian distributed in the same limit, with variances given by

$$\sigma_{\ln A_C,ijk\ell}^2 = \sigma_{\ln A,ij}^2 + \sigma_{\ln A,k\ell}^2 + \sigma_{\ln A,ik}^2 + \sigma_{\ln A,j\ell}^2. \tag{12}$$

This Gaussian approximation for logarithmic closure amplitudes holds well (i.e., the mean and standard deviation are biased by less than 10%) for $\sigma_{\ln A_C} \lesssim 2.0$; at least 87% of our logarithmic closure amplitude data for any day and band meet this criterion.

The likelihood function for a set of logarithmic closure amplitudes also depends on the covariances between individual measurements, which in general are not independent. We can construct a covariance matrix $\Sigma_A$ that captures the combined likelihood via

$$\mathcal{L}_{\ln A_C} = \frac{1}{\sqrt{(2\pi)^q \det(\Sigma_A)}} \exp\left(-\frac{1}{2}\boldsymbol{A}^\top \Sigma_A^{-1} \boldsymbol{A}\right), \tag{13}$$

where $\boldsymbol{A}$ is an ordered list of logarithmic closure amplitude residuals, and $q$ is the number of non-redundant closure amplitudes; for a fully connected array with $N_{el}$ elements, $q = N_{el}(N_{el} - 3)/2$ (TMS). The covariance between logarithmic closure amplitude measurements $\ln A_{C,1234}$ and $\ln A_{C,1235}$ is (Lannes 1990a; L. Blackburn et al. 2019, in preparation)

$$\text{Cov}(\ln A_{C,1234}, \ln A_{C,1235}) = \sigma_{\ln A,12}^2 + \sigma_{\ln A,13}^2 \tag{14}$$

in the Gaussian limit; here, we have used the fact that $\sigma_{ij} = \sigma_{ji}$ to simplify notation.

The distribution of measured visibility phases, $\phi$, corresponds to the projection of a symmetric bivariate normal random variable onto the unit circle, which once again reduces to a Gaussian distribution in the high-S/N limit. Closure phases, $\psi_C$, in this limit will also be Gaussian distributed with variances given by

$$\sigma_{\psi_C,ijk}^2 = \sigma_{\phi,ij}^2 + \sigma_{\phi,jk}^2 + \sigma_{\phi,ki}^2, \tag{15}$$

where $\sigma_{\phi,ij}^2$ is the variance in the visibility phase measurement, $\phi_{ij}$. The Gaussian approximation for closure phases is unbiased in the mean with respect to the true closure phase distribution, and the standard deviation is biased by less than 10% for $\sigma_{\psi_C} \lesssim 1.5$; this criterion is satisfied for at least 92% of our closure phase data on any day and band.

Closure phase measurements are also generally covariant, and for a covariance matrix $\Sigma_\psi$ the joint likelihood is given by

$$\mathcal{L}_{\psi_C} = \frac{1}{\sqrt{(2\pi)^t \det(\Sigma_\psi)}} \exp\left(-\frac{1}{2}\boldsymbol{\psi}^\top \Sigma_\psi^{-1} \boldsymbol{\psi}\right), \tag{16}$$

where $\boldsymbol{\psi}$ is an ordered list of closure phase residuals, and $t$ is the number of non-redundant closure phases; for a fully connected array containing $N_{el}$ elements, $t = (N_{el} - 1)(N_{el} - 2)/2$ (TMS). The covariance between two closure phase measurements, $\psi_{C,123}$ and $\psi_{C,124}$, is given by (Kulkarni 1989; Lannes 1990b; L. Blackburn et al. 2019, in preparation)

$$\text{Cov}(\psi_{C,123}, \psi_{C,124}) = \sigma_{\phi,12}^2. \tag{17}$$

As described in Section 2.2, a non-redundant subset of closure phases can be selected to maximize independence and thus ensure the near-diagonality of $\Sigma_\psi$. In this case the closure phase measurements can be treated as individually Gaussian,

$$\mathcal{L}_{\psi_C,ijk} = \frac{1}{\sqrt{2\pi\sigma_{\psi_C,ijk}^2}} \exp\left[-\frac{(\psi_{C,ijk} - \hat{\psi}_{C,ijk})^2}{2\sigma_{\psi_C,ijk}^2}\right], \tag{18}$$

where $\hat{\psi}_{C,ijk}$ is the modeled closure phase. The joint likelihood is then

$$\mathcal{L}_{\psi_C} = \prod \mathcal{L}_{\psi_C,ijk}, \tag{19}$$

where the product is taken over all closure phases in the selected minimal subset. Because closure phases wrap around the unit circle, we always select the branch of $\psi_C - \hat{\psi}_C$ such that the difference lies between $-180°$ and $180°$.

### 4.2. Parameter Space Exploration Techniques

We utilize three independent algorithms for parameter space exploration. For the geometric crescent model fitting presented in Section 5, we use both Markov chain Monte Carlo (MCMC) and nested sampling (NS) algorithms. The MCMC modeling scheme explores model fits to the visibility amplitude and closure phase data, while the NS scheme fits to the closure phase and logarithmic closure amplitude data. For the GRMHD model fitting in Section 6, we use both MCMC and a genetic algorithm to fit the visibility amplitude and closure phase data.

#### 4.2.1. THEMIS

THEMIS is an EHT-specific analysis framework for generating and comparing models to both EHT and ancillary data.

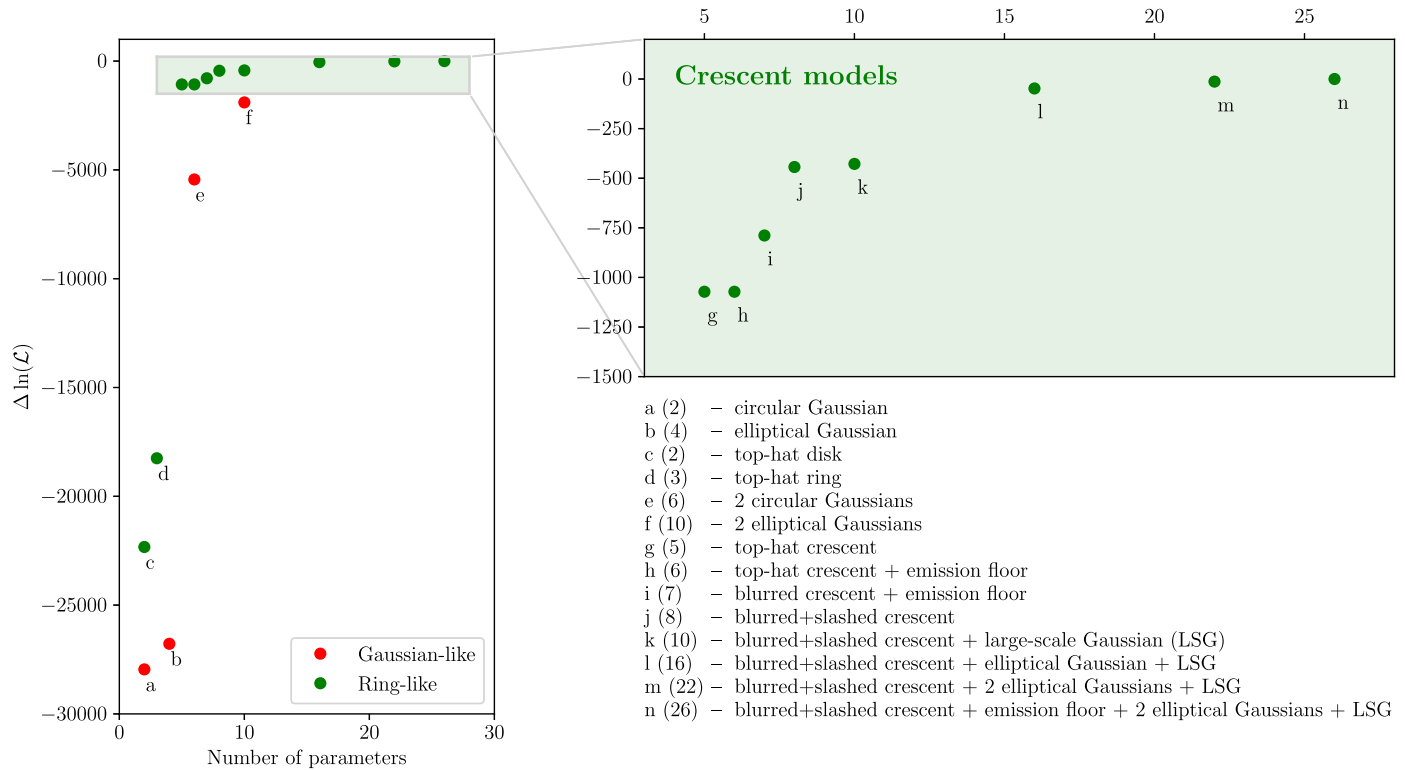

**Figure 2.** Relative log-likelihood values for different geometric models fit to the M87 data as a function of nominal model complexity; the number of parameters is given in parenthesis for each model. April 5 is shown here, and all days and bands show the same trend. The models shown in this figure are strict subsets of the "generalized crescent model" (labeled here as model "n"; see Section 5.1), and they have been normalized such that the generalized crescent fit has a value of $\mathcal{L} = 1$; the reduced-$\chi^2$ for the generalized crescent fit is 1.24 (see Table 2). We find that the data overwhelmingly prefer crescent models over, e.g., symmetric disk and ring models, and that additional Gaussian components lead to further substantial improvement. Note that a difference of ~5 on the vertical axis in this plot is statistically significant.

THEMIS is written in C++ and parallelized via the Message Passing Interface (MPI) standard. THEMIS implements a differential evolution MCMC algorithm, and it utilizes parallel tempering based on the algorithm described in Nelson et al. (2014) and Braak (2006). In particular, THEMIS uses the adaptive temperature ladder prescription from Vousden et al. (2016). All sampling techniques, validation tests, and implementation details for THEMIS are described in detail in A. E. Broderick et al. (2019, in preparation).

In Sections 5 and 6, we use THEMIS to model the visibility amplitude and closure phase data, using the corresponding likelihood functions given in Equations (10) and (19), respectively. Gain amplitude terms are incorporated as model parameters (see Equation (9)) and are marginalized as described in Appendix A.

### 4.2.2. dynesty

In Section 5, we also use an NS technique, developed by Skilling (2006) primarily to evaluate Bayesian evidence integrals. We use the Python code dynesty (Speagle & Barbary 2018) as a sampler for the NS analyses presented in this Letter. The NS algorithm estimates the Bayesian evidence, $\mathcal{Z}$, by replacing the multidimensional integral over $\Theta$ (see Equation (8)) with a 1D integral over the prior mass contained within nested isolikelihood contours. We permit dynesty to run until it estimates that less than 1% of the evidence is left unaccounted for.

Our NS analyses employ a likelihood function constructed exclusively from closure quantities; we account for data covariances in the likelihood function using Equation (16) for closure phases and Equation (13) for logarithmic closure amplitudes. Additionally, the use of logarithmic closure amplitudes in our NS fits removes information about the total flux density.

### 4.2.3. GENA

In addition to the above sampling algorithms, which seek to reconstruct a posterior distribution, we also employ an optimization procedure for comparing GRMHD simulations to data in Section 6. The optimization code, GENA (Fromm et al. 2019), is a genetic algorithm written in Python and parallelized using MPI. GENA minimizes a $\chi^2$ statistic on visibility amplitudes and closure phases, using the gain calibration procedures in the eht-imaging (Chael et al. 2016, 2018, 2019a) Python package to solve for the gain amplitudes. GENA implements the Non-dominated Sorting Genetic Algorithm II (NSGA-II; Deb et al. 2002) for parameter exploration and the differential evolution algorithm from Storn & Price (1997) for constrained optimization.

## 5. Geometric Modeling

As detailed in Paper IV, images reconstructed from the M87 data show a prominent and asymmetric ring ("crescent") structure. In this section we use the techniques described in Section 4 to fit the M87 data sets with a specific class of geometric crescent models.

We first quantify the preference for crescent structure in Figure 2, which summarizes the results of fitting a series of increasingly complex geometric models to the M87 data. We see that simple azimuthally symmetric models (e.g., uniform disks, rings) do a poor job of fitting the data; indeed, the strong detection of nonzero closure phases alone precludes such models. Models that allow for a central flux depression and a degree of asymmetry (e.g., double Gaussians or crescents) show significantly better performance. The most substantial gain in fit quality occurs for the crescent family of models. A top-hat crescent model (the difference of two uniform disks with the inner disk shifted, described by five parameters; see Kamruddin & Dexter 2013, Appendix B) performs vastly better than a top-hat ring model (three parameters). It also significantly outperforms the sum of two circular Gaussians (six parameters) and even the sum of two elliptical Gaussians (10 parameters). Adding parameters to the simplest crescent model continues to result in statistically significant, but comparatively modest, improvements.

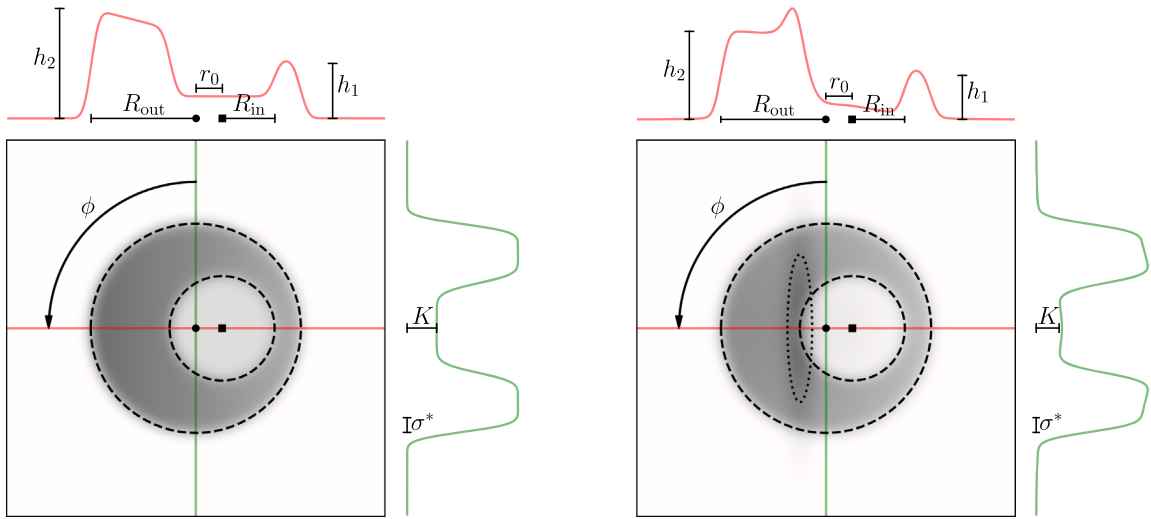

**Figure 3.** Schematic diagrams illustrating the crescent components of the xs-ring (left panel) and xs-ringauss (right panel) models. Dashed lines outline the inner and outer circular disk components that are differenced to produce the crescent models, and for the xs-ringauss model the FWHM of the fixed Gaussian component is additionally traced as a dotted line. The red and green curves above and to the right of each panel show cross-sectional plots of the intensity through the corresponding horizontal and vertical slices overlaid on the images. The circular and square markers indicate the centers of the outer and inner disks, respectively. The labeled parameters correspond to those described in Section 5.1. Both crescents are shown at an orientation of $\phi = \tilde{\phi} = 90°$.

### 5.1. Generalized Crescent Models

Among the large number of potential crescent-like models, we aim for one having the simplest geometry that is capable of both adequately fitting the M87 data and constraining several key observables. The geometric parameters of interest are the crescent diameter, its width and orientation, the sharpness of the inner edge, and the depth of any flux depression interior to the crescent.

With these key features in mind, we use an augmented version of the "slashed crescent" construction from Benkevitch et al. (2016) to provide the basis for a family of "generalized crescent" (GC) models. We refer to the two variants of the GC model that we use to fit the M87 data as xs-ring and xs-ringauss. Both GC models can be constructed in the image domain using the following procedure (see Figure 3).

1. Starting with a uniform circular disk of emission with radius $R_{out}$, we subtract a smaller uniform disk with radius $R_{in}$ that is offset from the first by an amount $r_0$. The resulting geometry is that of a "top-hat crescent."

2. We apply a "slash" operation to the top-hat crescent, which imposes a linear brightness gradient along the symmetry axis. The brightness reaches a minimum of $h_1$ and a maximum of $h_2$.

3. We add a "floor" of brightness $K$ to the central region of the crescent. For the xs-ring model this floor takes the form of a circular disk, while for the xs-ringauss model we use a circular Gaussian with flux density $V_F$ and width $\sigma_F$. The total flux density of the crescent plus floor component for the xs-ring model is denoted as $V_0$.

4. For the xs-ringauss model, we add an elliptical Gaussian component with flux density $V_1$ whose center is fixed to the inner edge of the crescent at the point where its width is largest (see the right panel of Figure 3), and whose orientation is set to align with that of the crescent. This fixed Gaussian component is inspired by the "xringaus" model from Benkevitch et al. (2016), which in turn sought to reproduce image structure seen in simulations from Broderick et al. (2014).

5. The image is smoothed by a Gaussian kernel of FWHM $\sigma^*$.

6. The image is rotated such that the widest section of the crescent is oriented at an angle $\phi$ in the counterclockwise direction (i.e., east of north).

The xs-ring model is described by eight parameters, while the xs-ringauss model is described by 11 parameters.

Though it is useful to conceptualize the GC models via their image domain construction, in practice we fit the models using their analytic Fourier domain representations. The Fourier domain construction of both models is described in Appendix B, along with a table of priors used for the model parameters. Throughout this Letter we fit the xs-ring model using the dynesty sampling code, while the xs-ringauss model is implemented as part of THEMIS. Below, we define the various desired key quantities within the context of this GC model parameterization.

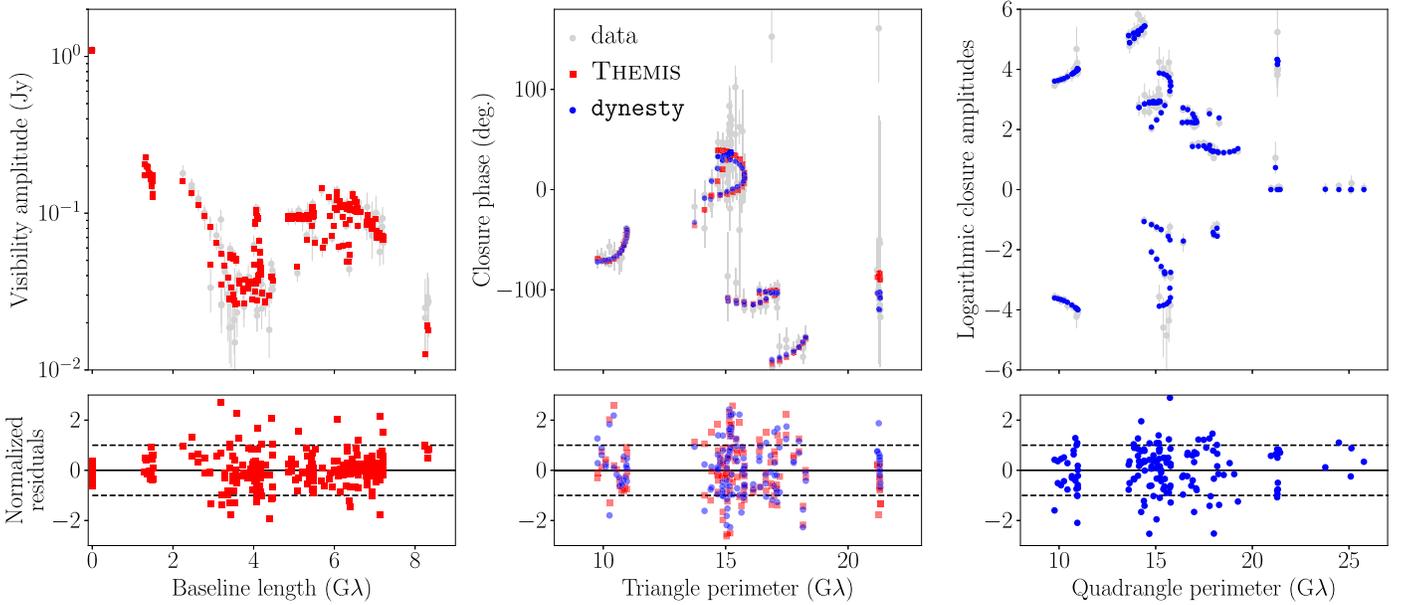

**Figure 4.** Modeled data (top panels) and residuals (bottom panels) for GC model fits to the April 6 high-band data set, with the data plotted in gray; we show results for the median posterior fit. The panels show visibility amplitude (left panels), closure phase (middle panels), and logarithmic closure amplitude (right panels) data. The xs-ringauss model, shown in red, is fit to the visibility amplitudes and closure phases using THEMIS; the `dynesty`-based xs-ring model, plotted in blue, comes from a fit to closure phases and logarithmic closure amplitude. Because both models fit to closure phases, the center panel shows two sets of models and residuals. All residuals are normalized by the associated observational noise values.

We define the crescent diameter $\hat{d}$ to be twice the average of the inner and outer crescent radii,

$$\hat{d} \equiv R_{\rm out} + R_{\rm in}. \tag{20}$$

The fractional crescent width $\hat{f}_{\rm w}$ is defined in a similar manner, as the mean difference between the outer and inner radii (normalized to the diameter) plus a term to account for the FWHM, $\sigma^* \equiv 2\sigma\sqrt{2\ln(2)}$, of the smoothing kernel:

$$\hat{f}_{\rm w} \equiv \frac{R_{\rm out} - R_{\rm in} + \sigma^*}{\hat{d}}. \tag{21}$$

The sharpness $\hat{s}$ is the ratio of the FWHM of the smoothing kernel to the crescent diameter, i.e.,

$$\hat{s} \equiv \frac{\sigma^*}{\hat{d}}. \tag{22}$$

The fourth quantity of interest is the ratio, $\hat{f}_{\rm c}$, between the brightness of the emission floor (interior to the crescent) and the mean brightness of the crescent,

$$\hat{f}_{\rm c} \equiv \frac{\text{brightness of emission floor}}{\text{mean brightness of crescent}}. \tag{23}$$

The different specifications for the xs-ring and xs-ringauss models (see Appendix B) means that these brightness ratios must be computed differently,

$$\hat{f}_{\rm c} = \begin{cases} \dfrac{V_F[(2R_{\rm out} + \sigma^*)^2 - (2R_{\rm in} - \sigma^*)^2]}{8\sigma_F^2(V_0 + V_1)}, & \text{xs-ringauss} \\[3mm] \dfrac{\pi K[(2R_{\rm out} + \sigma^*)^2 - (2R_{\rm in} - \sigma^*)^2]}{8V_0}, & \text{xs-ring}. \end{cases} \tag{24}$$

Finally, we determine the orientation of the crescent directly from the $\phi$ parameter, such that $\hat{\phi} = \phi$.

In addition to the crescent component of the GC model, we also include a small number (two to three) of additional "nuisance" elliptical Gaussian components intended to capture extraneous emission around the primary ring and to mitigate other unmodeled systematics; the parameterization and behavior of these additional Gaussian components are described in Appendix B.2. GRMHD simulations of M87 often exhibit spiral emission structures in the region immediately interior and exterior to the photon ring (see Paper V), and M87 is known to have a prominent jet that extends down to scales of several Schwarzschild radii (e.g., Hada et al. 2016, Kim et al. 2018). Such extra emission is unlikely to be adequately captured by the crescent component alone, and the nuisance Gaussian components serve as a flexible way to model generic emission structures. We define the total compact flux density, $\hat{\rm CF}$, of the model to be equal to the summed contributions from the crescent and nuisance Gaussian components,

$$\hat{\rm CF} \equiv V_0 + V_1 + V_F + \sum_i V_{g,i}, \tag{25}$$

where $V_0$ is the crescent flux density, $V_1$ is the flux density of the fixed Gaussian component, $V_F$ is the flux density of the central emission floor, and $V_{g,i}$ are the flux densities of the nuisance Gaussian components.

### 5.2. M87 Fit Results

We carry out independent fits to all days and bands using both the THEMIS- and `dynesty`-based codes. We show example GC model fits to the April 6 high-band data in Figure 4, with their corresponding image domain representations shown in Figure 5. There are no apparent systematic trends in the normalized residuals for either model, and we see similar behavior in the residuals from fits across all data sets. The corresponding THEMIS gain reconstructions are described in Appendix A.

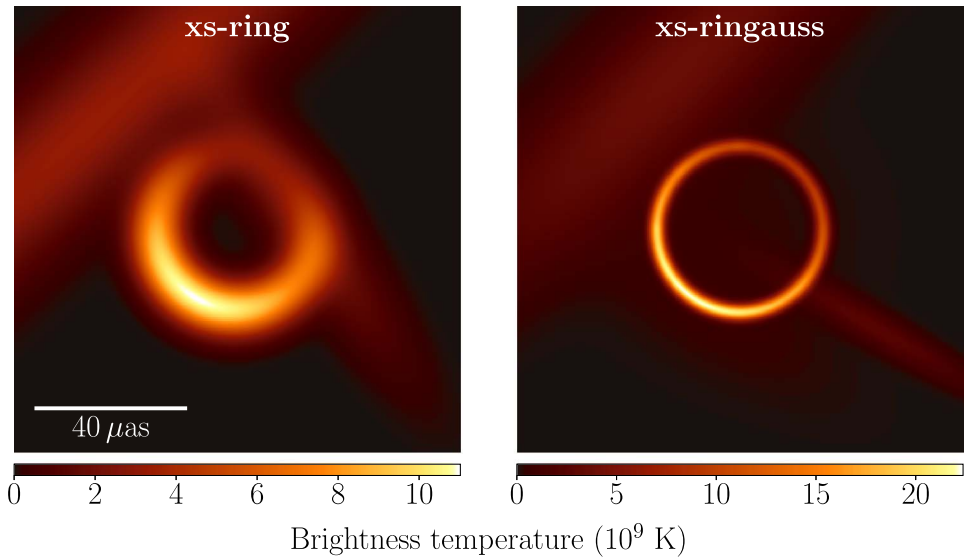

**Figure 5.** Image domain representations of a random posterior sample from the xs-ring (left panel) and xs-ringauss (right panel) model fits to the April 6 high-band data set; note that these are representative images drawn from the posteriors, and thus do not represent maximum likelihood or other "best-fit" equivalents. The xs-ring model fit uses only closure quantities, so we have scaled the total flux density to be equal to the 1.0 Jy flux density of the xs-ringauss model fit.

**Table 2**
Reduced-chi2 Statistics for the GC Model Fits

| Data Set | | Model | | | | | |
|---|---|---|---|---|---|---|---|
| | | xs-ringauss | | | xs-ring | | |
| Day | Band | $\chi^2_A$ | $\chi^2_{\psi C}$ | $\chi^2_{A+\psi C}$ | $\chi^2_{\psi C}$ | $\chi^2_{\ln Ac}$ | $\chi^2_{\psi C+\ln Ac}$ |
| April 5 | HI | 1.66 | 1.48 | 1.24 | 1.30 | 1.30 | 1.02 |
| | LO | 1.33 | 1.32 | 1.05 | 1.43 | 1.35 | 1.10 |
| | HI+LO | 1.16 | 1.07 | 1.01 | ... | ... | ... |
| April 6 | HI | 1.02 | 1.46 | 1.12 | 1.44 | 0.96 | 1.06 |
| | LO | 1.57 | 1.36 | 1.32 | 1.37 | 1.52 | 1.29 |
| | HI+LO | 1.16 | 1.35 | 1.19 | ... | ... | ... |
| April 10 | HI | 1.57 | 1.52 | 1.05 | 1.44 | 1.08 | 0.82 |
| | LO | 2.23 | 2.74 | 1.53 | 1.95 | 1.50 | 1.28 |
| | HI+LO | 1.11 | 1.32 | 1.04 | ... | ... | ... |
| April 11 | HI | 1.40 | 1.37 | 1.20 | 1.37 | 1.02 | 1.02 |
| | LO | 1.34 | 1.16 | 1.07 | 1.18 | 0.92 | 0.89 |
| | HI+LO | 1.32 | 1.14 | 1.15 | ... | ... | ... |

**Note.** Reduced-$\chi^2$ values corresponding to the maximum likelihood posterior sample for individual M87 data sets from both fitting codes, split by data type and calculated as described in Appendix C. The xs-ringauss values are from fits to visibility amplitudes (reduced-$\chi^2$ given by $\chi^2_A$) and closure phases (reduced-$\chi^2$ given by $\chi^2_{\psi C}$), with the joint visibility amplitude and closure phase reduced-$\chi^2$ denoted as $\chi^2_{A+\psi C}$. The xs-ring values are from fits to closure phases and logarithmic closure amplitudes (reduced-$\chi^2$ given by $\chi^2_{\ln Ac}$), with the joint closure phases and logarithmic closure amplitude reduced-$\chi^2$ denoted as $\chi^2_{\psi C+\ln Ac}$.

While our model-fitting procedures formally optimize a likelihood function modified by a prior (see Section 4.1), reduced-$\chi^2$ values serve as general-purpose distance metrics between model and data. In our case, reduced-$\chi^2$ values have the added benefit of enabling cross-comparisons with the images produced in Paper IV. The reduced-$\chi^2$ expressions we use are detailed in Appendix C, and Table 2 lists their values for all M87 fits.

In general we find joint reduced-$\chi^2$ values for the fits of $\sim$0.9–1.5. The number of degrees of freedom in the data ranges from $\sim$20 to 120, corresponding to an expected reduced-$\chi^2$ deviation from unity of $\sim$0.06–0.16 if our likelihoods follow a $\chi^2$-distribution. The implication is that while we see little evidence for overfitting, there are instances in which the residual values are distributed more broadly than the data error budget would nominally permit. The fact that such models sometimes underfit the data then indicates that we need to empirically determine our uncertainties, as posterior widths alone may not be reliable in the face of a statistically poor fit. We describe the empirical determination of the "observational uncertainties" in Appendix D.3, and we list the derived values in Table 3.

An illustrative pairwise parameter correlation diagram for both the xs-ring and xs-ringauss GC model fits to the April 5 high-band data set is shown in Figure 6, with single-parameter posteriors plotted along the diagonal. Figure 7 shows constraints on the GC model crescent component parameters, split up by data set. We also list the best-fit crescent parameters, as defined in Section 5.1, in Table 3.

Despite the differences in data products, sampling procedures, and model specifications, we find broad agreement between the derived posteriors from the two fitting codes. We note that the systematically wider posteriors from the xs-ring fits (by anywhere from a few to several tens of percent) are seen across all data sets and are an expected consequence of the use of closure amplitudes rather than visibility amplitudes.

Our primary parameter of interest is the diameter of the crescent, $\hat{d}$. The weighted mean xs-ringauss value of 43.4 $\mu$as is in excellent agreement with the corresponding xs-ring value of 43.2 $\mu$as, with an rms scatter in the measurements of 0.64 $\mu$as and 0.69 $\mu$as, respectively, across all days and bands. This remarkable consistency provides evidence for the diameter measurement being robustly recoverable. The posterior widths of diameter measurements for individual data sets are typically at the $\sim$1% level, but our empirically determined uncertainties associated with the changing ($u$, $v$)-coverage and other

**Table 3**
Best-fit GC Model Parameters for All Data Sets and Both Models, as Defined in Section 5.1; Median Posterior Values are Quoted with 68% Confidence Intervals

| Data Set | | | | Parameter | | | | | |
| Day | Band | Code | Model | $\hat{d}$ ($\mu$as) | $\hat{f}_w$ | $\log_{10}(\hat{s})$ | $\log_{10}(\hat{f}_c)$ | $\hat{\phi}$ (deg.) | $\hat{CF}$ (Jy) |
| --- | --- | --- | --- | --- | --- | --- | --- | --- | --- |
| April 5 | HI | THEMIS | xs-ringauss | $43.1^{+0.35}_{-0.36}$ | $0.12^{+0.07}_{-0.06}$ | $-1.35^{+0.30}_{-0.54}$ | $-1.60^{+0.26}_{-0.37}$ | $160.6^{+2.3}_{-1.6}$ | $0.75^{+0.16}_{-0.17}$ |
| | | dynesty | xs-ring | $42.9^{+0.59}_{-0.54}$ | $0.39^{+0.06}_{-0.07}$ | $-1.07^{+0.23}_{-0.48}$ | $-1.97^{+0.40}_{-0.50}$ | $160.5^{+3.4}_{-3.1}$ | $\cdots$ |
| | LO | THEMIS | xs-ringauss | $43.5^{+0.27}_{-0.28}$ | $0.09^{+0.06}_{-0.04}$ | $-1.41^{+0.29}_{-0.51}$ | $-1.76^{+0.29}_{-0.37}$ | $160.9^{+1.5}_{-1.9}$ | $0.72^{+0.17}_{-0.15}$ |
| | | dynesty | xs-ring | $43.5^{+0.44}_{-0.41}$ | $0.20 \pm 0.06$ | $-1.23^{+0.26}_{-0.50}$ | $-2.15^{+0.29}_{-0.48}$ | $157.9^{+1.7}_{-1.8}$ | $\cdots$ |
| | HI+LO | THEMIS | xs-ringauss | $43.3^{+0.22}_{-0.23}$ | $0.09^{+0.05}_{-0.04}$ | $-1.62^{+0.33}_{-0.55}$ | $-1.74^{+0.24}_{-0.31}$ | $160.7^{+0.8}_{-0.9}$ | $0.75 \pm 0.15$ |
| April 6 | HI | THEMIS | xs-ringauss | $44.1^{+0.23}_{-0.30}$ | $0.16^{+0.04}_{-0.06}$ | $-0.94^{+0.12}_{-0.30}$ | $-1.96^{+0.39}_{-0.58}$ | $146.4^{+2.6}_{-3.5}$ | $0.99 \pm 0.04$ |
| | | dynesty | xs-ring | $43.3^{+0.44}_{-0.43}$ | $0.34^{+0.05}_{-0.06}$ | $-0.70^{+0.07}_{-0.11}$ | $-2.23^{+0.36}_{-0.56}$ | $149.1 \pm 1.5$ | $\cdots$ |
| | LO | THEMIS | xs-ringauss | $43.5 \pm 0.14$ | $0.18^{+0.03}_{-0.04}$ | $-0.87^{+0.09}_{-0.16}$ | $-2.14^{+0.43}_{-0.62}$ | $153.0^{+2.0}_{-2.4}$ | $1.07^{+0.05}_{-0.04}$ |
| | | dynesty | xs-ring | $43.4^{+0.27}_{-0.26}$ | $0.20^{+0.07}_{-0.06}$ | $-1.31^{+0.31}_{-0.52}$ | $-2.63^{+0.41}_{-0.60}$ | $148.5^{+1.4}_{-1.2}$ | $\cdots$ |
| | HI+LO | THEMIS | xs-ringauss | $43.7^{+0.10}_{-0.11}$ | $0.19^{+0.03}_{-0.02}$ | $-0.88^{+0.05}_{-0.07}$ | $-2.28^{+0.48}_{-0.61}$ | $151.8^{+1.6}_{-1.7}$ | $1.03 \pm 0.03$ |
| April 10 | HI | THEMIS | xs-ringauss | $42.9^{+1.09}_{-0.86}$ | $0.46 \pm 0.06$ | $-1.03^{+0.28}_{-0.51}$ | $-1.12^{+0.32}_{-0.52}$ | $199.8^{+4.1}_{-4.5}$ | $0.78^{+0.18}_{-0.17}$ |
| | | dynesty | xs-ring | $43.3^{+0.79}_{-0.86}$ | $0.50 \pm 0.06$ | $-0.97^{+0.26}_{-0.50}$ | $-1.89^{+0.47}_{-0.53}$ | $194.3^{+4.1}_{-4.6}$ | $\cdots$ |
| | LO | THEMIS | xs-ringauss | $43.6^{+1.50}_{-1.92}$ | $0.41 \pm 0.06$ | $-1.08^{+0.32}_{-0.60}$ | $-1.57^{+0.42}_{-0.53}$ | $204.2^{+4.4}_{-4.5}$ | $0.71^{+0.20}_{-0.15}$ |
| | | dynesty | xs-ring | $44.4^{+0.84}_{-0.87}$ | $0.41^{+0.08}_{-0.07}$ | $-1.02^{+0.26}_{-0.48}$ | $-1.87^{+0.42}_{-0.57}$ | $204.0^{+3.9}_{-4.2}$ | $\cdots$ |
| | HI+LO | THEMIS | xs-ringauss | $43.9^{+0.69}_{-0.86}$ | $0.42 \pm 0.05$ | $-1.16^{+0.29}_{-0.54}$ | $-1.52^{+0.40}_{-0.60}$ | $203.1^{+3.0}_{-3.4}$ | $0.69^{+0.17}_{-0.14}$ |
| April 11 | HI | THEMIS | xs-ringauss | $41.8^{+0.46}_{-0.43}$ | $0.35 \pm 0.04$ | $-1.41^{+0.31}_{-0.51}$ | $-1.38^{+0.32}_{-0.52}$ | $207.4^{+1.8}_{-1.9}$ | $0.50 \pm 0.03$ |
| | | dynesty | xs-ring | $43.4^{+0.74}_{-0.68}$ | $0.44^{+0.05}_{-0.06}$ | $-1.00^{+0.22}_{-0.48}$ | $-1.69^{+0.37}_{-0.49}$ | $180.1^{+2.3}_{-1.8}$ | $\cdots$ |
| | LO | THEMIS | xs-ringauss | $42.2^{+0.43}_{-0.41}$ | $0.35^{+0.05}_{-0.04}$ | $-1.27^{+0.26}_{-0.51}$ | $-1.69^{+0.41}_{-0.61}$ | $201.1^{+2.6}_{-2.5}$ | $0.50^{+0.04}_{-0.03}$ |
| | | dynesty | xs-ring | $41.6^{+0.51}_{-0.46}$ | $0.50^{+0.04}_{-0.05}$ | $-0.95^{+0.17}_{-0.42}$ | $-1.69^{+0.36}_{-0.59}$ | $175.9^{+2.1}_{-2.0}$ | $\cdots$ |
| | HI+LO | THEMIS | xs-ringauss | $42.4^{+0.34}_{-0.33}$ | $0.34 \pm 0.04$ | $-1.35^{+0.27}_{-0.48}$ | $-1.80^{+0.47}_{-0.62}$ | $198.1^{+1.9}_{-1.8}$ | $0.49 \pm 0.03$ |
| Observational uncertainty | | THEMIS | xs-ringauss | $^{+2.51\%}_{-1.66\%}$ | $^{+30.1\%}_{-47.1\%}$ | $^{+0.41}_{-0.57}$ | $^{+0.47}_{-0.71}$ | $^{+24.8}_{-25.9}$ | $^{+57.9\%}_{-35.2\%}$ |
| | | dynesty | xs-ring | $^{+1.75\%}_{-1.69\%}$ | $^{+21.3\%}_{-22.1\%}$ | $^{+0.21}_{-0.46}$ | $^{+0.46}_{-0.58}$ | $^{+21.9}_{-11.4}$ | $\cdots$ |

**Note.** Observational uncertainties are listed at the bottom of the table and have been determined as described in Appendix D.3. Note that recovery of the total compact flux density $\hat{CF}$ is not possible for the `dynesty`-based fits, which only make use of closure quantities.

observational systematics (see Appendix D.3) place a more realistic uncertainty of ~2% on this measurement. This more conservative uncertainty value is consistent with the magnitude of the scatter observed between measurements across different days and bands.

The fractional width $\hat{f}_w$ of the crescent is considerably less well constrained than the diameter, with a scatter between data sets that is comparable to the magnitude of individual measurements. Furthermore, we see systematically larger fractional width measurements from the xs-ring model than from the xs-ringauss model (see Figure 7). Some of this offset is expected from differences in model specifications. A larger contributor to the discrepancy is likely to be the different data products being fit by the two models. For the data sets in which the xs-ring model prefers a significantly larger $\hat{f}_w$ than the xs-ringauss model, we find that the smoothing kernel (described by $\hat{s}$) for the xs-ring model is also systematically wider. This smoothing by a Gaussian has no effect on the modeled closure phases. The additional gain amplitude degrees of freedom permitted in the xs-ringauss model fits can thus compensate for the smoothing in a manner that is not possible for the xs-ring model, which is fit only to closure amplitudes. Such smoothing affects the inferred diameter in a correlated manner that is well understood (see Section 7 and Paper IV, their Appendix G).

In both models, and across all data sets, we consistently measure a value for $\hat{f}_w$ that is significantly smaller than unity. This rules out a filled-in disk structure at high confidence. We find instead that the emission must be concentrated in a relatively thin annulus, with a fractional width of $\lesssim 0.5$, indicating the presence of a central flux depression. The brightness ratio $\hat{f}_c$ in this hole is also well constrained: we consistently measure $\hat{f}_c \lesssim 0.1$ (and often $\hat{f}_c \ll 0.1$). This value corresponds to a brightness contrast between the crescent and hole of at least a factor of 10.

We find a sharpness $\hat{s} \lesssim 0.1$, indicating that the smoothing kernel stays smaller than ~20% of its diameter. The inner and outer edges of the crescent are therefore well defined, even if their locations are uncertain due to the large uncertainty in the width measurement.

We find that the crescent position angle $\hat{\phi}$ consistently confines the brightest portion of the crescent to be located in the southern half. We see some evidence for a net shift in orientation from ~150°–160° (southeast) on April 5–6 to ~180°–200° (south/southwest) on April 10–11, which amounts to a difference of ~20–50 degrees between the two pairs of days. The direction of this shift is consistent with structural changes seen in the images from Paper IV, though the magnitude is a factor of ~2 larger.

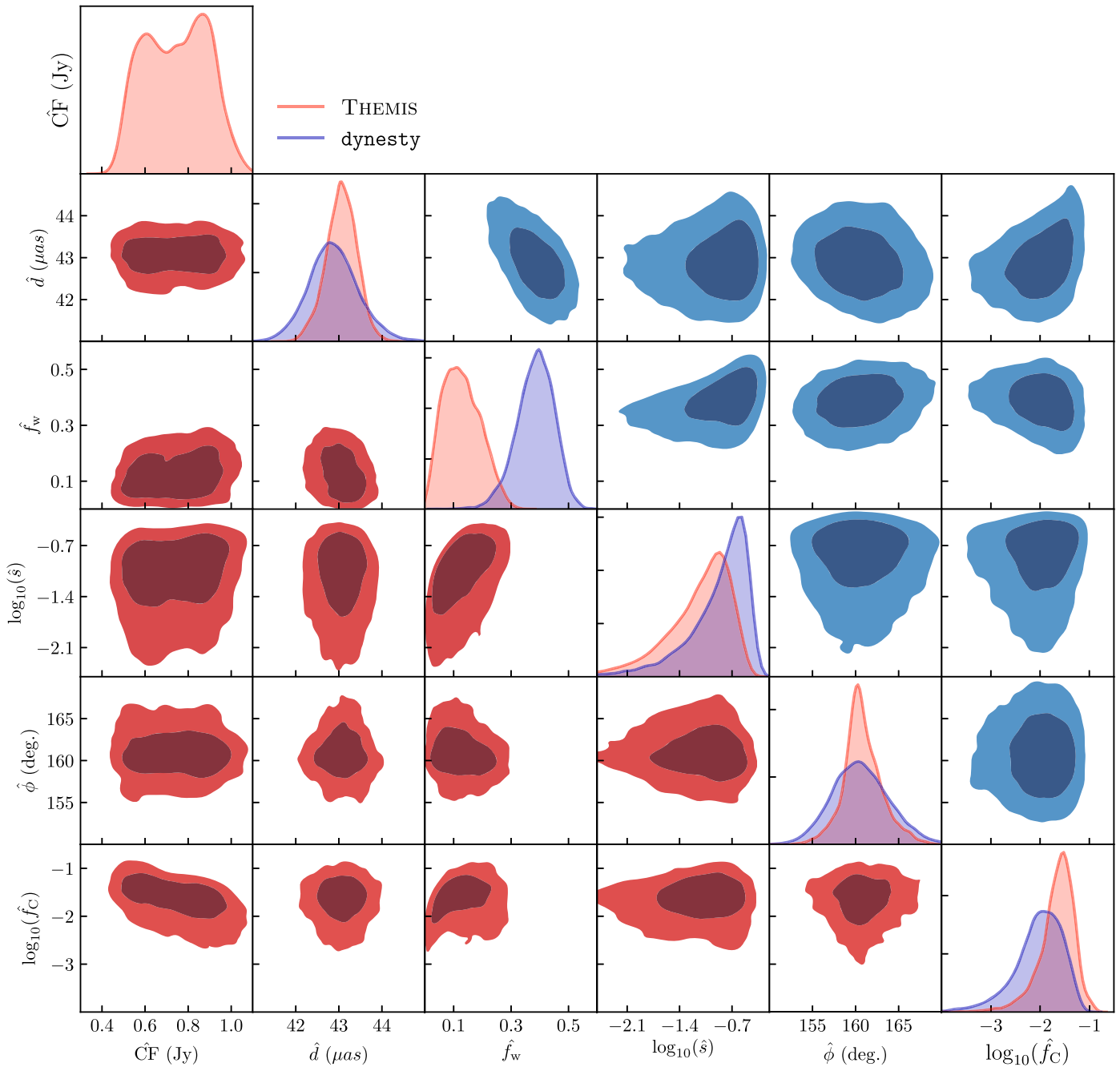

**Figure 6.** Joint posteriors for the key physical parameters derived from GC model fits to the April 5, high-band data set. Blue contours (upper-right triangle) show xs-ring posteriors obtained from the `dynesty`-based fitting scheme, while red contours (lower-left triangle) show xs-ringauss posteriors obtained using THEMIS. Contours enclose 68% and 95% of the posterior probability. Note that recovery of the total compact flux density $\hat{C}F$ is not possible for the `dynesty`-based fits, which only make use of closure quantities.

The xs-ringauss model fits find a typical compact flux density value of $\hat{C}F \approx 0.75$ Jy. The inter-day measurement scatter is at the ~50% level, which is consistent with the expected magnitude of the observational uncertainties as predicted by synthetic data in Appendix D.3. We find that the modeled $\hat{C}F$ value is less well-constrained than, but in good agreement with, the $0.66^{+0.16}_{-0.10}$ Jy determined in Paper IV from consideration of both EHT and multi-wavelength constraints.

### 5.3. Calibrating the Crescent Diameter to a Physical Scale Using GRMHD Simulations

Though the GC models have been constructed to fit the M87 data well, the geometric parameters describing these models do not directly correspond to any physical quantities governing the underlying emission. Our primary physical parameter of interest is the angular size corresponding to one gravitational radius,

$$\theta_g = \frac{GM}{c^2 D}. \qquad (26)$$

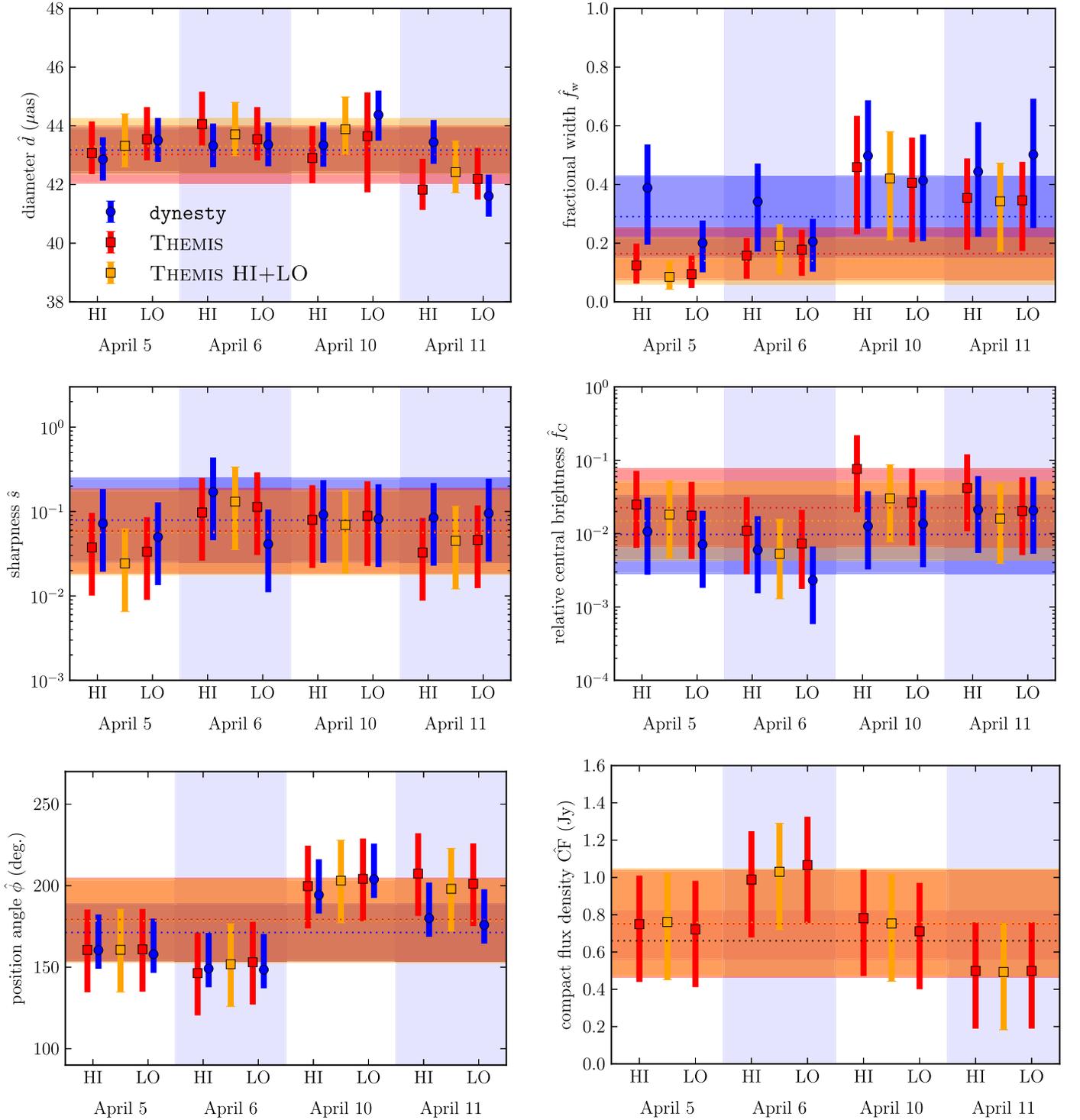

**Figure 7.** Posterior medians and 68% confidence intervals for selected parameters derived from GC model fitting for all observing days and bands. Blue circular points indicate xs-ring fits using the `dynesty`-based fitting scheme applied to individual data sets (i.e., a single band on a single day). Red square points indicate xs-ringauss fits using THEMIS applied to individual data sets, while orange square points show THEMIS-based xs-ringauss fits to data sets that have been band-combined. All plotted error bars include the systematic "observational uncertainties" estimated from simulated data in Appendix D; these uncertainties are listed in the bottom row of Table 3. Note that recovery of the total compact flux density $\hat{CF}$ is not possible for the `dynesty`-based fits, which use only closure quantities. The light purple band in the lower-right panel is the range inferred in Paper IV.

The gravitational radius sets the physical length scale of the emission region. Most of the observed 230 GHz emission is expected to originate near the photon ring (see, e.g., Dexter et al. 2012), whose scaling with $\theta_g$ is known for a given black hole mass and spin $a_*$ (Bardeen 1973; Chandrasekhar 1983). The crescent component in the GC models does not necessarily correspond to the photon ring itself. If, however, the crescent component is formed by lensed emission near the horizon, then

**Table 4**
Calibrated Scaling Factors, $\alpha$, and Corresponding $\theta_g$ Measurements

| Model | $\hat{d}$ ($\mu$as) | Calibration | $\alpha$ | $\theta_g$ ($\mu$as) | $\sigma_{\rm stat}$ ($\mu$as) | $\sigma_{\rm obs}$ ($\mu$as) | $\sigma_{\rm thy}$ ($\mu$as) |
|---|---|---|---|---|---|---|---|
| xs-ring | 43.2 | MAD+SANE | 11.56 | 3.74 | (+0.064, −0.063) | (+0.064, −0.069) | (+0.42, −0.43) |
| | | MAD only | 11.13 | 3.88 | (+0.057, −0.055) | (+0.050, −0.060) | (+0.32, −0.25) |
| | | SANE only | 12.06 | 3.58 | (+0.073, −0.073) | (+0.089, −0.096) | (+0.44, −0.51) |
| xs-ringauss | 43.4 | MAD+SANE | 11.35 | 3.82 | (+0.038, −0.038) | (+0.078, −0.077) | (+0.44, −0.36) |
| | | MAD only | 11.01 | 3.94 | (+0.040, −0.039) | (+0.092, −0.10) | (+0.25, −0.24) |
| | | SANE only | 11.93 | 3.64 | (+0.036, −0.036) | (+0.061, −0.050) | (+0.54, −0.55) |

**Note.** We use the angular diameter measurements ($\hat{d}$) from Section 5.2 and combine them with the calibrated scaling factors ($\alpha$) from Section 5.3 to arrive at our measurements of $\theta_g$. We list scaling factors calibrated using both magnetically arrested disks (MAD) and standard and normal evolution (SANE) GRMHD simulations (see Section 6.1), as well as ones calibrated using only MAD and only SANE simulations; for the final measurement presented in the text we have used the MAD +SANE calibration. The various uncertainty components are described in Appendix D.2. We quote median values for all measurements and the associated 68% confidence intervals for the different categories of uncertainty.

its size should obey a similar scaling with $\theta_g$,

$$\hat{d} = \alpha\theta_g. \tag{27}$$

For emission at the photon ring, $\alpha \simeq 9.6$–$10.4$ depending on the black hole spin parameter. For a realistic source model, the emission is not restricted to lie exactly at the photon ring. The value of this scaling factor $\alpha$ and its uncertainty are therefore unknown a priori.

We have measured $\alpha$ and its uncertainty for both GC models using a suite of synthetic data sets generated from snapshots of GRMHD simulations from the GRMHD image library (Paper V). The full calibration procedure, including properties of the selected GRMHD simulations, the synthetic data generation process, and the calibration uncertainty quantification, is detailed in Appendix D. By fitting each calibration image with a GC model, and then comparing the corresponding $\hat{d}$ measurement to the input value of $\theta_g$ for the simulation that produced the image, we determine the value of $\alpha$ for that image. Combining the results of such fits from a large number of GRMHD simulations yields a calibration (and uncertainty) for $\alpha$. For the xs-ring model we find a mean value of $\alpha = 11.55$, while for the xs-ringauss model we find a nearly identical $\alpha = 11.50$. Both of these values are somewhat larger than the $\alpha \approx 10$ expected for the photon ring itself, indicating that the GC models are accounting for emission in the GRMHD model images that preferentially falls outside of the photon ring.

Figure 8 shows the $\theta_g$ values obtained as a result of applying our calibrated scaling factor to the crescent diameter values measured for each day and band, and Table 4 lists the results from combining the measurements from all data sets. There is excellent agreement between the two GC models, resulting in an averaged value of $\theta_g = 3.77^{+0.45}_{-0.40}$ $\mu$as. We note that the 12% uncertainty in the $\theta_g$ measurement is dominated by the diversity of GRMHD models used in the primary calibration; the quantification of this "theoretical" uncertainty component from the GRMHD simulations is described in Appendix D.2.

### 6. Direct Comparison with GRMHD Models

EHT data have the power to directly constrain GRMHD simulation based models of M87 and to estimate the physical properties of the black hole and emitting plasma. Such a direct comparison is challenging due to stochastic structure in the models.

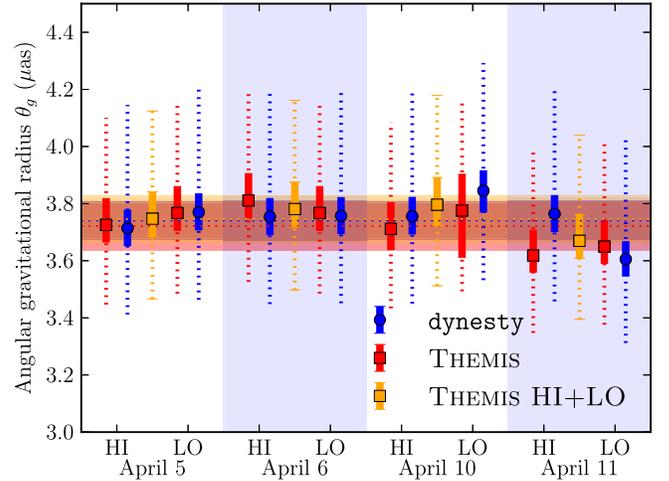

**Figure 8.** Constraints on $\theta_g$ arising from the GRMHD simulation calibrated GC model fits, by day and band. Solid error bars indicate 68% confidence intervals, while dashed error bars indicate the systematic uncertainty in the calibration procedure. Circular blue points indicate independent analyses for each band and on each day in the context of the xs-ring GC model fit using the dynesty-based method. Square points indicate independent analyses for each band (red) and band-combined analyses (orange) for each day using the xs-ringauss GC model with THEMIS. Colored bands around dashed lines (right) indicate the combined constraint across both bands and all days, neglecting the systematic uncertainty in the calibration procedure.

The EHT 2017 data span a very short time frame for the M87 source structure. Its characteristic variability timescale at 230 GHz is $\simeq$50 days (Bower et al. 2015), much longer than our observing run. Thus, although the entire ensemble of model snapshots taken from a given simulation captures both the persistent structure as well as the statistics of the stochastic components, we do not yet have enough time coverage for M87 itself to measure its structural variations.

Model images from GRMHD simulations show a dominant, compact, asymmetric ring structure resulting from strong gravitational lensing and relativistic gas motions (Paper V). Hence, they capture the qualitative features found by image reconstructions in Paper IV and by geometric crescent models in Section 5. This motivates a direct comparison of the GRMHD model images with the EHT data.

In this section we summarize the GRMHD image library (Section 6.1) and fit individual simulation snapshot images (Sections 6.2 and 6.3) in a similar fashion to past work on GRMHD fitting to mm-VLBI data of Sgr A* (e.g., Dexter et al. 2010; Kim et al. 2016). We further develop and apply a method for testing the consistency of the M87 data with the simulation models (Section 6.4) and find that the majority of the simulation library models is consistent with the data. We use the results to estimate physical parameters, including the black hole angular radius $GM/Dc^2$ (Section 6.5). The implications of our results for the physical properties of the emission region are discussed in more detail in Paper V. From EHT data alone it is difficult to rule out many of the broad range of possible models for the black hole and plasma properties. However, in combination with other data (especially the observed jet power), ultimately more than half of the models can be excluded.

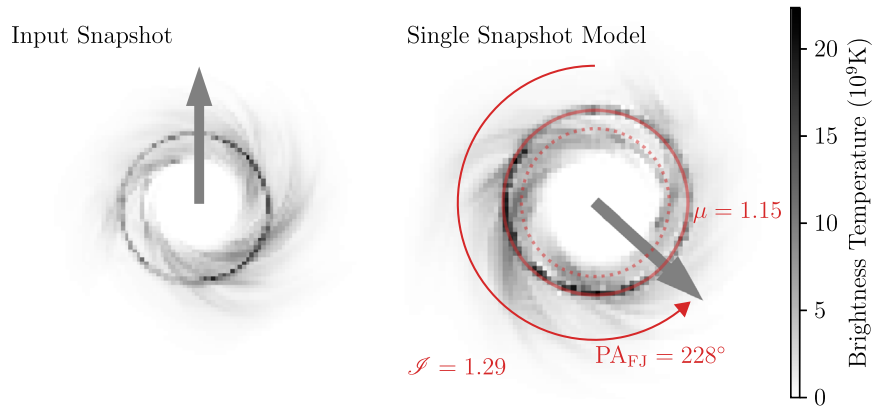

**Figure 9.** Illustration of the parameters of the SSM described in Section 6.2. Both the original GRMHD simulation (left) and the corresponding SSM for an arbitrary set of parameter values, (flux rescaling, stretching of the image, and rotation; right) are shown. In both panels, the gray arrow indicates the orientation of the forward jet.

### 6.1. Summary of Simulations

As described in detail in Paper V, we have constructed a large image library of horizon-scale synchrotron emission images at 230 GHz computed from GRMHD simulations. We summarize the broad features of this library here and direct the reader to Paper V for more information. The GRMHD simulations cover a wide range of black hole spins as well as initial magnetic field geometries and fluxes. These result in images associated with a variety of accretion flow morphologies and degrees of variability. The magnetic flux controls the structure of the accretion flow near the black hole. Low magnetic fluxes produce the standard and normal evolution (SANE) disks characterized by low-efficiency jet production. In contrast, magnetically arrested disks (MAD) are characterized by large magnetic fluxes, set by the ram pressure of the confining accretion flow.

From these models, families of between 100 and 500 snapshot images were produced assuming synchrotron emission from an underlying thermal electron population (see Section 3.2 of Paper V). The snapshot image generation introduces additional astrophysical parameters associated with the intrinsic scales in the radiative transfer. These parameters include the black hole mass, the viewing inclination, $i$, and a model for the electron thermodynamics: $T_i/T_e \approx R_{high}$ in gas-pressure dominated regions and is unity otherwise (Mościbrodzka et al. 2016), where $T_i$ and $T_e$ are the ion and electron temperatures.

The number density of emitting electrons is scaled independently for each simulation such that the typical 230 GHz flux density is ~0.5 Jy. The temporal separation between snapshots is selected such that adjacent snapshots are weakly correlated.

### 6.2. Single Snapshot Model (SSM)

Each snapshot image generates a three-parameter SSM defined by the total compact flux (CF), angular scale ($\theta_g$), and orientation (defined to be the position angle of the forward jet measured east of north, $PA_{FJ}$). Variations in these parameters approximately correspond to variations in the accretion rate, black hole mass, and orientation of the black hole spin, respectively. Variations in mass are associated with changes in the diameter of the photon ring, a generic feature found across all of the images in the GRMHD image library.

Each snapshot image is characterized by a nominally scaled, normalized intensity map of the image, $\hat{I}(x, y)$, with a corresponding nominal total intensity $\hat{CF}$, gravitational radius $\hat{\theta}_g$, and forward jet position angle $PA_{FJ} = 0°$; associated with the intensity map are complex visibilities $\hat{V}(u, v)$. The SSM is then generated by rescaling, stretching, and rotating $\hat{V}(u, v)$:

$$V_{SSM}(u, v; \mathscr{I}, \mu, PA_{FJ}) = \mathscr{I}\hat{V}(\mu u', \mu v') \qquad (28)$$

where $\mathscr{I} \equiv CF/\hat{CF}$, $\mu \equiv \theta_g/\hat{\theta}_g$, and $(u', v')$ are counter-rotated from $(u, v)$ by the angle $PA_{FJ}$. This procedure is illustrated in Figure 9.

We show in Paper V that these approximations generally hold for flux and mass for rescalings by factors of $\lesssim 2$ from their fiducial values.

### 6.3. Fitting Single Snapshots to EHT Data

For both model selection and parameter estimation, the first step is fitting an SSM to the EHT data set described in detail in Section 2.1. The only difference here is that intra-site baselines are excluded. These probe angular scales between $0''.1$ and $10''$, at which unmodeled large-scale features, e.g., HST-1, contribute substantially (see Section 4 of Paper IV). We verify after the fact that the reconstructed compact flux estimates are consistent with the upper limits necessarily implied by these baselines. The fitting process is complicated by large structural variations between snapshots resulting from turbulence in the simulations (see Section 6.4).

We employ two independent methods to fit SSMs to the EHT data sets. Both employ the likelihoods constructed as described in Section 4.1 for visibility amplitudes and closure phases. The two methods, THEMIS and GENA, are utilized as described in Section 4, producing posterior estimates or best-fit estimates, respectively, for CF, $\theta_g$, and PA$_{FJ}$.

We adopt a uniform prior on the total compact flux density between 0.1 and 10 Jy for THEMIS and between 0.1 Jy and 4 Jy for GENA; both of these priors cover ranges that substantially exceed the limits placed by Paper IV. While the position angle (PA) of the large-scale radio jet in M87 is well known, we permit the PA of the horizon-scale jet to be unconstrained, setting a uniform prior from [0°, 360°). Finally, we place a flat prior on $\theta_g$ ranging from 0.1 to 100 $\mu$as for THEMIS and between 0 and 10 $\mu$as for GENA, again substantially exceeding the physically relevant ranges.

### 6.4. Model Selection and Average Image Scoring (AIS)

Quantitatively assessing the quality of SSM fits presents unique challenges because of the presence of stochastic image features due to turbulence in the underlying GRMHD simulations that produce large variations in image structure. The structural variability leads to changes in EHT observables that are much larger than the observational errors. It is therefore not feasible to generate a sufficient number of images from existing GRMHD simulations to have a significant probability of finding a formally adequate (i.e., $\chi^2 \approx 1$) fit to the data (see the discussion in Paper V). Nevertheless, the ensemble of snapshots from a given simulation provides a natural way to characterize the impact of these stochastic features on the inferred SSM parameters.

We can thus assess individual GRMHD simulations, effectively comprised of many (100–500) snapshot images, by comparing the quality of SSM fits to a numerically constructed distribution of reduced $\chi^2$ values. Note that this procedure is conceptually identical to the normal fitting procedure, in which a $\chi^2$ is interpreted relative to the standard $\chi^2$-distribution and thus a reduced $\chi^2 \approx 1$ is a "good" fit, with the exception that an additional source of noise has been effectively introduced as part of the underlying model, altering the anticipated distribution of $\chi^2$ values. In practice, this comparison is executed via the AIS method described in Appendix F using the THEMIS SSM fitting pipeline.

We estimate the appropriate $\chi^2$-distribution by performing multiple fits of an SSM constructed from the arithmetic average snapshot image to simulated data generated from each snapshot image within the underlying GRMHD model. Finally, the average snapshot image is compared to the EHT data, and the resulting $\chi^2$ is assessed. Thus, THEMIS-AIS is effectively determining if the EHT data are consistent with being drawn from the GRMHD simulation. The result is characterized by a simulation $p$-value, which we call $p_{AIS}$. Because of the significant variations in data quality and baseline coverage across days, this procedure must be repeated independently for each day.

It is possible for a model to be ruled out by the AIS procedure both by the average image SSM having a $\chi^2$ that is too high (the data is "further" from the average snapshot image than typical for the simulation) and by the average image SSM having a $\chi^2$ that is too low (the data is "closer" to the average snapshot image than typical for the simulation, usually a consequence of too much variability in the GRMHD model). These are similar to finding a reduced $\chi^2$ that is much larger

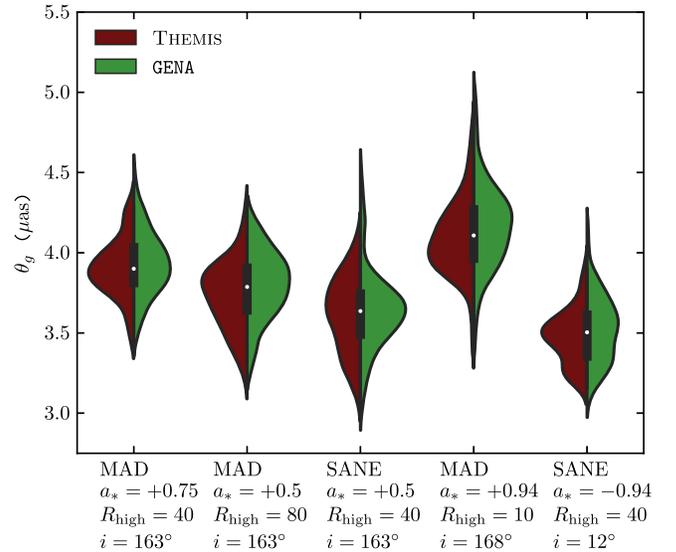

**Figure 10.** Distributions of recovered $\theta_g$ from five representative GRMHD simulations as measured by the THEMIS (left; maroon) and GENA (right; green) pipelines. Only those snapshots for which the likelihood is above the median (THEMIS), or for which the combined $\chi^2$ statistic is below the median (GENA) are included. For a wide range of simulations, the recovered $\theta_g$ are consistent with each other. All of the simulations shown are deemed acceptable by AIS ($p_{AIS} > 0.01$; see Section 6.4 for details).

and smaller than unity, respectively, in traditional fits. Both occur in practice.

The result of the THEMIS-AIS procedure is limited by the number of snapshots from each model, and thus is currently capable of excluding models only at the 99% level for a given day and band (i.e., $p_{AIS} < 0.01$). Due to the long dynamical timescale of M87—GRMHD snapshots exhibit correlations extending up to $20GM/c^3 \approx 1$ week—this is not significantly improved by combining bands and/or days. Therefore, unless otherwise noted, we set a threshold of $p_{AIS} > 0.01$, below which we deem a GRMHD model unacceptable and exclude it from consideration.

### 6.5. Ensemble-based Parameter Estimation

The posteriors for the SSM parameters for a given GRMHD model are estimated using both the observational errors in the EHT data and the stochastic fluctuations in the snapshot images themselves; among these "noise" terms, the latter significantly dominates. This is evidenced by the variations among SSM fit parameters from snapshots within a GRMHD model, which typically exceed the formal fit errors from the SSM fit alone (see Appendix G). Some care must be taken in extracting and interpreting parameter estimates from these.

Only SSM fits with likelihoods among the highest 10% within each model are used for parameter estimation. The results for $\theta_g$ are insensitive to the precise value of this fit-quality cut after it passes 50%. The measured values are consistent with the range of GRMHD models explored (see Figure 10 for five examples). The consistency is likely a result of the dominant strong lensing and relativistic motion that are common to all models.

The full posteriors are then obtained by marginalizing over all GRMHD models that are found to be acceptable via the THEMIS-AIS procedure. At this point we also include the ancillary priors on jet power, X-ray luminosity, and emission efficiency described in Paper V (see their Table 2). This procedure corresponds to simply summing the posteriors obtained for each acceptable GRMHD model.

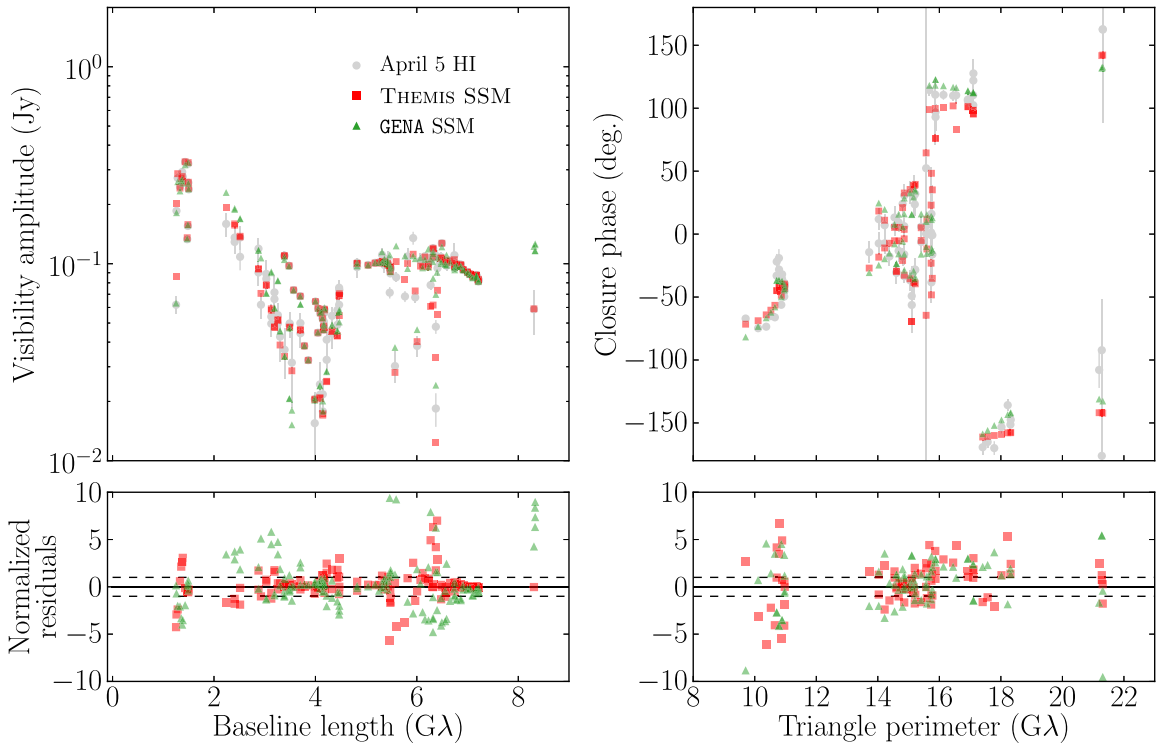

**Figure 11.** Visibility amplitude and closure phase residuals for an SSM fit to the April 5, high-band data for a "good" snapshot image frame from a MAD simulation with $a_* = 0$, $i = 167°$, and $R_{high} = 160$. The reduced-$\chi^2$ values for the fits are 5.9 (THEMIS) and 7.3 (GENA). All residuals are normalized by their corresponding estimated observational errors.

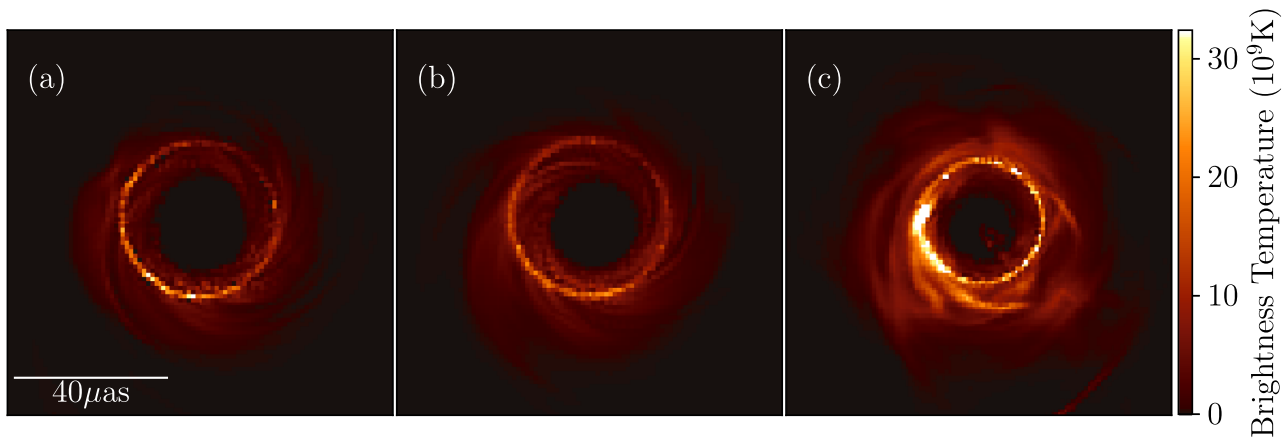

**Figure 12.** Three sample SSM fits to April 6 high-band data. All are from models with THEMIS-AIS $p_{AIS} > 0.1$. In panels (a) and (b), the prominent photon ring places a strong constraint on $\theta_g$. In panel (c), the extended disk emission results in a smaller $\theta_g$ estimate.

In principle, it is possible for the posteriors constructed in this fashion to suffer from biases induced by the treatment of the high values of $\chi^2$ found when including only observational errors. We have attempted to estimate this bias using a large number of mock analyses. In those, simulated data were generated from a snapshot within a GRMHD model. Posteriors were then generated for that model given the simulated data and compared to their known SSM parameters; no significant biases were found. We have further conducted two posterior mock analyses for the full suite of GRMHD models, independently for the THEMIS and GENA pipelines. In these the impact of including "incorrect" GRMHD models was a key difference with the previous tests. See Appendix G for more details.

The results of both of these experiments only hold if the GRMHD models used as synthetic data provide a good description of M87. As multiple observation epochs become available, it will be possible to explicitly measure the statistics of the stochastic fluctuations, empirically addressing this assumption. It will also enable direct ensemble-to-ensemble comparison like that described in Kim et al. (2016). A promising, alternative approach would be to perform a principal component analysis (PCA) decomposition of the snapshots within each model

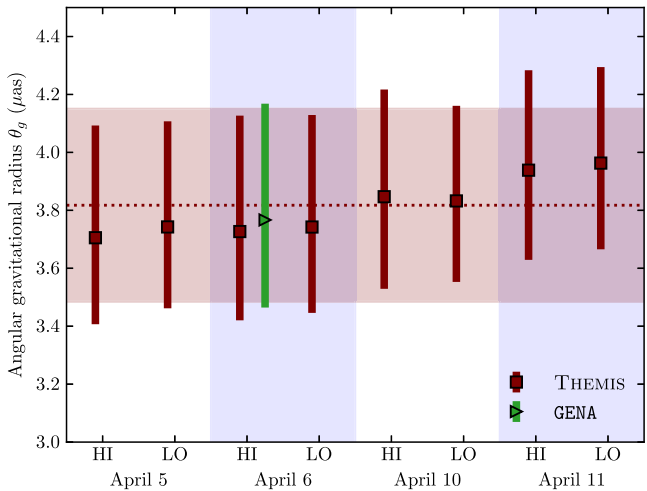

**Figure 13.** Constraints on $\theta_g$ arising from the GRMHD model fitting procedure, by day and band. The maroon squares and green triangles are the constraints arising form the THEMIS and GENA pipelines. Solid error bars indicate the 68% confidence levels about the median. The maroon colored band indicates the combined constraint across both bands and all days.

**Table 5**
Best-fit Parameters for All Data Sets from Direct GRMHD
Simulation Fitting, as Defined in Section 6.2

| Data Set | | | Parameter | | |
|---|---|---|---|---|---|
| Day | Band | Code | $\theta_g$ ($\mu$as) | CF (Jy) | PA$_{FJ}$ (deg.) |
| April 5 | HI | THEMIS | $3.71^{+0.39}_{-0.30}$ | $0.53^{+0.18}_{-0.11}$ | $233^{+35}_{-37}$ |
| | LO | THEMIS | $3.74^{+0.36}_{-0.28}$ | $0.53^{+0.16}_{-0.10}$ | $228^{+33}_{-36}$ |
| April 6 | HI | THEMIS | $3.73^{+0.40}_{-0.31}$ | $0.52^{+0.16}_{-0.10}$ | $223^{+34}_{-35}$ |
| | HI | GENA | $3.77^{+0.40}_{-0.30}$ | $0.42^{+0.08}_{-0.07}$ | $220^{+37}_{-40}$ |
| | LO | THEMIS | $3.74^{+0.39}_{-0.30}$ | $0.56^{+0.16}_{-0.11}$ | $232^{+33}_{-35}$ |
| April 10 | HI | THEMIS | $3.85^{+0.37}_{-0.32}$ | $0.55^{+0.13}_{-0.10}$ | $232^{+52}_{-62}$ |
| | LO | THEMIS | $3.83^{+0.33}_{-0.28}$ | $0.57^{+0.14}_{-0.10}$ | $238^{+49}_{-61}$ |
| April 11 | HI | THEMIS | $3.93^{+0.35}_{-0.30}$ | $0.58^{+0.14}_{-0.10}$ | $264^{+36}_{-52}$ |
| | LO | THEMIS | $3.96^{+0.33}_{-0.30}$ | $0.60^{+0.14}_{-0.10}$ | $261^{+36}_{-50}$ |

**Note.** Median posterior values are quoted with 68% confidence intervals.

(Medeiros et al. 2018), and fit images generated by varying the weights of the PCA components to the data to mimic the set of possible realizations of the turbulence.

### 6.6. M87 Fit Results

An example SSM fit is shown for both the THEMIS and GENA methods in Figure 11. As anticipated, the presence of stochastic features within the images results in a poor formal fit quality for all simulation snapshots, that is, reduced $\chi^2 \gtrsim 2$. Nevertheless, broad features of the visibilities and closure phases are accurately reproduced. These include the deep visibility amplitude minimum near 3.4 G$\lambda$ and amplitude of the following bump at 6.0 G$\lambda$, both of which are key constraints on the image structure (see Section 3). Significant structure in the residuals, as seen in Figure 11, is a natural consequence of the presence of stochastic model components, but may also signify that the underlying model is insufficient to fully explain the EHT data. Representative fits obtained via this process are shown in Figure 12 for the best

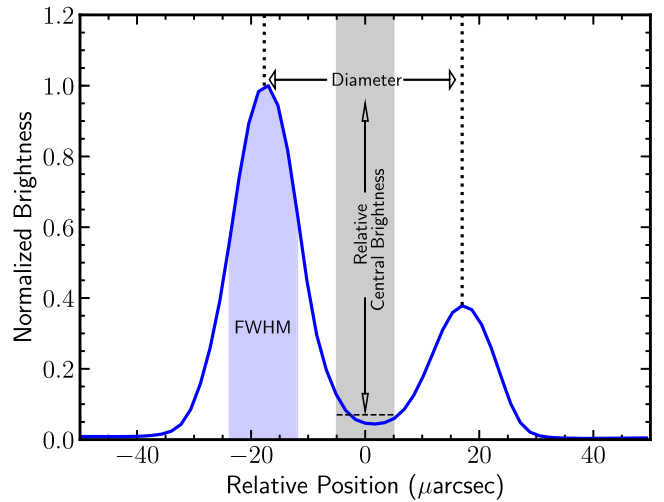

**Figure 14.** A sample image cross section showing the definitions of the image domain measures that we use in identifying and comparing features in the images (see Section 7).

SSM from a selection of models that are not ruled out by THEMIS-AIS.

Even aggressive model selection via THEMIS-AIS produces a very weak cut on the GRMHD models—upon choosing $p_{AIS} > 0.01$ and $p_{AIS} > 0.1$ only 9.7% and 28.5% of models are excluded. After applying the additional observational constraints, we discard 65.3% of the models (for details of the model selection, see Paper V). Here we focus on estimates of the compact flux density, $\theta_g$ and PA$_{FJ}$, after marginalizing over the acceptable models that remain.

The GRMHD-based SSM models produce compact flux estimates between 0.3 and 0.7 Jy. This is at the lower end of the range reported in Paper IV ($0.66^{+0.16}_{-0.10}$ Jy). The PA$_{FJ}$ obtained via comparison of the GRMHD snapshots and the $\hat{\phi}$ found from fitting the GC models are generally consistent after accounting for the average 90° offset between the location of the forward jet in the GRMHD simulation and the location of the brightest region in the crescent image (Paper V). This appears marginally consistent with the mas-scale forward-jet PA measured at 3 and 7 mm of 288° (e.g., Walker et al. 2018), though see Paper V for further discussion on this point.

The posteriors for $\theta_g$ are broadly consistent among days and bands, as illustrated in Figure 13 and Table 5. The combined value for both the THEMIS and GENA analyses is $\theta_g = 3.77^{+0.51}_{-0.54}$ $\mu$as. Finally, note that the process by which this estimate is arrived at differs qualitatively from the analysis presented in Section 5. Here GRMHD snapshot images are fit directly to data, whereas in Section 5 GRMHD snapshots were used to calibrate the geometric models. These subsets are independent—the set of images used to calibrate the geometric models are not used in the GRMHD analyses. Nevertheless, a systematic correlation between these estimates may remain as a result of the use of the same set of underlying GRMHD simulations.

### 7. Image Domain Feature Extraction

In the previous sections, we fit geometric and numerical models in the visibility domain to measure the properties of features in the models that give rise to the observed interferometric data. In Paper IV, we performed direct image reconstruction using two RML methods (eht-imaging and SMILI; Akiyama et al. 2017a, 2017b) and one CLEAN method (DIFMAP; Shepherd 1997). Here we present a first analysis of ring properties seen in reconstructed M87 images. We describe the method used to extract parameters (Section 7.1). We then show that the extracted ring diameter is consistent across imaging methods and compare the ring diameter and width with that measured from geometric modeling (Section 7.2). We convert the diameter measurements to the physical scale around the black hole, $\theta_g$, by calibrating with GRMHD simulations (Section 7.3). Finally, we measure the image circularity (Section 7.4).

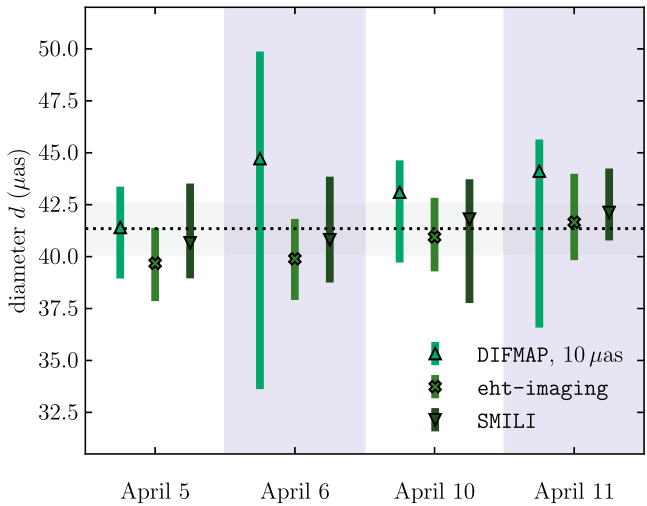

**Figure 15.** Image domain measurements of the ring diameter of M87 over all observing days comparing three image reconstruction methods. The error bars show the full range of results for the Top Set images using low-band data. The gray dashed line and band show the weighted average and uncertainty across methods and observing days.

### 7.1. Measuring the Parameters of Image Features

We describe the M87 image morphology using five quantities that are similar in spirit but not identical to those introduced in Equations (20)–(23) in Section 5.1: the diameter $d$ of the ring/crescent, its fractional width $f_w$, the relative brightness $f_c$ of its center with respect to its rim, and the PA $\phi$. Figure 14 shows a schematic diagram of these parameters. Here we focus on the first two parameters. Measurements of the remaining parameters in the various image reconstructions and more details on the method can be found in Paper IV.

The reconstructions obtained in Paper IV use a pixel size of 2 $\mu$as ($\simeq$1/10 of the EHT 2017 beam), but the GC image models considered below often show narrower structures. We interpolate both the GC models and image reconstructions onto a grid with pixel size of 0.5 $\mu$as before applying the feature extraction methods. When necessary, we use a 2D linear interpolation between adjacent pixels for finer sampling.

The radial brightness profile is characterized by a peaked distribution that declines toward the center. We first locate the (arbitrary) center of the ring. We measure the position of the radial brightness profile maximum along different azimuthal angles. For a trial center position, we calculate the dispersion of the radii defined by these maxima. The center is then chosen iteratively as the location that minimizes this dispersion. Once a center position is chosen, the mean diameter $d$ of the image is defined as twice the distance to the peak, averaged over azimuth.

We define the ring width to be the FWHM of the bright region along each radial profile. The fractional width $f_w$ of the image is the average of the FWHM over azimuthal directions divided by its mean diameter. In the following we first smooth the image with a 2 $\mu$as Gaussian. In Appendix H we show that variations on this method produce small (sub-pixel) differences in the results.

### 7.2. M87 Image Ring Diameters and Widths

Figure 15 shows the mean diameters of images reconstructed using the low-band data during all four days of M87 observations with three image reconstruction methods (eht-imaging, SMILI, DIFMAP; see Paper IV). Image samples are reconstructed for each data set and for a wide range of weights of the regularizers ($\sim$2000 images for eht-imaging and SMILI and $\sim$30 for DIFMAP, see Paper IV). Diameters measured from the full set of images that produce acceptable fits to the visibility data (the "Top Set") are shown here. The diameters of the ring features found across all methods and all days span the narrow range $\sim$38–44 $\mu$as.

Figure 16 shows the fractional width versus mean diameter for the Top Set images from eht-imaging and SMILI compared with those from geometric models for the low-band M87 data of April 6 (Section 5). For the two visibility domain methods, the points correspond to the diameters and widths obtained from a sample of 100 images from the xs-ring and xs-ringauss model, chosen randomly from their posteriors. Those model images have been analyzed in the same way as the reconstructed ones.

The fractional widths for the reconstructed images are $\lesssim$0.5, which are consistent with the results of geometric crescent models in Section 5. While both the image reconstructions and model fits show a large uncertainty in fractional width, the widths measured from the image reconstructions in the Top Sets are $\gtrsim$10 $\mu$as. Images of rings with narrower widths are consistent with the data and can be produced by the imaging algorithms; however, the parameters in the imaging algorithms (e.g., regularizer weights) that determine the Top Set images were trained on synthetic data from sources smoothed with a 10 $\mu$as beam (Paper IV, their Section 6.1). This may help explain the differences in the fractional widths measured with the different techniques.

An anti-correlation between diameter and fractional width for these image domain results is clearly present in three of the four days, but is less clear in the data of April 11. We consider two manifestations for this anti-correlation, based on the location of the first visibility amplitude minimum in simple ring models. First, we use the geometric crescent model of Kamruddin & Dexter (2013), and derive the following approximate fitting formula to the exact expression relating the mean ring diameter $d$ and $\psi$:

$$d \simeq d_0 \frac{1 - \psi/2}{1 - 0.48\psi + 0.11\psi^5}, \quad (29)$$

where $d_0$ is the diameter of an infinitesimally thin ring that produces a null at baseline length $b_1$ (Equation (6)). The fractional width is $\hat{f}_w = \psi/(2 - \psi)$. The blue shaded region in Figure 16 shows the expected diameter and fractional width anti-correlation for a visibility minimum occurring at a range of baseline lengths 3.5–4.0 G$\lambda$, similar to that seen in

**Table 6**
Measured Diameters, $d$, Calibrated Scaling Factors, $\alpha$, and Corresponding Gravitational Radii, $\theta_g$, for the Image Domain Analysis Presented in Section 7

| Imaging Method | $d$ ($\mu$as) | $\sigma_d$ ($\mu$as) | $\alpha$ | $\theta_g$ ($\mu$as) | $\sigma_{obs}$ ($\mu$as) | $\sigma_{thy}$ ($\mu$as) |
|---|---|---|---|---|---|---|
| `eht-imaging` | 40.5 | 0.5 | 10.67 | 3.79 | (+0.06, −0.06) | (+0.42, −0.37) |
| `SMILI` | 41.5 | 0.4 | 10.86 | 3.82 | (+0.04, −0.05) | (+0.40, −0.38) |
| `DIFMAP` | 42.5 | 0.8 | 11.01 | 3.84 | (+0.09, −0.10) | (+0.42, −0.32) |

**Note.** We quote median values and 68% confidence intervals. The angular diameters $d$ are averages across the four observing days, weighted by the range within each day. Their uncertainties $\sigma_d$ are the standard deviation of the mean over the four days. We combine the angular diameter measurements with the calibrated scaling factors ($\alpha$) to arrive at measurements of $\theta_g$. The various uncertainty components of $\theta_g$ are obtained using synthetic data, as described in Appendix E.

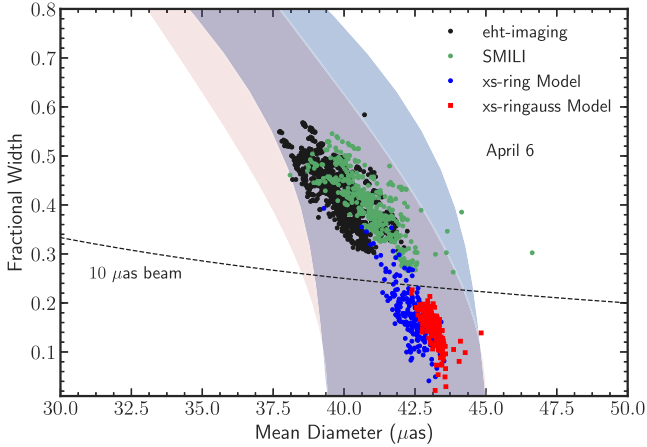

**Figure 16.** Diameters and fractional widths inferred from image (black and green) and visibility (red and blue) domain measurements on April 6. The visibility domain measurements are from GC model fitting (see Section 5), while the reconstructed images are from Paper IV. The filled regions show diameter and width anti-correlations expected in simple ring models (Section 7.2). The anti-correlation between mean diameter and width helps to explain the small (≃5%) offset in mean diameter found between methods.

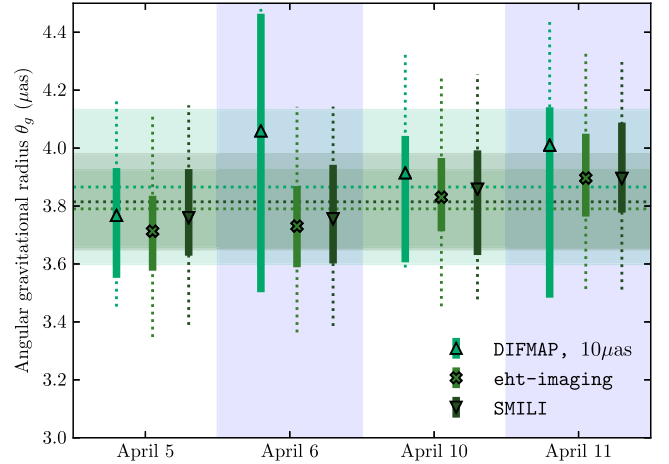

**Figure 17.** Constraints on $\theta_g$ from GRMHD simulation calibrated image domain feature extraction, by day and band. The solid lines show the full range of diameter measurements from the Top Set images. The dotted lines show the systematic calibration uncertainty. The shaded regions show the weighted average and uncertainty over observing days for each method.

the M87 data (Section 3, Paper III). This corresponds to $d_0 \simeq 39.5$–45 $\mu$as.

We repeat this exercise for an infinitesimally thin ring convolved with a Gaussian. Here the mean diameter $d$ can be approximated for FWHM $w \ll d$ as (Appendix G of Paper IV):

$$d \simeq d_0 \left(1 - \frac{w^2}{4 \ln 2 d_0^2}\right). \tag{30}$$

Here the fractional width $\hat{f}_w = w/d$. The pink shaded area in Figure 16 shows the shaded region for this model, which follows a similar trend.

The measured properties of the images and source models inferred by all methods generally fall within the expected bands. At least part of the systematic differences in our diameter measurements may be attributed to the relatively large uncertainty in width, as a result of their weak anti-correlation.

### 7.3. From Image Diameter to Angular Gravitational Radius

We can convert the diameters measured from the reconstructed images to the black hole angular gravitational radius using a scaling factor, $\alpha$ (Equation (27)), following the same procedure used in the calibration of crescent model diameters to GRMHD images discussed in Section 5.3. We calculate a separate value of $\alpha$ for each image reconstruction method (`eht-imaging`, `SMILI`, `DIFMAP`) as described in

Appendix E and listed in Table 6. The image domain methods do not report posteriors. We therefore estimate the statistical along with the observational component of the uncertainty in the calibration procedure using synthetic data. As found for the geometric crescent models, the theoretical uncertainty dominates the final error budget. The total observational uncertainty is similar to but slightly larger than the statistical spread of the median diameter measurement between days.

After applying the calibrated scaling factor to the image domain diameter measurements, we find consistent results across all observing days and reconstruction methods (Figure 17). This is further indication that the statistical component of the calibration error is sub-dominant. The combined value from all methods is $\theta_g = 3.82^{+0.42}_{-0.38}$ $\mu$as. Despite small differences in the measured mean diameters, the physical scale $\theta_g$ is remarkably consistent with that found earlier from both geometric and GRMHD model fitting. This is because the corresponding calibration factors $\alpha$ obtained from synthetic data show the same trends as the measured diameters.

### 7.4. The Circular Shapes of the M87 Images

Our reconstructed M87 images appear circular. We quantify their circularity by measuring the fractional spread in the inferred diameters measured along different orientations for each of the reconstructed images. Here we define the fractional spread as the standard deviation of the diameters measured along different orientations divided by the mean diameter. For each image, the diameters along different orientations measure the distance of the location of peak brightness from the center of the image along each orientation. A small fractional spread suggests that the locus of peak brightness on each image has a circular shape.

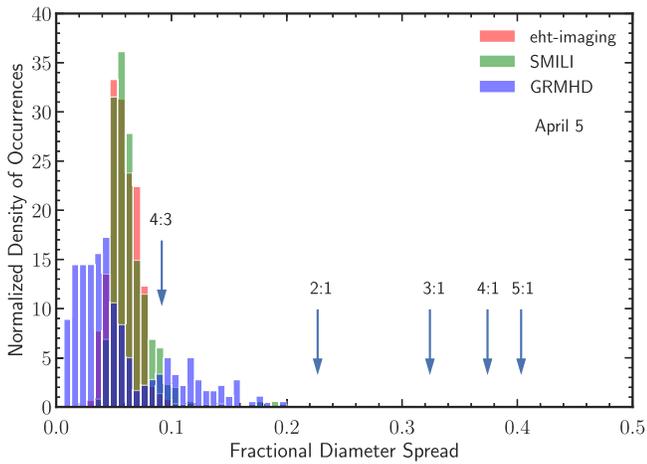

**Figure 18.** Distributions of fractional spreads in the inferred diameters of the M87 images measured along different orientations, for images reconstructed using two methods for the April 5 data set. The blue histogram is the distribution of fractional spreads for a subset of images from the GRMHD library. The values for elliptical shapes with different axes ratios are indicated with arrows (a value of 0 is an axis ratio of 1:1). The distribution for reconstructed images peaks at values ∼0.05–0.06, which is consistent with the range found from GRMHD model images.

Figure 18 shows the distribution of the fractional spread in diameters for the Top Set images of M87 reconstructed using two image domain methods (eht-imaging and SMILI) for the April 5 low-band data set. The figure also shows the fractional spread in diameters of a subset of ∼300 images among those in the GRMHD image library discussed in Paper V that provide acceptable fits according to AIS. Note that these comparison images have a much higher intrinsic resolution than the reconstructed images of M87. The distributions of fractional spreads for the reconstructed images peak at a fractional spread of ∼0.05–0.06. This is within the range found in the GRMHD images. The GRMHD models mostly show circular image structure (peak ≲0.1 in blue distribution in Figure 18). This corresponds to the bulk of the images where the photon ring is the dominant feature (see Paper V). The tail to larger fractional diameter spread may in part be the result of occasional bright filaments that light up in a small azimuthal range and then quickly fade. During such snapshots, the locus of points with the brightest emission around the shadow becomes non-circular and the fractional spread increases. Arrows in Figure 18 indicate the expected fractional diameter spread for elliptical models. Our results suggest an image axial ratio of ≲4:3.

## 8. Prior Mass and Distance Estimates

In this section we briefly review prior estimates of the mass-to-distance ratio for M87's central black hole for comparison with our new EHT results. For a more detailed discussion, see Appendix I.

### 8.1. Distance

M87 is the central galaxy of the Virgo cluster, the closest galaxy cluster to the Milky Way. This proximity allows a high-precision measurement of its distance using primary or secondary standard candles available only for nearby galaxies within ∼20 Mpc. Among numerous distance estimators obtained for M87, we base the distance that we use here on two estimators of the absolute distance modulus ($\mu$). The first uses as a primary candle the tip of the red giant branch (TRGB method). The second uses a well-calibrated secondary standard candle, the surface brightness fluctuations (SBF method) from resolved stellar distributions observed at high angular resolution.

We combine one TRGB measurement (Bird et al. 2010) with two SBF measurements (Blakeslee et al. 2009; Cantiello et al. 2018a) to arrive at a combined distance measurement. We consider the three distance measurements to be independent and calculate the combined posterior as their product. Based on these, we adopt a distance of $D = 16.8^{+0.8}_{-0.7}$ Mpc to M87. A more detailed description, together with the summary of each measurement, can be found in Appendix I.

### 8.2. Mass-to-distance Ratio and Related Quantities

For the mass of the central black hole in M87 we focus on the two latest dynamical measurements of the kinematic properties of surrounding stars and gas (Gebhardt et al. 2011; Walsh et al. 2013) that have high-quality data and state-of-the-art modeling approaches.

*Stellar dynamical measurements.* Gebhardt et al. (2011) used infrared spectroscopic observations of stellar absorption lines over the central 2″ (∼150 pc) of M87 and modeled the observed stellar kinematics to infer a mass of (6.6 ± 0.4) × $10^9 M_\odot$, assuming a distance of 17.9 Mpc. They compared the full line-of-sight stellar velocity distributions at all points with the projected kinematics of axisymmetric models of the galaxy. The models allowed for adjustable but radially constant mass-to-light ratios of the central stellar population and the central effects of the M87 dark matter halo. The analysis demonstrates that kinematics cannot be fitted without the inclusion of a central compact non-luminous mass.

*Gas dynamical measurements.* Walsh et al. (2013) used *Hubble Space Telescope* (*HST*) Space Telescope Imaging Spectrograph observations of Hα and [N II] emission lines from the central gas-disk within the M87 nucleus to map its kinematic field within 40 pc of the black hole. The velocity field implies a compact mass of $(3.5^{+0.9}_{-0.7}) \times 10^9 M_\odot$, assuming a distance of 17.9 Mpc, on the assumption that the velocity field is Keplerian and is not subject to non-gravitational forces such as shocks and winds. Several factors could account for the lower mass inferred via this technique (see, e.g., Kormendy & Ho 2013; Jeter et al. 2018).

## 9. Inferred Black Hole Parameters and Discussion

### 9.1. Lensed Emission Around the Black Hole Shadow

The results of fitting analytic models (Section 5) and GRMHD image snapshots (Section 6) to visibility domain data, as well as image reconstructions (Paper IV, Section 7) strongly favor an asymmetric ring (crescent) source morphology with a deep central brightness depression. This morphology is robust among the different analysis methods and is consistent with characteristic features in the interferometric visibilities (Section 3).

Fitting geometric models to the data and extracting feature parameters in the image domain both allow us to quantify the following properties of the crescents: (i) a compact flux density of ≃0.7 Jy, which is ≳50% of the total at arcsecond scale (see also Paper III and Paper IV); (ii) a mean emission diameter of ~42 ± 3 μas; (iii) a fractional width of ≲0.5; (iv) a deep central brightness depression with a brightness contrast ratio of ≳10; and (v) a persistent asymmetry with the brightest region to the South (PA of 150°–200°).

All of these features support the interpretation that we are seeing emission from near the event horizon that is gravitationally lensed into a crescent shape near the photon ring. The central flux depression is the result of photons captured by the black hole: the black hole shadow (Falcke et al. 2000). The asymmetry is a result of Doppler beaming due to the rotation of plasma at relativistic speeds. The peak brightness location is expected to be oriented ≃90° away from the jet PA (Dexter et al. 2012; Mościbrodzka et al. 2016; Ryan et al. 2018; Chael et al. 2019b; Paper V), roughly north–south as we find here.

In many active galactic nuclei, the radio VLBI jet core on parsecs and larger scales is inferred to be significantly offset from the black hole (e.g., Marscher et al. 2008). For M87, this is disfavored by past VLBI observations. The jet and counter-jet angular separation inferred by VLBA observations (Ly et al. 2004, 2007; Kovalev et al. 2007) locates the radio core close to the black hole. VLBA astrometric observations further tightly constrain the jet apex to be ≃41 ± 12 μas upstream of the VLBI core at 43 GHz (Hada et al. 2011, 2013; Nakamura & Asada 2013), suggesting that the mm emission originates very close to the black hole.

The observed EHT image morphology also disfavors an origin offset from the black hole. Models of M87 where the emission predominantly arises in the forward jet have found a larger size and blob-like Gaussian structure (e.g., Broderick & Loeb 2009) due to Doppler beaming resulting from the acceleration of the jet along the line of sight. Additionally, the stable source size over several years of mm-VLBI (Doeleman et al. 2012; Akiyama et al. 2015), despite changes in compact flux density by factors of ≃2–3 (Paper IV), favors the extreme gravitational lensing scenario. Future EHT observations will provide a sensitive additional test of this paradigm by measuring the source morphology on year timescales.

### 9.2. The Black Hole Mass and Comparison to Prior Dynamical Measurements

We have used several techniques to arrive at $\theta_g$ estimates from the EHT 2017 M87 data, and in Table 7 we list the result for each technique after averaging over observing days and bands (see also Figure 19). We find a striking level of consistency across all measurement methods, with GC modeling, direct GRMHD fitting, and image domain feature extraction procedures converging on a characteristic value of $\theta_g = 3.8 \pm 0.4$ μas as the angular size subtended by one gravitational radius. The measurement techniques themselves are entirely independent of one another. The mutual agreement that we see among the multiple measurements indicates that the $\theta_g$ value we have converged upon is robust to the many different choices that can be made regarding data products, model specifications, and parameter space exploration algorithms.

All of the individual $\theta_g$ estimates use the GRMHD simulation library, either through directly fitting GRMHD snapshots to the data (Section 6) or through calibration of diameters resulting from geometric models or reconstructed

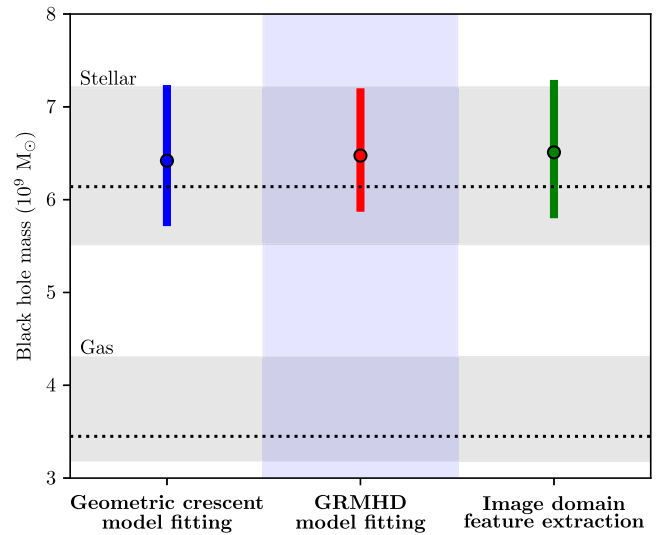

**Figure 19.** Estimates of the mass of the central black hole in M87 for the three different measurement techniques employed in this Letter (see also Table 7). Gray horizontal bands around dashed lines correspond to the 68% confidence levels for the stellar (top) and gas (bottom) dynamical mass measurements (see Table 9).

**Table 7**
Summary of $\theta_g$ and $M$ Measurements

| Measurement Method | $\theta_g$ (μas) | $M$ ($10^9\,M_\odot$) |
|---|---|---|
| GC model fitting | $3.77^{+0.45}_{-0.40}$ | $6.42^{+0.82}_{-0.71}$ |
| GRMHD model fitting | $3.80^{+0.39}_{-0.31}$ | $6.48^{+0.73}_{-0.61}$ |
| Image domain feature extraction | $3.83^{+0.42}_{-0.36}$ | $6.51^{+0.78}_{-0.71}$ |
| Gas dynamics (Walsh et al. 2013) | $2.05^{+0.48}_{-0.16}$ | $3.45^{+0.85}_{-0.26}$ |
| Stellar dynamics (Gebhardt et al. 2011) | $3.62^{+0.60}_{-0.34}$ | $6.14^{+1.07}_{-0.62}$ |
| Combined measurements from this work | $3.8 \pm 0.4$ | $6.5 \pm 0.7$ |

**Note.** Measurements made in this work (top) and from the literature (middle) are quoted as median values with 68% confidence intervals. The final combined measurements from this work are listed in bottom row of the table, rounded to two significant figures. The distance used to compute $M$ from $\theta_g$ is 16.8±0.8 Mpc (see Section 8).

images (Sections 5 and 7). A degree of caution is therefore warranted. The measurements rely on images generated from GRMHD simulations and should be understood within that context (see also Paper V). We have also used only a small subset of 100 such randomly chosen frames (with randomly assigned SSM parameters) out of a much larger library. A simple (but poorly motivated) alternative to the full calibration presented here would be to assign the mean emission diameter to the size of the photon ring itself ($\alpha \simeq 9.6$–10.4 for all methods). That would give a nearly identical $\theta_g$ result for the image domain estimates and only a ≃10% increase for the geometric models, within its current systematic uncertainty.

The mass measurements that we have carried out with EHT data show a high level of consistency, converging to an average value of $M = 6.5 \times 10^9\,M_\odot$. All measurement techniques share a large source of systematic uncertainty arising from the GRMHD calibration, with an average value of $\sigma_{sys} = 0.7 \times 10^9\,M_\odot$. The geometric crescent modeling and image domain measurements further enable an estimate of the systematic uncertainty associated with the angular diameter measurement, which manifests in the mass measurement with an average, rounded value of $\sigma_{\mathrm{stat}} = 0.2 \times 10^9 M_\odot$. As can be readily seen in Figure 19, our mass measurements are also in excellent agreement with those made by Gebhardt et al. (2011) using stellar dynamics (Section 8.2). Our mass measurement is inconsistent with the gas dynamics measurement of Walsh et al. (2013) at the ~99% confidence level.

### 9.3. Implications for the Black Hole Environment

The measurement of a large black hole mass in M87 has a number of implications for the relationship between the black hole and its immediate environment. The observed line-of-sight velocities of atomic line emission produced within the gas flow at 10–100 pc place strong constraints on the dynamical state of gas on these scales and its relationship with the horizon-scale and milliarcsecond-scale features.

In combination with the stellar dynamics mass estimate, the mass $M$ reported here provides an upper limit on the extended dark mass enclosed within the smallest scale probed by Gebhardt et al. (2011), $0''25$ or approximately 20 pc, set by the upper limit on the difference between the two masses: $M_{\mathrm{G11}} - M = -0.3^{+1.3}_{-0.9} \times 10^9 M_\odot$, where $M_{\mathrm{G11}}$ is the stellar dynamics mass in Table 7 and the 68% confidence intervals have simply been added in quadrature. Thus, at 95% confidence this comparison implies that inside 20 pc $\lesssim 2.3 \times 10^9 M_\odot$, or $\lesssim 36\%$ of the central black hole mass, can exist as a dark extended component. This limit is not yet sufficient to constrain the presence of a density "spike" of dark matter associated with the formation of the central black hole (see, e.g., Lacroix 2015).

### 9.4. Evidence for a Horizon

The constraint on the size of the emission region, coupled with multi-wavelength limits on the emission from within the shadow region, provide evidence for the existence of event horizons. These limits arise from looking for the boundary-layer emission that should be present if a photosphere is visible, i.e., not hidden from view by an event horizon (Narayan & McClintock 2008; Broderick et al. 2015). The source of such emission is often called a "surface," though we note that it need not be so in practice.

The results reported in the previous section strengthen this argument. First, they provide an accurate reconstruction of the source structure, which now locates the central mass to within the photon orbit. Second, as discussed in Paper V, the observed image is broadly consistent with models of jet launching driven by electromagnetic processes. Jets are collimated by accreting material; thus a significant boundary-layer emission component would be expected if M87 is not a black hole (Broderick et al. 2015).

The spectrum of this additional boundary-layer emission component is generally thermal when the photon escape fraction is small, in which case this emission effectively couples the photosphere to itself on a timescale that depends only logarithmically on redshift (though see Lu et al. 2017 regarding the possibility of longer equilibrium times). The color temperature of the emission is set uniquely by the accretion power and *apparent* size of the emission region. The former can be related to within factors of a few to the total jet power, $\sim 10^{42}\,\mathrm{erg\,s^{-1}}$ (see the discussion in Paper V). Note that this lower limit is an order of magnitude smaller than the lower end of the range considered in Broderick et al. (2015). The resulting putative photospheric emission would have a color temperature of $T_C \lesssim 10^4\,\mathrm{K}$ and peak in the near-infrared/optical.

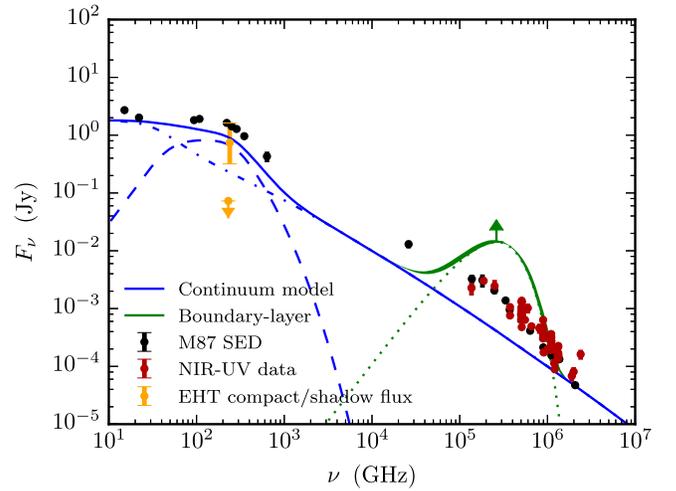

**Figure 20.** Spectral energy distribution of a putative photosphere in comparison to EHT and multi-wavelength data. The lower limit for the thermal bump associated with boundary-layer emission in the absence of a horizon is shown for a compact object with the size reported here (dark green band and arrow) and a jet power $\gtrsim 10^{42}\,\mathrm{erg\,s^{-1}}$. The photospheric component has been added to a jet synchrotron emission model (blue lines, M. Lucchini et al. 2019, in preparation). EHT compact flux densities and limits (orange points) are insensitive to a putative boundary-layer component. Its presence is ruled out by near-infrared, optical, and ultraviolet measurements (dark red and black points, see Broderick et al. 2015; Prieto et al. 2016).

While the EHT is not sensitive to this component, its existence can be excluded by near-infrared, optical, and ultraviolet flux measurements (see, e.g., Broderick et al. 2015; Prieto et al. 2016). We show this excess in Figure 20 for the mass measurement reported here.

### 9.5. Implications for the Kerr Nature of the Black Hole

The detection of a substantial central brightness depression in the image of M87 that we identify with the shadow of the black hole allows us, in principle, to perform two tests of the nature of the compact object and its spacetime. One is related to the size of the shadow, and the other to its shape.

There are two properties of spinning black holes that, depending on spin and observer inclination, could alter the size and shape of the black hole shadow. The spacetime quadrupole $q$ tends to lend an oblate shape to the shadow, while frame dragging due to the spin $a$ acts to compress the shadow in the direction perpendicular to its spin. When the spacetime quadrupole satisfies the Kerr condition $q = -a^2$, the two effects nearly cancel each other out, leaving a quasi-circular shadow with a mean diameter that lies in the very small range (9.6–10.4) $GM/c^2$. This shape is expected to be nearly circular for the low inclination of M87 (e.g., Takahashi 2004; Chan et al. 2013).

For a black hole of known mass-to-distance ratio, the angular size of the shadow on the sky is, therefore, fixed and can be used as a null-hypothesis test of the Kerr nature of the compact object (Psaltis et al. 2015). Alternatively, we can infer the mass-to-distance ratio of the black hole using the size of its shadow image, as we do here, and compare it to the mass-to-distance ratio measured dynamically at much larger distances. Demonstrating agreement between the two inferences results in an equivalent null-hypothesis test.

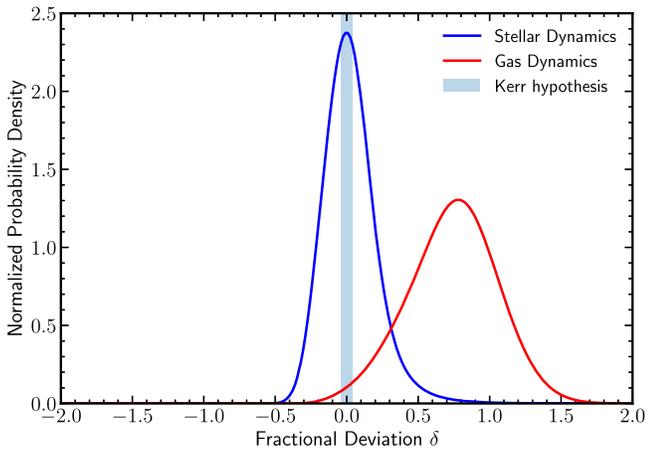

**Figure 21.** Posterior in the fractional difference $\delta$ between the inferred angular gravitational radius of the black hole in M87 based on the EHT data and the values measured using stellar and gas dynamics. The shaded region corresponds to the maximum ±4% discrepancy between the measurements that can be due to the weak dependence of the black hole shadow on the spin magnitude and orientation of a Kerr black hole. The fact that at least one prior mass measurement is consistent with the EHT measurement reported here provides a null hypothesis test of the Kerr nature of the central black hole.

The null hypothesis in this case consists of three ingredients: that the dynamical measurements provide an accurate determination of the black hole mass, that the brightness depression we detect in the EHT image of M87 is indeed the black hole shadow, and that the spacetime of the black hole is described by the Kerr solution. Demonstrating a violation of this null hypothesis would imply that one or more of these assumptions is invalid.

We have used the 2017 EHT observations of M87 to infer the posterior probability $P_{\mathrm{obs}}(\theta_g)$ of the angular gravitational radius at the distance of M87. Earlier dynamical measurements also provide us with the prior probability $P_{\mathrm{dyn}}(\theta_{\mathrm{dyn}})$ for the same quantity. We can, therefore, measure the fractional difference $\delta$ between the predicted and measured mass-to-distance ratio for the black hole as

$$\delta \equiv \frac{\theta_g}{\theta_{\mathrm{dyn}}} - 1 \qquad (31)$$

and its posterior as

$$P(\delta) = \int \theta_{\mathrm{dyn}} P_{\mathrm{dyn}}(\theta_{\mathrm{dyn}}) P_{\mathrm{obs}}(\theta_{\mathrm{dyn}}(\delta+1)) d\theta_{\mathrm{dyn}}. \qquad (32)$$

Figure 21 shows the posterior on the fractional deviation between the EHT and the dynamical inferences of the mass-to-distance ratio, when we assume the prior measurements based on stellar and gas dynamics. We find $\delta = -0.01 \pm 0.17$ (68% credible intervals) for the stellar and $\delta = 0.78 \pm 0.3$ for the gas dynamics priors. The fact that our measurement $\theta_g$ is consistent with one of the prior measurements $\theta_{\mathrm{dyn}}$ allows us to conclude that our null hypothesis has not been violated.

The second property of the black hole shadow that is affected by potential deviations of its spacetime from the Kerr metric is its shape. Shadows of Kerr black holes are nearly circular with a maximum asymmetry of $\lesssim 10\%$, as defined in Johannsen & Psaltis (2010). For the case of the $\sim 17°$ inclination of M87 (and assuming that the large-scale jet points along the spin axis), the maximum asymmetry is $\lesssim 2\%$ for all values of black hole spin (e.g., Chan et al. 2013). Measuring a shadow shape that deviates from circular to a degree larger than this amount would require that the spacetime of the compact object is not described by the Kerr metric (Johannsen & Psaltis 2010).

Figure 18 shows that the circularity inferred from reconstructed images of M87 is consistent with that found in model images from GRMHD simulations in the Kerr metric. The images used in our analysis in Section 7 measure the shape of the emitting region. In future work a similar analysis can be done using images from models that violate the no-hair theorem, where the photon ring/shadow shape can be highly prolate, oblate, or even amorphous (Broderick et al. 2014; Giddings & Psaltis 2018).

## 10. Conclusions

The horizon-scale emission of M87 at 1.3 mm exhibits a robust crescent-like structure. This is seen in image reconstructions (Paper IV) and supported by features in the visibility data from the 2017 EHT campaign (Paper III). Geometric crescent models are overwhelmingly preferred to similarly complex models that we have explored that do not have a central flux depression.

The crescent structure is well defined, with a rapid decline to a dark interior with a brightness $\gtrsim 10\times$ lower than the average of that in the annular region. The average diameter of the crescent is well constrained by multiple methods with a combined range of $42 \pm 3$ $\mu$as. This is consistent among all observation days and bands.

Based on geometric models, GRMHD simulations, and image domain feature extraction, the angular size of the gravitational radius is $\theta_g = 3.8 \pm 0.4$ $\mu$as. The uncertainty of this value is currently dominated by a systematic component associated with the location of the mean emission diameter in GRMHD simulations that is used to convert from an angular to physical scale. Accounting for this uncertainty explicitly, the resulting black hole mass is $M = 6.5 \pm 0.2|_{\mathrm{stat}} \pm 0.7|_{\mathrm{sys}} \times 10^9 M_\odot$. This mass is consistent with that measured from stellar but not gas dynamics.

The crescent morphology, rapid drop to a deep interior flux depression, and broad consistency among days, methods, and the stellar dynamics measurement all point to the emission structure from M87 being due to strong gravitational lensing around a central black hole. The consistency between the 10 pc-scale stellar dynamics and EHT mass measurements provides a null test of the Kerr metric. The size constraint and qualitative support of the standard jet formation paradigm argues for the presence of a horizon. Together, our results strongly support the hypothesis that the central object in M87 is indeed a Kerr black hole, and provide new evidence for the long-believed connection between AGNs and SMBHs.

The authors of this Letter thank the following organizations and programs: the Academy of Finland (projects 274477, 284495, 312496); the Advanced European Network of E-infrastructures for Astronomy with the SKA (AENEAS) project, supported by the European Commission Framework Programme Horizon 2020 Research and Innovation action under grant agreement 731016; the Alexander von Humboldt

Stiftung; the Black Hole Initiative at Harvard University, through a grant (60477) from the John Templeton Foundation; the China Scholarship Council; Comisión Nacional de Investigación Científica y Tecnológica (CONICYT, Chile, via PIA ACT172033, Fondecyt 1171506, BASAL AFB-170002, ALMA-conicyt 31140007); Consejo Nacional de Ciencia y Tecnología (CONACYT, Mexico, projects 104497, 275201, 279006, 281692); the Delaney Family via the Delaney Family John A. Wheeler Chair at Perimeter Institute; Dirección General de Asuntos del Personal Académico–Universidad Nacional Autónoma de México (DGAPA–UNAM, project IN112417); the European Research Council Synergy Grant "BlackHoleCam: Imaging the Event Horizon of Black Holes" (grant 610058); the Generalitat Valenciana postdoctoral grant APOSTD/2018/177; the Gordon and Betty Moore Foundation (grants GBMF-3561, GBMF-5278); the Istituto Nazionale di Fisica Nucleare (INFN) sezione di Napoli, iniziative specifiche TEONGRAV; the International Max Planck Research School for Astronomy and Astrophysics at the Universities of Bonn and Cologne; the Jansky Fellowship program of the National Radio Astronomy Observatory (NRAO); the Japanese Government (Monbukagakusho: MEXT) Scholarship; the Japan Society for the Promotion of Science (JSPS) Grant-in-Aid for JSPS Research Fellowship (JP17J08829); JSPS Overseas Research Fellowships; the Key Research Program of Frontier Sciences, Chinese Academy of Sciences (CAS, grants QYZDJ-SSW-SLH057, QYZDJ-SSW-SYS008); the Leverhulme Trust Early Career Research Fellowship; the Max-Planck-Gesellschaft (MPG); the Max Planck Partner Group of the MPG and the CAS; the MEXT/JSPS KAKENHI (grants 18KK0090, JP18K13594, JP18K03656, JP18H03721, 18K03709, 18H01245, 25120007); the MIT International Science and Technology Initiatives (MISTI) Funds; the Ministry of Science and Technology (MOST) of Taiwan (105-2112-M-001-025-MY3, 106-2112-M-001-011, 106-2119-M-001-027, 107-2119-M-001-017, 107-2119-M-001-020, and 107-2119-M-110-005); the National Aeronautics and Space Administration (NASA, Fermi Guest Investigator grant 80NSSC17K0649); the National Institute of Natural Sciences (NINS) of Japan; the National Key Research and Development Program of China (grant 2016YFA0400704, 2016YFA0400702); the National Science Foundation (NSF, grants AST-0096454, AST-0352953, AST-0521233, AST-0705062, AST-0905844, AST-0922984, AST-1126433, AST-1140030, DGE-1144085, AST-1207704, AST-1207730, AST-1207752, MRI-1228509, OPP-1248097, AST-1310896, AST-1312651, AST-1337663, AST-1440254, AST-1555365, AST-1715061, AST-1614868, AST-1615796, AST-1716327, OISE-1743747, AST-1816420); the Natural Science Foundation of China (grants 11573051, 11633006, 11650110427, 10625314, 11721303, 11725312, 11873028, 11873073, U1531245, 11473010); the Natural Sciences and Engineering Research Council of Canada (NSERC, including a Discovery Grant and the NSERC Alexander Graham Bell Canada Graduate Scholarships-Doctoral Program); the National Youth Thousand Talents Program of China; the National Research Foundation of Korea (grant 2015-R1D1A1A01056807, the Global PhD Fellowship Grant: NRF-2015H1A2A1033752, and the Korea Research Fellowship Program: NRF-2015H1D3A1066561); the Netherlands Organization for Scientific Research (NWO) VICI award (grant 639.043.513) and Spinoza Prize SPI 78-409; the New Scientific Frontiers with Precision Radio Interferometry Fellowship awarded by the South African Radio Astronomy Observatory (SARAO), which is a facility of the National Research Foundation (NRF), an agency of the Department of Science and Technology (DST) of South Africa; the Onsala Space Observatory (OSO) national infrastructure, for the provisioning of its facilities/observational support (OSO receives funding through the Swedish Research Council under grant 2017-00648) the Perimeter Institute for Theoretical Physics (research at Perimeter Institute is supported by the Government of Canada through the Department of Innovation, Science and Economic Development Canada and by the Province of Ontario through the Ministry of Economic Development, Job Creation and Trade); the Russian Science Foundation (grant 17-12-01029); the Spanish Ministerio de Economía y Competitividad (grants AYA2015-63939-C2-1-P, AYA2016-80889-P); the State Agency for Research of the Spanish MCIU through the "Center of Excellence Severo Ochoa" award for the Instituto de Astrofísica de Andalucía (SEV-2017-0709); the Toray Science Foundation; the US Department of Energy (USDOE) through the Los Alamos National Laboratory (operated by Triad National Security, LLC, for the National Nuclear Security Administration of the USDOE (Contract 89233218CNA000001)); the Italian Ministero dell'Istruzione Università e Ricerca through the grant Progetti Premiali 2012-iALMA (CUP C52I13000140001); ALMA North America Development Fund; Chandra TM6-17006X. This work used the Extreme Science and Engineering Discovery Environment (XSEDE), supported by NSF grant ACI-1548562, and CyVerse, supported by NSF grants DBI-0735191, DBI-1265383, and DBI-1743442. XSEDE Stampede2 resource at TACC was allocated through TG-AST170024 and TG-AST080026N. XSEDE JetStream resource at PTI and TACC was allocated through AST170028. The simulations were performed in part on the SuperMUC cluster at the LRZ in Garching, on the LOEWE cluster in CSC in Frankfurt, and on the HazelHen cluster at the HLRS in Stuttgart. This research was enabled in part by support provided by Compute Ontario (http://computeontario.ca), Calcul Quebec (http://www.calculquebec.ca) and Compute Canada (http://www.computecanada.ca). We thank the staff at the participating observatories, correlation centers, and institutions for their enthusiastic support. This Letter makes use of the following ALMA data: JAO.ALMA#2016.1.01154. V. ALMA is a partnership of the European Southern Observatory (ESO; Europe, representing its member states), NSF, and National Institutes of Natural Sciences of Japan, together with National Research Council (Canada), Ministry of Science and Technology (MOST; Taiwan), Academia Sinica Institute of Astronomy and Astrophysics (ASIAA; Taiwan), and Korea Astronomy and Space Science Institute (KASI; Republic of Korea), in cooperation with the Republic of Chile. The Joint ALMA Observatory is operated by ESO, Associated Universities, Inc. (AUI)/NRAO, and the National Astronomical Observatory of Japan (NAOJ). The NRAO is a facility of the NSF operated under cooperative agreement by AUI. APEX is a collaboration between the Max-Planck-Institut für Radioastronomie (Germany), ESO, and the Onsala Space Observatory (Sweden). The SMA is a joint project between the SAO and ASIAA and is funded by the Smithsonian Institution and the Academia Sinica. The JCMT is operated by the East Asian Observatory on behalf of the NAOJ, ASIAA, and KASI, as well as the Ministry of Finance of China, Chinese Academy of Sciences, and the National Key R&D Program (No. 2017YFA0402700) of China. Additional funding support for the JCMT is provided by the Science and Technologies Facility Council (UK) and participating universities in the UK and Canada. The LMT project is a joint effort of the Instituto Nacional de Astrofísica, Óptica, y Electrónica (Mexico) and the University of Massachusetts at Amherst (USA). The IRAM 30-m telescope on Pico Veleta, Spain is operated by IRAM and supported by CNRS (Centre National de la Recherche Scientifique, France), MPG (Max-Planck-Gesellschaft, Germany) and IGN (Instituto Geográfico Nacional, Spain). The SMT is operated by the Arizona Radio Observatory, a part of the Steward Observatory of the University of Arizona, with financial support of operations from the State of Arizona and financial support for instrumentation development from the NSF. Partial SPT support is provided by the NSF Physics Frontier Center award (PHY-0114422) to the Kavli Institute of Cosmological Physics at the University of Chicago (USA), the Kavli Foundation, and the GBMF (GBMF-947). The SPT hydrogen maser was provided on loan from the GLT, courtesy of ASIAA. The SPT is supported by the National Science Foundation through grant PLR-1248097. Partial support is also provided by the NSF Physics Frontier Center grant PHY-1125897 to the Kavli Institute of Cosmological Physics at the University of Chicago, the Kavli Foundation and the Gordon and Betty Moore Foundation grant GBMF 947. The EHTC has received generous donations of FPGA chips from Xilinx Inc., under the Xilinx University Program. The EHTC has benefited from technology shared under open-source license by the Collaboration for Astronomy Signal Processing and Electronics Research (CASPER). The EHT project is grateful to T4Science and Microsemi for their assistance with Hydrogen Masers. This research has made use of NASA's Astrophysics Data System. We gratefully acknowledge the support provided by the extended staff of the ALMA, both from the inception of the ALMA Phasing Project through the observational campaigns of 2017 and 2018. We would like to thank A. Deller and W. Brisken for EHT-specific support with the use of DiFX. We acknowledge the significance that Maunakea, where the SMA and JCMT EHT stations are located, has for the indigenous Hawaiian people.

*Facility:* EHT.

*Software:* THEMIS, `dynesty`, `eht-imaging`, DIFMAP, SMILI, GENA.

## Appendix A
## THEMIS Station Gain Amplitude Reconstruction

Both THEMIS and `eht-imaging` provide tools for analyzing EHT data within the visibility domain. Both address the reconstruction of individual station gains, though in substantially different ways. Here, we briefly describe the procedure employed by THEMIS and compare the reconstructed gains with the results reported in Paper IV.

We reconstruct gains on a scan-by-scan basis. Thus, for EHT 2017 data, the gain amplitudes constitute between 40 and 143 additional nuisance parameters per data set; more if sub-scan variations are necessary. For the GC models this represents a proliferation of parameters by 200%–700%.

Even with efficient sampling techniques, treating these identically with the other model parameters would result in a substantial (super-linear) increase in the computation expense needed to adequately sample the likelihood. Thus, in THEMIS, these are efficiently addressed instead by directly marginalizing the likelihood in Equation (9).

This is done by numerically maximizing $\mathcal{L}_{A,ij}$ over the $|g_i|$ (subject to Gaussian priors on the $|g_i|$) for each scan and then marginalizing over an expansion around this maximum (see A. E. Broderick et al. 2019, in preparation, for more details). Within this procedure it is assumed that gains from neighboring scans are independent. For all analyses presented here, the standard deviation of these priors for the LMT and all other stations are 100% and 20%, respectively. While the latter substantially exceeds the anticipated uncertainty stated in Table 3 of Paper III, the reconstructed station gains are typically very well constrained and close to unity.

Typically, the station gains on a given scan are significantly overdetermined. Thus, calibrating the station gains in this way produces model-independent patterns, driven primarily by the need for consistency among a given observation.

In Figure 22 we show the reconstructed LMT gains for the best-fit GC model model on each day (low-band only) in comparison to those obtained by image-domain methods for the LMT. This station was selected because it exhibits large gain excursions; the other reconstructed station gains are generally much closer to unity.

The two different procedures based on different intrinsic source models find remarkable agreement over all days. Importantly, there are no large gain offsets that would produce a significant impact on the reconstruction of the total compact flux.

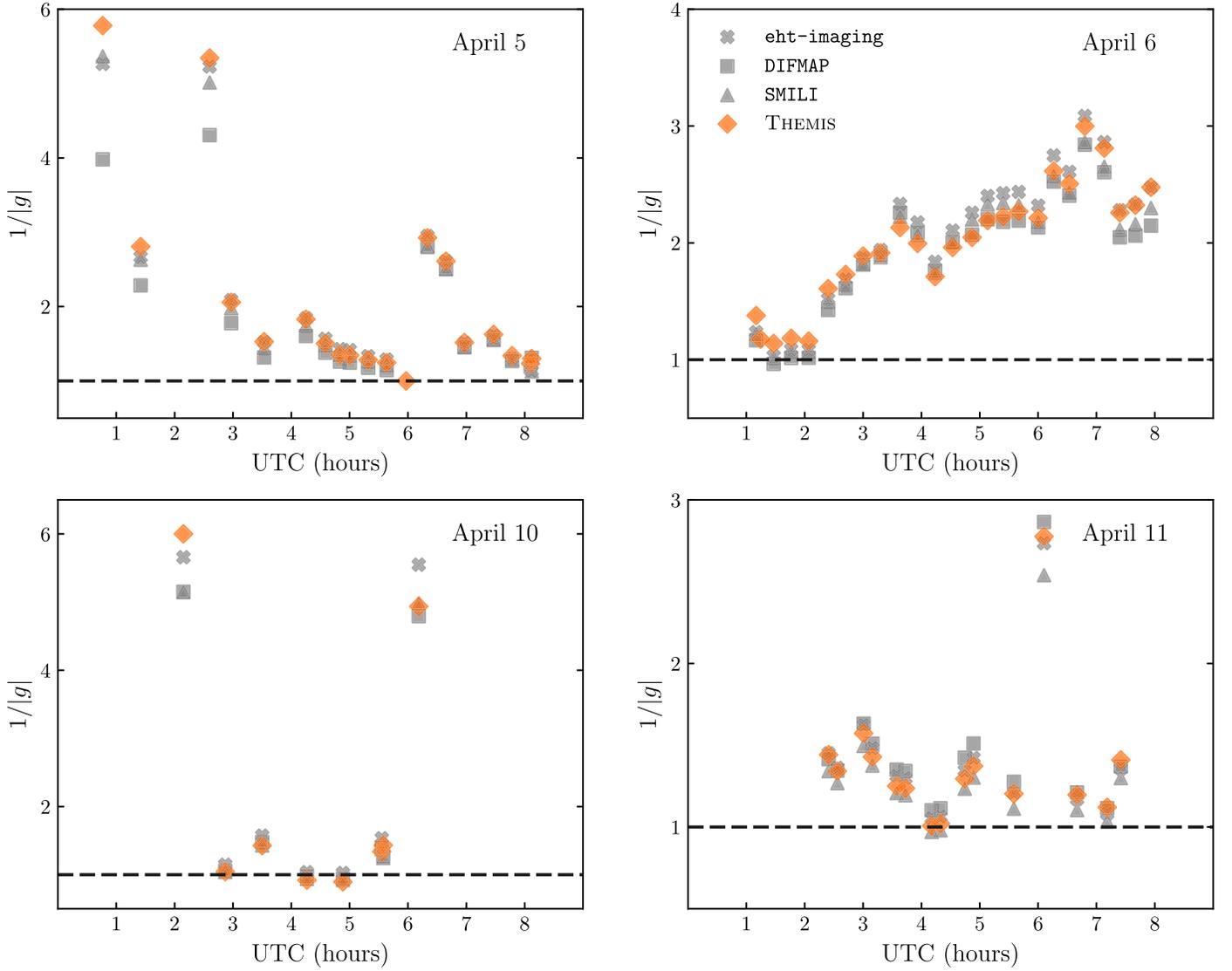

**Figure 22.** Reconstructed LMT station gains by THEMIS from GC model fits to the April 5, 6, 10, and 11 low-band data sets. All other station gains are very close to unity, and are consistent with those from image reconstruction. The inferred LMT gains are consistent with those found in Paper IV across all days.

## Appendix B
## Generalized Crescent Models

In this appendix we describe the GC models used in Section 5 to fit the M87 data. Descriptions of the model parameters and their associated priors are listed in Table 8.

### B.1. Fourier-domain Construction of the GC models

Following Kamruddin & Dexter (2013), we start with a unit circular "top-hat" function circ(r), defined in the image domain to be

$$\text{circ}(r) = \begin{cases} 1, & r \leqslant 1, \\ 0, & r > 1. \end{cases} \quad (33)$$

A disk of radius $R$ can then be expressed as $D(r) = \text{circ}(r/R)$. The Fourier transform of such a disk is given by

$$\tilde{D}(\rho) = \frac{R}{\rho} J_1(2\pi\rho R), \quad (34)$$

where $\rho = \sqrt{u^2 + v^2}$ is the Fourier-domain radial coordinate and $J_1(x)$ is the Bessel function of the first kind of order 1. We can construct a circular crescent C via the difference of two such disks after shifting the smaller, inner disk by $r_0$ in the negative $x$-direction prior to subtracting it from the outer disk. A spatial shift in the image domain corresponds to a frequency shift in the Fourier domain, such that

$$\tilde{C}(u, v) = \tilde{D}_{\text{out}} - e^{2\pi i r_0 u} \tilde{D}_{\text{in}}$$
$$= \frac{R_{\text{out}}}{\rho} J_1(2\pi\rho R_{\text{out}}) - e^{2\pi i r_0 u} \frac{R_{\text{in}}}{\rho} J_1(2\pi\rho R_{\text{in}}), \quad (35)$$

where $R_{\text{out}}$ and $R_{\text{in}}$ are the outer and inner disk radii, respectively.

Similar to Benkevitch et al. (2016), we define a "slash" operation $s(x)$ that imposes a linear brightness gradient across the image. The slash operator can be expressed in the image

**Table 8**
Parameters and Priors for the GC Models

| Parameter | Description | Prior xs-ringauss | Prior xs-ring | Units |
|---|---|---|---|---|
| $V_0$ | Flux density of crescent component | ... | $\mathscr{U}(0, 2)$ | Jy |
| $V_0 + V_1$ | Flux density of crescent component (including fixed Gaussian) | $\mathscr{U}(0, 2)$ | ... | Jy |
| $R_{out}$ | Outer radius of crescent | $\mathscr{U}(0, 100)$ | | μas |
| $\phi$ | PA | $\mathscr{U}(0, 360)$ | | degrees |
| $\psi$ | Fractional thickness | $\mathscr{U}(0, 1)$ | | unitless |
| $\tau$ | Structural asymmetry parameter | $\mathscr{U}(0, 1)$ | | unitless |
| $\beta$ | Flux asymmetry parameter | $\mathscr{U}(0, 1)$ | $\mathscr{L}_e(-5, 5)$ | unitless |
| $\sigma$ | Width of Gaussian smoothing kernel | $\mathscr{U}(0, 100)$ | | μas |
| $\gamma$ | Fractional amplitude of central emission floor | ... | $\mathscr{U}(0, 1)$ | unitless |
| $V_F$ | Flux density of central Gaussian emission floor | $\mathscr{U}(0, 2)$ | ... | Jy |
| $\sigma_F$ | Width of central Gaussian emission floor | $\mathscr{U}(0, 25)$ | ... | μas |
| $V_G$ | Flux density of large-scale Gaussian | $\mathscr{U}(0, 10)$ | | Jy |
| $\sigma_G$ | Width of large-scale Gaussian | $\mathscr{L}_{10}(-2, 1)$ | | arcsec |
| $V_1/V_0$ | Ratio of fixed Gaussian flux density to crescent flux density | $\mathscr{U}(0, 1)$ | ... | unitless |
| $\sigma_x/R_{out}$ | Ratio of fixed Gaussian width to outer radius of crescent | $\mathscr{U}(0, 3)$ | ... | unitless |
| $\sigma_y/\sigma_x$ | Axis ratio of fixed Gaussian | $\mathscr{U}(0, 100)$ | ... | unitless |
| $V_g$ | Flux density of additional Gaussian component | $\mathscr{U}(0, 2)$ | | Jy |
| $x_0$ | Central x-coordinate of additional Gaussian component | $\mathscr{U}(-200, 200)$ | ... | μas |
| $y_0$ | Central y-coordinate of additional Gaussian component | $\mathscr{U}(-200, 200)$ | ... | μas |
| $\sigma$ | Width of additional Gaussian component | $\mathscr{U}(0, 100)$ | ... | μas |
| $A$ | Anisotropy of additional Gaussian component | $\mathscr{U}(0, 0.99)$ | ... | unitless |
| $\theta$ | PA of additional Gaussian component | $\mathscr{U}(0, 180)$ | $\mathscr{U}(0, 90)$ | degrees |
| $x_{0,0}$ | Central x-coordinate of first additional Gaussian component | ... | $\mathscr{U}(-100, 100)$ | μas |
| $\Delta x$ | Change in x-coordinates of adjacent Gaussian components | ... | $\mathscr{U}(0, 100)$ | μas |
| $y_{0,0}$ | Central y-coordinate of first additional Gaussian component | ... | $\mathscr{U}(-100, 100)$ | μas |
| $\Delta y$ | Change in y-coordinates of adjacent Gaussian components | ... | $\mathscr{U}(-100, 100)$ | μas |
| $\sigma_x$ | Major axis width of additional Gaussian component | ... | $\mathscr{U}(0, 100)$ | μas |
| $\sigma_y$ | Minor axis width of additional Gaussian component | ... | $\mathscr{U}(0, 100)$ | μas |

**Note.** The top section of the table corresponds to crescent component parameters, and the bottom section corresponds to additional Gaussian component parameters (see Appendix B.2). $\mathscr{U}(a, b)$ denotes a uniform distribution spanning the range $[a, b]$, $\mathscr{L}_x(a, b)$ denotes a log-uniform distribution in base $x$ spanning the range $[x^a, x^b]$.

domain as

$$s(x) = \frac{h_1 + h_2}{2} + \left(\frac{h_1 - h_2}{2R_{out}}\right)x, \quad (36)$$

where $h_1$ and $h_2$ are the amplitudes of the function at $x = R_{out}$ and $x = -R_{out}$, respectively. We apply s(x) to the crescent via multiplication in the image domain, which corresponds to taking a derivative in the Fourier domain

$$\tilde{s}(u) = \frac{h_1 + h_2}{2} + \frac{i}{2\pi}\left(\frac{h_1 - h_2}{2R_{out}}\right)\frac{d}{du}. \quad (37)$$

Application of the slash is aided by the fact that the derivative of a Bessel function can be expressed in terms of other Bessel functions

$$\frac{d}{d\rho}J_1(2\pi\rho R) = \pi R[J_0(2\pi\rho R) - J_2(2\pi\rho R)], \quad (38)$$

and we denote the slashed crescent model as sC.

### B.1.1. xs-ring Model

We now seek to add an emission "floor," F, to the center of sC to account for nonzero central emission. In the xs-ring construction, F takes the form of a circular disk with amplitude $K$ and radius $R_{in}$ that is shifted by $r_0$ along the x-axis,

$$\tilde{F}(u, v) = \frac{KR_{in}}{\rho}e^{2\pi i r_0 u}J_1(2\pi\rho R_{in}). \quad (39)$$

In the image domain, we can express the slashed crescent plus emission floor as [sC+F] as

$$[sC + F](x, y) = s(x)[D_{out}(x, y) - D_{in}(x - r_0, y)] + F(x - r_0, y). \quad (40)$$

The model thus corresponds to the difference of two spatially offset circular disks having a linear brightness gradient across them, to which a central emission floor is then added. We then further allow the entire model to be rotated by some PA $\phi$, which amounts to replacing $u \rightarrow u\cos(\phi) + v\sin(\phi)$ and $v \rightarrow -u\sin(\phi) + v\cos(\phi)$. Finally, we convolve the model in the image domain with a circular Gaussian kernel of width $\sigma$ and then scale it to have a total flux density of $V_0$. The compact component of the xs-ring model can thus be written in the

Fourier domain as

$$
\begin{aligned}
&\widetilde{\text{xs-ring}}(u, v) \\
&= \frac{V_0 e^{-2\pi^2 \sigma^2 \rho^2} \{\tilde{D}_{\text{out}} - e^{2\pi i \rho R (1-\tau)[u\cos(\phi) + v\sin(\phi)]} (\tilde{D}_{\text{in}} - \tilde{F})\}}{\frac{1}{2}\pi R_{\text{out}}^2 \{(1+\beta) - (1-\psi)^2[(1+\beta) - \psi(1-\tau)(1-\beta) - 2\gamma\beta]\}},
\end{aligned}
\tag{41}
$$

where

$$
\begin{aligned}
\tilde{D}_{\text{out}} = {} & \frac{R_{\text{out}}(1+\beta)}{2\rho} J_1(X_{\text{out}}) - \frac{i(1-\beta)}{4\pi\rho^2} \\
& \times \left\{ \pi R_{\text{out}} [J_0(X_{\text{out}}) - J_2(X_{\text{out}})] - \frac{1}{\rho} J_1(X_{\text{out}}) \right\} \\
& \times [u\cos(\phi) + v\sin(\phi)],
\end{aligned}
\tag{42}
$$

$$
\begin{aligned}
\tilde{D}_{\text{in}} = {} & \frac{R_{\text{out}}(1-\psi)}{2\rho} J_1(X_{\text{in}})[(1+\beta) + \psi(1-\tau)(1-\beta)] \\
& - \frac{i(1-\psi)(1-\beta)}{4\pi\rho^2} \\
& \times \left\{ \pi R_{\text{out}}(1-\psi)[J_0(X_{\text{in}}) - J_2(X_{\text{in}})] - \frac{1}{\rho} J_1(X_{\text{in}}) \right\} \\
& \times [u\cos(\phi) + v\sin(\phi)],
\end{aligned}
\tag{43}
$$

$$
\tilde{F} = \frac{\gamma\beta R_{\text{out}}(1-\psi)}{\rho} J_1(X_{\text{in}}),
\tag{44}
$$

and we have followed Kamruddin & Dexter (2013) in defining the quantities

$$
\begin{aligned}
\psi &\equiv 1 - \frac{R_{\text{in}}}{R_{\text{out}}} \\
\tau &\equiv 1 - \frac{r_0}{\psi R_{\text{out}}} \\
\gamma &\equiv \frac{K}{h_1} \\
\beta &\equiv \frac{h_1}{h_2} \\
h &\equiv h_2 \\
V_0 &\equiv \frac{1}{2}\pi h R_{\text{out}}^2 \{(1+\beta) - (1-\psi)^2[(1+\beta) \\
&\quad - \psi(1-\tau)(1-\beta) - 2\gamma\beta]\} \\
X_{\text{out}} &\equiv 2\pi\rho R_{\text{out}} \\
X_{\text{in}} &\equiv X_{\text{out}}(1-\psi).
\end{aligned}
$$

The xs-ring model has eight free parameters describing the compact emission: $V_0$, $R_{\text{out}}$, $\phi$, $\sigma$, $\psi$, $\tau$, $\gamma$, and $\beta$. The $\psi$, $\tau$, and $\gamma$ parameters are restricted to lie between 0 and 1. The $\psi$ parameter describes the typical fractional thickness of the crescent; as $\psi \to 1$ the crescent reduces to a filled disk, and as $\psi \to 0$ the crescent reduces to an infinitesimally thin ring (i.e., delta function in radius). The $\tau$ parameter is a measure of the crescent's structural symmetry; as $\tau \to 1$ the crescent reduces to a circular ring, and $\tau \to 0$ when the inner and outer disks meet at one edge. The $\gamma$ parameter describes the amplitude of the emission floor, with $\gamma = 0$ indicating no central emission. The $\beta$ parameter describes the strength of the emission gradient across the crescent, with $\beta = 1$ indicating no gradient.

In addition to the compact component, we add a large-scale circular Gaussian, G, to the model

$$
\tilde{G}(\rho) = V_G e^{-2\pi^2 \sigma_G^2 \rho^2}.
\tag{45}
$$

This component is permitted to have very large widths, $\sigma_G \gg 1$ arcsec, to account for short-spacing flux that might be missed by the bulk of the array. The free parameters in G are its flux density ($V_G$) and width ($\sigma_G$), bringing the total number of parameters for the xs-ring model to 10. An example of the xs-ring model is shown in the left panel of Figure 3.

### B.1.2. xs-ringauss Model

The xs-ringauss model differs from the xs-ring model in two key respects. The first difference is that the xs-ringauss model follows Benkevitch et al. (2016) and includes a fixed elliptical Gaussian component on top of the crescent. This component is centered at $(x, y) = (r_0 - R_{\text{in}}, 0)$ prior to applying the PA rotation, and its orientation is fixed to match that of the crescent. The additional free parameters provided by this Gaussian component are thus its widths ($\sigma_x$, $\sigma_y$) and flux density $V_1$. The second difference is that the emission floor $F$ in the xs-ringauss model takes the form of a circular Gaussian rather than a disk,

$$
\tilde{F}(\rho) = V_F e^{-2\pi^2 \sigma_F^2 \rho^2} = 2\pi\sigma_F^2 K e^{-2\pi^2 \sigma_F^2 \rho^2}.
\tag{46}
$$

In addition to the amplitude parameter $K$, this form of the emission floor has a width parameter $\sigma_F$ that also enters into the model.

The remaining parameters in the xs-ringauss model match those in the xs-ring model, for a total of 14 free parameters: 10 of these parameters are shared with the xs-ring model, and the additional four are $V_1$, $\sigma_x$, $\sigma_y$, and $\sigma_g$. An example of the xs-ringauss model is shown in the right panel of Figure 3.

### B.2. Nuisance Gaussian Components

In addition to the crescent components of the GC models, we also fit a small number (two to three) of additional "nuisance" Gaussian components (see Section 5.1). These additional components are intended primarily to capture extraneous amorphous emission surrounding the main ring (see, e.g., Figure 12) and secondarily to provide additional flexible degrees of freedom that allow the model to absorb systematic uncertainties.

### B.2.1. Parameterization

We parameterize the Gaussian components in one of two ways. The first of these, used when fitting the xs-ring model, is given by

$$
\tilde{G}_1(u, v) = V_{g,1} e^{2\pi i (ux_0 + vy_0)} e^{-4\pi^2 \sigma_x^2 \sigma_y^2 (cu^2 - buv + av^2)},
\tag{47}
$$

where

$$
a = \frac{\cos^2(\theta)}{2\sigma_x^2} + \frac{\sin^2(\theta)}{2\sigma_y^2},
\tag{48}
$$

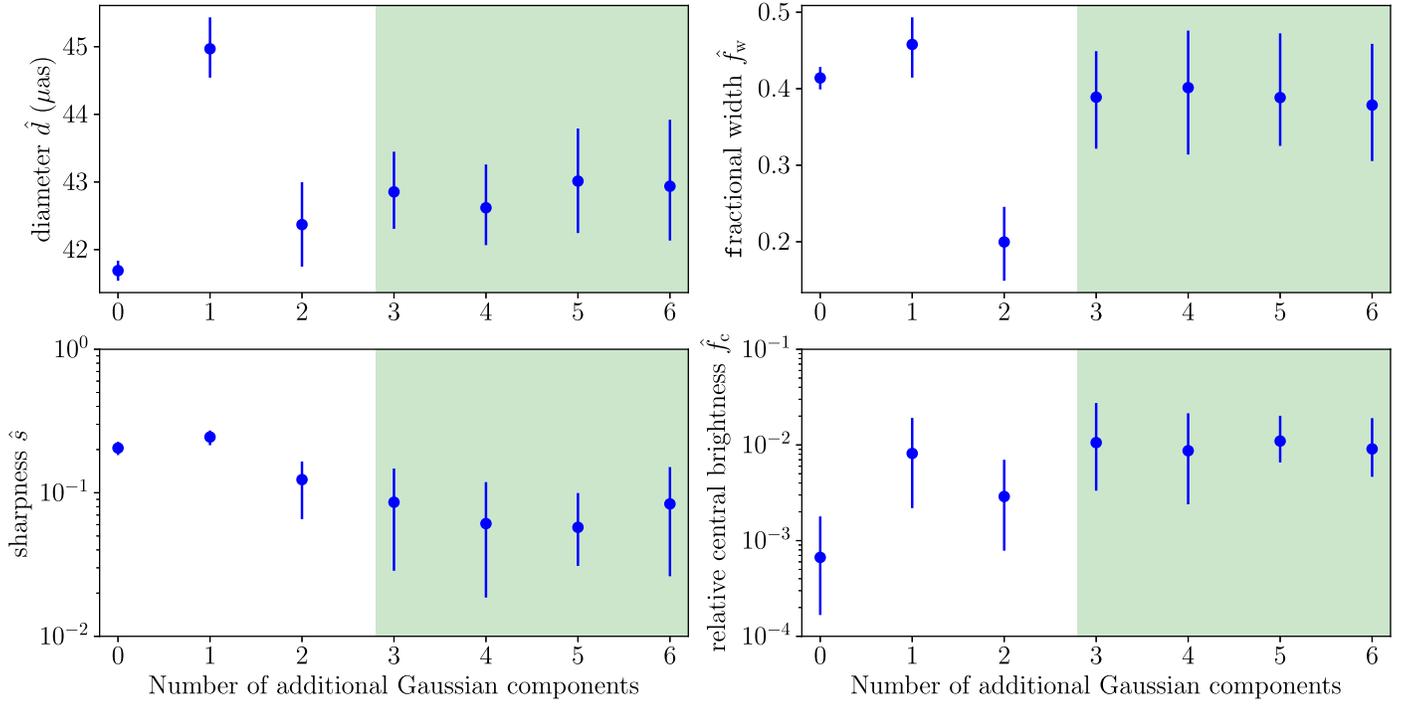

**Figure 23.** Measured crescent parameter values from fitting the xs-ring model with an increasing number of nuisance Gaussian components, shown for the April 5 high-band data set. We see that after a small number of components is reached (three for this data set, marked in green), adding additional components does not substantially modify the parameter posteriors.

$$b = \frac{\sin(2\theta)}{2\sigma_x^2} - \frac{\sin(2\theta)}{2\sigma_y^2}, \qquad (49)$$

and

$$c = \frac{\sin^2(\theta)}{2\sigma_x^2} + \frac{\cos^2(\theta)}{2\sigma_y^2}. \qquad (50)$$

Here, $V_{g,1}$ is the total flux density of the Gaussian, $(x_0, y_0)$ are its central coordinates, $(\sigma_x, \sigma_y)$ are the Gaussian widths along the two principal axes, and $\theta$ is the PA. A circular Gaussian model is constructed by setting $\sigma_x = \sigma_y$ and $\theta = 0$.

When fitting multiple Gaussian components in the context of the xs-ring model, we only parameterize the first component, as described above, with a central coordinate position of $(x_{0,0}, y_{0,0})$. All additional components have positions that are referenced to that of the previous component, such that $(x_i, y_i) = (x_{i-1} + \Delta x_i, y_{i-1} + \Delta y_i)$. Such a parameterization enables the imposition of a spatial ordering by treating the $\Delta x$ or $\Delta y$ parameters as slack variables, thereby breaking the multimodal labeling degeneracy otherwise caused by pairwise swaps of different Gaussian components.

The second Gaussian parameterization, used by the xs-raiguuss model, follows the form described in Broderick et al. (2011). Specifically,

$$\tilde{G}_2(u, v) = V_{g,2} e^{2\pi i(ux_0 + vy_0)} e^{-\frac{\rho^2}{2\sigma^2}[1 - A\cos(\psi - \theta)]}, \qquad (51)$$

where $(\rho, \psi)$ are cylindrical polar coordinates in the $(u, v)$-plane. Here,

$$\sigma = \left(\frac{1}{\sigma_x^2} + \frac{1}{\sigma_y^2}\right)^{-1/2} \qquad (52)$$

is a width parameter and

$$A = \frac{\sigma^2}{2}\left(\frac{1}{\sigma_x^2} - \frac{1}{\sigma_y^2}\right) \qquad (53)$$

is an anisotropy parameter. For small numbers ($\lesssim 5$) of Gaussian components, tempering is sufficient to ensure that the MCMC exploration can navigate all posterior modes; because the xs-raiguuss model never utilizes more than two additional Gaussian components, no mode collapsing is necessary.

### B.2.2. Behavior in Fits to the M87 Data

In principle, it is possible to add a large number of nuisance Gaussian components to a model and thereby to fit the data arbitrarily well. However, our primary goal with model fitting is not to exactly reproduce all details of the emission structure, but rather to make robust measurements of the crescent component model parameters relevant for constraining the lensed part of the image. Once a sufficient number of Gaussian components have been added to the model, further additions do not substantially modify the derived crescent parameters (see Figure 23). We find that two nuisance Gaussians for the xs-raiguuss GC model (for a total of 26 free parameters), and three for the xs-ring GC model (for a total of 28 free parameters), are the threshold values that most generally satisfy this criterion. These values are thus used for all GC model fits presented in this Letter.

When fit alongside the GC models to the M87 data, we find that the nuisance Gaussian components account for ~10%–70% of the compact flux density. Assessing consistency in the best-fit parameter values for the nuisance Gaussian components across models is complicated by the differing model specifications. In general we find less consistency in the properties of nuisance Gaussian components for a given day across models (within a single data set) than we do across bands (within a single model).

Within the context of a particular GC model, we find that the nuisance Gaussian components approximately fall into two categories: (1) components that reside near or on the crescent component and whose location and structure are consistent across bands within a single day, and (2) components that tend to reside far ($\gtrsim 40\,\mu$as) outside the crescent component and whose location and structure may vary across bands and days. We associate the first class of nuisance Gaussian components with attempts by the model to account for a real flux distribution that is too complex for the crescent component alone. The second class of nuisance Gaussians acts in a similar manner to the artifacts frequently seen in image reconstructions (see Paper IV). We thus associate this second class of components with the image-domain manifestation of systematic errors in the data products. We emphasize, however, that we have no mechanism for enforcing a categorization such as that described on the nuisance Gaussian components, and in general we expect that any single component will be influenced both by real source emission and by systematic uncertainties.

### Appendix C
### $\chi^2$ Statistics

In this appendix we define the various $\chi^2$ statistics used as diagnostics of model fit quality in Section 5. We note that the expressions presented here do not have any bearing on the fitting process itself; the likelihood functions used during fitting are described in Section 4.1.

The visibility amplitude reduced-$\chi^2$ is given by

$$\chi_A^2 = \frac{1}{N_A - N_{\rm p} - N_g}\sum\left(\frac{\hat{A} - A}{\sigma}\right)^2, \tag{54}$$

where $N_A$ is the number of visibility amplitudes (see Table 1), $N_{\rm p}$ is the number of GC model parameters, $N_g$ is the number of independent gain terms, $\sigma$ is the uncertainty in visibility amplitude, and $\hat{A}$ and $A$ are the modeled and measured visibility amplitudes, respectively. For the xs-ring GC model fits using dynesty we have $N_{\rm p} = 28$ model parameters, while for the xs-ringauss GC model fits using THEMIS we have $N_{\rm p} = 26$.

Analogous to Equation (54), the closure phase reduced-$\chi^2$ is

$$\chi_{\psi_C}^2 = \frac{1}{N_{\psi_C} - N_{\rm p}}\sum\left(\frac{\hat{\psi}_C - \psi_C}{\sigma_{\psi_C}}\right)^2, \tag{55}$$

where $N_{\psi_C}$ is the number of closure phases, $\sigma_{\psi_C}$ is the closure phase uncertainty, and $\hat{\psi}_C$ and $\psi_C$ are the modeled and measured closure phases, respectively. As in Section 4.1, we avoid phase wrapping by selecting a branch of the closure phase space such that differences always fall between $-180°$ and $180°$.

For the logarithmic closure amplitude reduced-$\chi^2$, we have

$$\chi_{\ln A_C}^2 = \frac{1}{N_{\ln A_C} - N_{\rm p}}\sum\left(\frac{\ln\hat{A}_C - \ln A_C}{\sigma_{\ln A_C}}\right)^2. \tag{56}$$

where $N_{\ln A_C}$ is the number of logarithmic closure amplitudes, $\sigma_{\ln A_C}$ is the logarithmic closure amplitude uncertainty, and $\ln\hat{A}_C$ and $\ln A_C$ are the modeled and measured logarithmic closure amplitudes, respectively.

For the THEMIS fits, we report the joint visibility amplitude and closure phase reduced-$\chi^2$

$$\chi_{A+\psi_C}^2 = \frac{1}{N_A + N_{\psi_C} - N_{\rm p} - N_g}$$
$$\times\left[\sum\left(\frac{\hat{A} - A}{\sigma}\right)^2 + \sum\left(\frac{\hat{\psi}_C - \psi_C}{\sigma_{\psi_C}}\right)^2\right]. \tag{57}$$

For the dynesty fits we have the joint closure phase and logarithmic closure amplitude reduced-$\chi^2$

$$\chi_{\psi_C+\ln A_C}^2 = \frac{1}{N_{\psi_C} + N_{\ln A_C} - N_{\rm p}}$$
$$\times\left[\sum\left(\frac{\hat{\psi}_C - \psi_C}{\sigma_{\psi_C}}\right)^2 + \sum\left(\frac{\ln\hat{A}_C - \ln A_C}{\sigma_{\ln A_C}}\right)^2\right]. \tag{58}$$

The values for each of these statistics are reported in Table 2 for both fitting codes and for all data sets.

### Appendix D
### Using GRMHD Simulations to Calibrate the Generalized Crescent Model

To create a mapping between the GC model diameter and the angular gravitational radius $\theta_g$, we require data sets for which we know both quantities. To this end, we take the GRMHD simulation library described in Paper V to provide our "ground truth" images. These simulations are necessarily limited in the range and resolution of physical parameter space that they probe, but they critically provide a large sample of plausible emission structures for which we know the underlying physical parameters. By treating these GRMHD simulations as "ground truth" images, generating synthetic data from them, and then fitting that data with the GC models, we aim to (1) determine a calibration of the measured crescent diameter $\hat{d}$ in terms of $\theta_g$, and (2) estimate the magnitude of the uncertainty in this calibration.

#### D.1. GRMHD Image Selection and Synthetic Data Generation

For a given GRMHD simulation, individual "snapshots" are ray-traced to produce images of the 230 GHz emission as it would be seen from Earth. These images are then used as input for generating synthetic observational data with eht-imaging, which generates visibilities corresponding to the $(u, v)$-coverage from the 2017 EHT observations of M87 and corrupts this data with realistic levels of thermal noise and station-based systematics. A detailed description of the synthetic data generation pipeline is provided in Paper IV.

From the GRMHD simulation library we selected 103 separate snapshot images with which to generate synthetic data sets. These images were selected randomly from within the parameter space explored by the library, and thus included examples of emission structures that were incompatible with the M87 data. The images were split into two groups, from

**Part I**

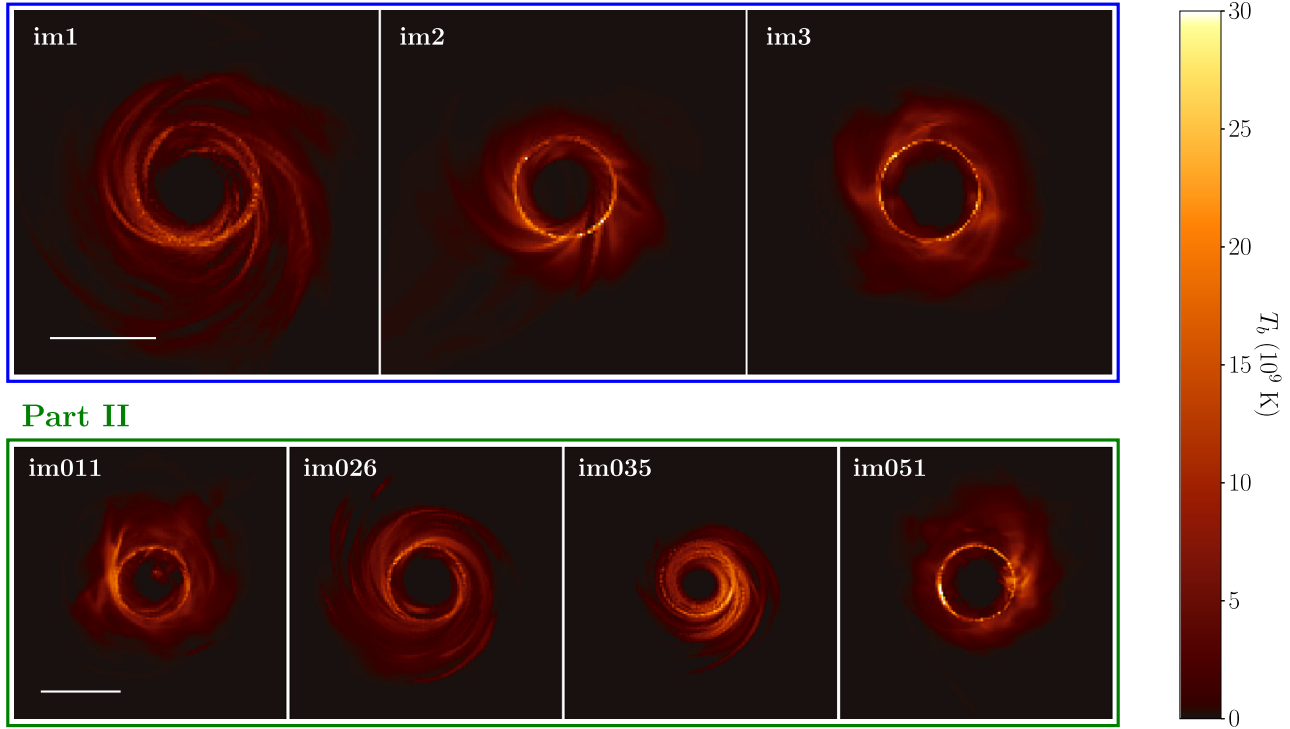

**Part II**

**Figure 24.** Examples of GRMHD snapshots used for calibration (see Appendix D). The top three panels show the images used to calibrate the "observational uncertainty" associated with changing $(u, v)$-sampling and systematics across days (Part I). The bottom four panels show example images taken from the pool of 100 that were used to calibrate the "theoretical uncertainty" caused by different physical parameters and turbulence realizations in the GRMHD simulations (Part II). 40 $\mu$as scale bars are shown in white.

which were generated two sets of synthetic data targeting different aspects of the calibration.

The first group of synthetic data sets, Part I, was created using only three simulated GRMHD images (shown in the top three panels of Figure 24). From each of these images 10 synthetic data sets were generated, corresponding to 10 different realizations of the observational noise (i.e., thermal noise and known systematics such as station gain fluctuations and polarization leakage) spanning the full range of $(u, v)$-coverage in the 2017 EHT M87 data and containing independent instances of thermal and systematic errors. The goal of the Part I data sets is to assess the impact of such observational corruptions on the recovered model parameters, which enters into the error budget for the calibration.

The second group of synthetic data sets, Part II, contains the remaining 100 GRMHD images (samples are shown in the bottom four panels of Figure 24). Only a single synthetic data set was generated for each of these images, using a randomly selected realization of the $(u, v)$-coverage from among the four days of 2017 M87 observations. The goal of the Part II data sets is to provide a sample of images for which we know both the underlying values of $\theta_g$ and the best-fit values of the diameter $\hat{d}$ from the GC model fits.

### D.2. $\theta_g$ Calibration

We know the input value of the angular gravitational radius $\theta_g$ for each simulation in Parts I and II of the GRMHD calibration data set. The basic calibration task is then to fit each of the corresponding synthetic data sets with a GC model and determine the scaling factor $\alpha$ (see Equation (27)) relating the best-fit crescent diameter to $\theta_g$ for the underlying simulation.

The value of this scaling factor is equal to the diameter measured in units of $\theta_g$, which Figure 25 shows plotted for all simulations in the calibration data set. We obtain an estimate of the scaling factor from each individual fit in Parts I and II, and we use the mean of these fits as our final determination of $\alpha$; the uncertainty in this mean value is roughly an order of magnitude smaller than any of the calibration uncertainties we consider below, and we thus neglect it in our error budget. Table 4 lists the calibrated $\alpha$ values, which we determine separately for THEMIS and `dynesty`.

We can see in Figure 25 that each set of 10 synthetic data sets corresponding to a single simulation from Part I shows excellent agreement across all measured $d$ values. Comparing these to the measurements from Part II demonstrates that the inter-simulation scatter is substantially larger than the inter-observation scatter, and that the former will be the limiting factor in measuring $\theta_g$ from the M87 data.

### D.3. Uncertainty Budgeting for All Parameters

In general, the error budget for any single measurement of $\theta_g$ will have three contributions: (1) a "statistical uncertainty" associated with the width of the posterior, (2) an "observational uncertainty" associated with incomplete $(u, v)$-coverage and unmodeled systematics that change from one observation to the next, and (3) a "theoretical uncertainty" associated with the data being a single sample from a dynamic system whose specific physical properties we are largely ignorant of. For other measured parameters that lack a counterpart input to the GRMHD simulations, we only consider the first two uncertainty components.

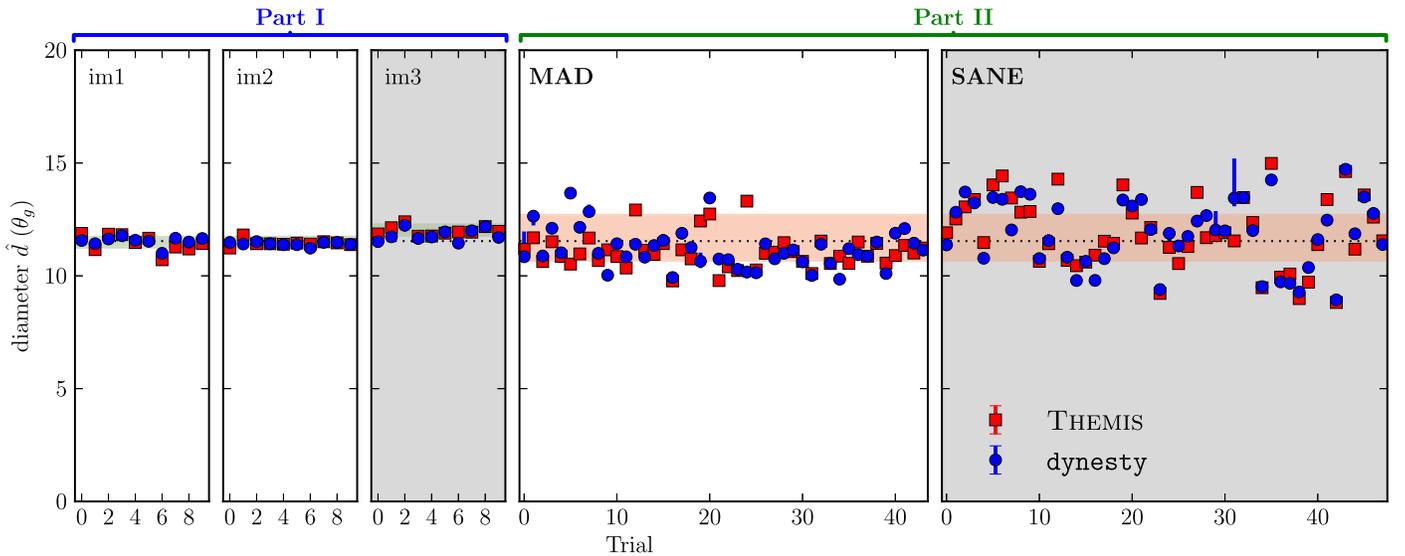

**Figure 25.** Reconstructed diameters using the GC models from synthetic data generated from GRMHD simulations. Green bands show the one-sided 68-percentile ranges within the Part I reconstructions (left). The red bands show the one-sided 68-percentile range of the combined Part I and II reconstructions (right). These are separated by GRMHD simulation type: MAD models are collected in white panels, while SANE models are collected in gray panels. The dotted lines indicate the MAD+SANE calibration factor, $\alpha$ for the xs-ringauss model (see Table 4).

To quantify the magnitude of these different uncertainty components, we employ a generalized lambda distribution (G$\lambda$D). The G$\lambda$D is an extension of the Tukey $\lambda$ distribution (Tukey 1962) that provides a compact parameterization for representing a diverse family of probability densities. The G$\lambda$D is also flexible and convenient, permitting highly asymmetric and heavy-tailed distributions to be analytically represented in terms of a function that depends only on their quantiles. We use the FMKL parameterization (Freimer et al. 1988), defined in terms of the inverse cumulative distribution function (CDF), which can be expressed using four parameters ($\lambda_1$, $\lambda_2$, $\lambda_3$, $\lambda_4$) as

$$Q_{\mathrm{FMKL}}(q) = \lambda_1 + \frac{1}{\lambda_2}\left[\frac{q^{\lambda_3}-1}{\lambda_3} - \frac{(1-q)^{\lambda_4}-1}{\lambda_4}\right], \quad (59)$$

where $q$ is the quantile of the distribution. The fits were performed using the GLDEX package in R (Su 2007a, 2007b).

The G$\lambda$D fits to all three components of the error budget are shown in Figure 26 for the $\theta_g$ measurements from each of the two fitting codes. We find that the theoretical uncertainty is dominant, being larger by a factor of ~4–5 than either the statistical or observational components. However, the magnitude of this theoretical uncertainty depends strongly on the category of simulation being used for calibration. We find that SANE simulations show nearly a factor of ~2 larger theoretical uncertainty than MAD simulations; the increased SANE scatter is visually apparent in Figure 25, and the effective decrease in calibration uncertainty that would result from using only MAD simulations is listed in Table 4. However, because such a preference for MAD over SANE is not well motivated from a theoretical standpoint (Paper V), we use the calibration factor determined from MAD+SANE for the measurements presented in this Letter.

The output of a MCMC or NS fitting run is a "chain" of samples from the posterior distribution. We determined the statistical uncertainty component for each parameter by fitting a single G$\lambda$D to the ensemble of chains from all 130 Part I and II data sets simultaneously. Each chain was first mean-subtracted so that the G$\lambda$D characterizes only the shape of the typical posterior distribution for a single GC model fit, rather than capturing variation in mean value between different fits.

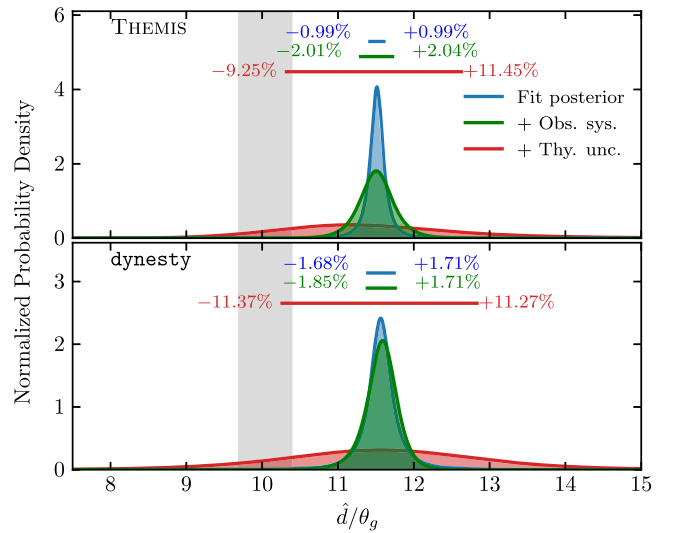

**Figure 26.** Sources and magnitudes of random and systematic uncertainty on the measurement of $\theta_g$ with GC models. These include the average posterior (blue), the impact of different realizations of unmodeled observational systematic errors (e.g., polarization leakage, ($u$, $v$)-coverage, etc.; green), and the impact of variations in the assumed underlying GRMHD simulation (red). (The latter two match the appropriately colored ranges shown in Figure 25.) These are similar for analyses that employ the xs-ringauss-based (Themis; top) and xs-ring-based (dynesty; bottom) GC models. The gray band shows the range of photon ring diameters for a Kerr black hole from spin 0 to 0.998 over all viewing inclination angles.

The observational uncertainty component was determined using Part I data sets only. For each of the three simulations in Part I, we combined the 10 chains corresponding to the individual realizations of observational uncertainty for that simulation. We then subtracted out the mean of each combined chain before further combining all three chains and fitting a G$\lambda$D to the resulting stack. This fitted distribution then describes the typical variation in individual parameter measurements that we expect to see across multiple observations of the same underlying emission structure. The derived observational uncertainties are listed in Table 3 for each parameter.

The theoretical uncertainty component was determined using synthetic data sets from Parts I and II, and applies only to $\theta_g$. We first made a fit-quality cut on the Part II data sets, removing all that had a measured diameter of $\hat{d} > 15\theta_g$; these outlying fits were deemed poor upon inspection, as they had arrived at large-diameter solutions by fitting the crescent portion of the model to extraneous emission while the Gaussian components were trying to form the ring. We found eight such fits out of 100. The chains for the remaining 92 data sets from Part II and three data sets from Part I (one randomly selected for each image) were then combined without performing any mean subtraction, and a G$\lambda$D was fit to the entire set.

Note that this is by far the largest component of the calibration uncertainty. That is, the emission distribution can vary significantly in position and structure. This is reflected both in its net shift away from the photon ring size ($\alpha \simeq 11.5$ instead of $\simeq 10$) and in its scatter between model images.

## Appendix E
## Using GRMHD Simulations to Calibrate Image Domain Feature Extraction

To create a mapping between image domain feature extraction methods and $\theta_g$, we follow an analogous method to that described in Appendix D. For the same simulated data sets used in Appendix D, we generate reconstructed images using the RML methods as implemented in `eht-imaging` and `SMILI` and the CLEAN algorithm as implemented in `DIFMAP` (Paper IV). The images are generated with the fiducial hyper-parameters determined from the M87 Top Set. The spread in diameters from varying the hyper-parameters can be significant (Figure 15). Still, in the final error budget this term is significantly smaller than the full calibration error measured here (Table 6).

The reconstructed images frequently exhibit the dominant ring structures found in the underlying GRMHD model images. Occasionally, however, significant non-ring features are present, e.g., a single compact component or multiple rings. We discard images with such pathologies. For `DIFMAP`, `eht-imaging`, and `SMILI`, this leaves all 30 images from Part I, and 70, 86, and 96 of the 100 images from Part II. The acceptance rates for `eht-imaging` and `SMILI` are similar to those found for the geometric crescent models (92/100). The large number of rejected `DIFMAP` frames could lead to a bias in either the resulting calibration factor $\alpha$ or to an underestimate of the theoretical uncertainty component in the error bar on $\theta_g$ from this method.

The image domain feature extraction method described in Section 7 and Paper IV is then used to estimate the ring diameters of the Part I and Part II images. We find a qualitative similarity between the diameter measurements from the image reconstructions and those obtained in Figure 25: the Part II images exhibit substantially more scatter than the Part I images, and within Part II the SANE models are less tightly constrained than MAD.

We generate an error budget for the image domain diameter measurements in a fashion similar to that employed in Appendix D. The primary difference between the two budgets arises from the lack of posterior distributions reported using the image reconstructions. We estimate the combined statistical and observational systematic uncertainty components from the comparisons in Part I. The analysis of Part II retains the same meaning as in Appendix D, providing a measurement of the observational uncertainties (statistical and systematic) combined with the theoretical contribution to the systematic uncertainty associated with variations in the underlying intrinsic image structure. Both sets of uncertainties appear in Table 6.

## Appendix F
## Average Image Scoring

Assessing the consistency of a given GRMHD model with the EHT observations is complicated by stochastic features in the image, associated with turbulence in the underlying accretion flow. As a result of these, no single snapshot image from a simulation is expected to provide a high-quality fit directly to the EHT observations when only observational errors are included. However, the self-consistency of a model may be directly assessed by numerically constructing the anticipated distribution of $\chi^2$ values using the ensemble of snapshot images from a GRMHD model.

Note that this procedure is identical in principle to that typically employed for assessing fit quality in the presence of Gaussian errors: the construction of a $\chi^2$ and its comparison to the standard $\chi^2$ distribution. In the case of interest here, the stochastic image fluctuations may be thought of as an additional noise term that is specified by the model. Apart from the need to numerically construct the appropriate $\chi^2$ distribution, its interpretation is very similar to the normal procedure.

The AIS method formalizes this process and makes use of key efficiencies permitted by the small observational errors in comparison to the stochastic fluctuations. In order to both formulate a concrete method and ensure computational feasibility, the THEMIS-AIS method employs a SSM based on the arithmetic average image from the simulation. Therefore, the procedure for each GRMHD model and observation day is as follows.

1. Generate an SSM based on the average snapshot image.
2. Numerically construct a CDF of reduced $\chi^2$ values by fitting this SSM to simulated data generated from each snapshot image associated with the GRMHD model of interest for the particular observation day of interest.
3. Finally, fit this SSM to the data, obtain the associated reduced $\chi^2$ by fitting to the corresponding EHT observations, and assess in light of the previously constructed CDF.

The second step is the most onerous, requiring the generation of many additional simulated data sets. However, the dominance of the stochastic image features over the observational noise simplifies this process considerably. The CDF constructed from fitting the average image to the frames is similar to the CDF constructed from fitting the frames to the average image. This has been verified explicitly, and the resulting CDFs constructed both ways are indistinguishable. For each GRMHD model and observation day this procedure must be repeated, resulting in a family of more than 1200 CDFs generated by the THEMIS-AIS procedure.

The final step, interpreting the fit to the observations, is quantified by the two-sided quantile of the fit reduced $\chi^2$:

$$p_{\mathrm{AIS}}(\mathcal{M}|\boldsymbol{D}) \approx \frac{N_{>|\chi^2-\chi^2_{\mathrm{med}}|}}{N}, \tag{60}$$

where $N$ and $N_{>|\chi^2-\chi^2_{\mathrm{med}}|}$ are the number of frames for a given simulation, $\mathcal{M}$, and the number of those with a reduced $\chi^2$ further from the median, $\chi^2_{\mathrm{med}}$ than that observed in the data. Note that this may be directly related to the likelihood of $\mathcal{M}$ via

$$P(\mathcal{M}|\boldsymbol{D}) = \int d\chi^2 \frac{dp_{\mathrm{AIS}}}{d\chi^2} P(\chi^2|\boldsymbol{D}, \mathcal{M})$$
$$\approx p_{\mathrm{AIS}}(\mathcal{M}|\boldsymbol{D}) \tag{61}$$

where $P(\chi^2|\boldsymbol{D}, \mathcal{M})$ is the distribution of fit reduced $\chi^2$ and we have assumed that the distribution of $\chi^2$ is broad and extends beyond the tails of the intrinsic $\chi^2$ distribution in the relevant direction.

A qualitative illustration of this interpretation that uses only three closure phases as a proxy for the total reduced $\chi^2$ is shown in Figure 27 for three GRMHD models that show the three potential outcomes. The direct comparisons to the full CDFs for each of these models are shown in Figure 28. For the first (top row), the typical deviation between the closure phases of the average image (red diamond) and those from the individual snapshot images (blue points) are similar to the deviation between the former and the EHT data (green triangle); this model has a $p_{\mathrm{AIS}}$ near unity.

In the second (middle row), the EHT data are further from the average image than the typical snapshot image; this model has a $p_{\mathrm{AIS}} < 0.01$. This is a model for which no snapshots are likely to be an adequate description of the data, regardless of computational limits. In the third (bottom row), the EHT data are *closer* to the average image than is typical; again the $p_{\mathrm{AIS}} = 0.01$, and this model is excluded. In this case it is likely that a snapshot could be found, given a sufficiently large number of images from this GRMHD model, that fits the EHT observations well. However, the fact that the reduced $\chi^2$ is much less than that expected implies that the "noise" model, the distribution of stochastic image fluctuations, is wrong. That is, this GRMHD model is too variable. This is akin to finding a reduced $\chi^2$ much less than unity in the standard fitting process.

Finally, note that the "true" model is necessarily accepted by the THEMIS-AIS procedure. That is, were a GRMHD simulation to produce the true image on the sky, the probability for it to be assigned a score of $p_{\mathrm{AIS}}$ is identically $p_{\mathrm{AIS}}$. Given a choice of the minimum acceptable $p_{\mathrm{AIS}}$, the probability of excluding the true model is then exactly that value (e.g., here 1%).

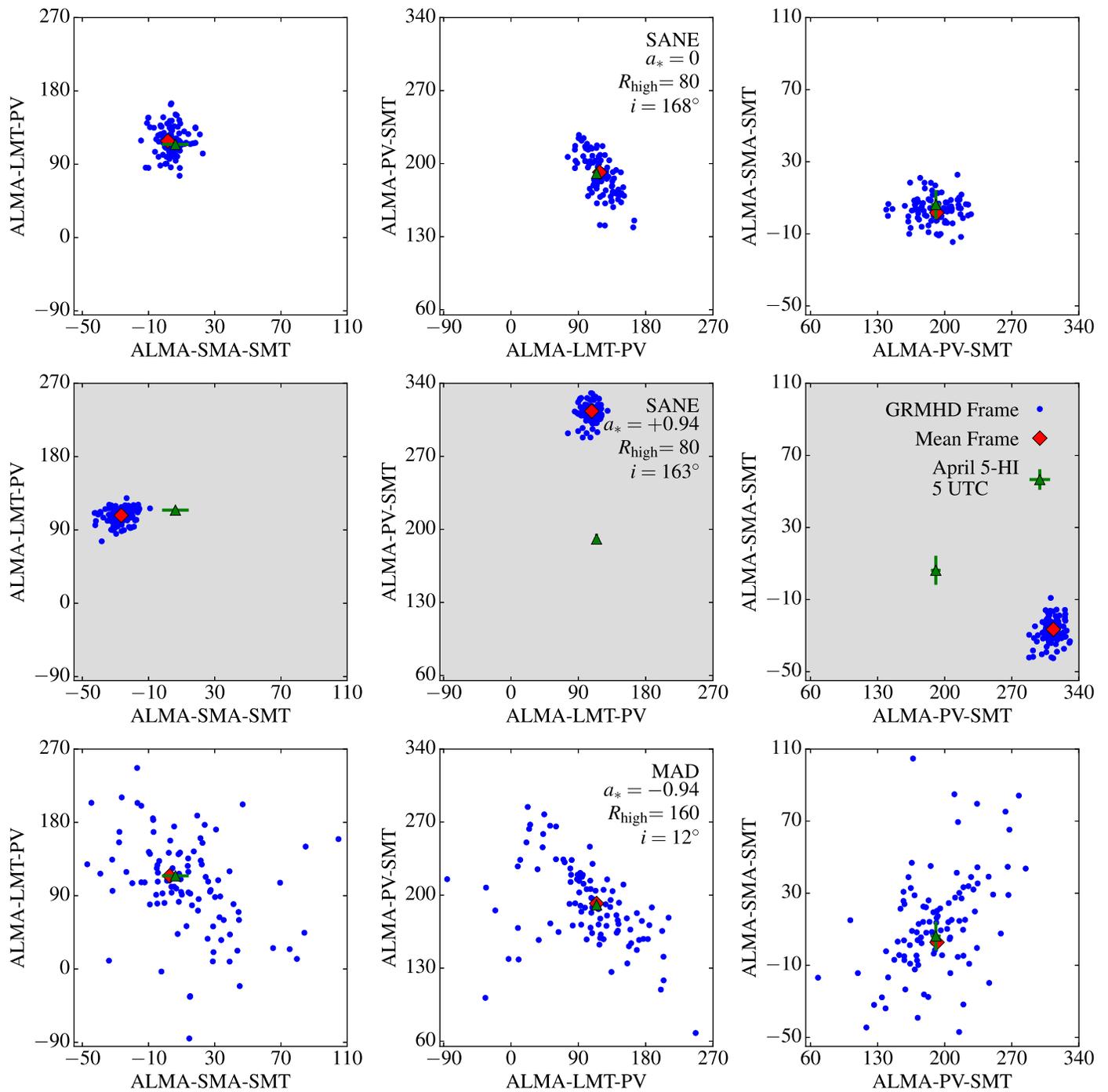

**Figure 27.** Illustration of "good" (top panels) and two "bad" (middle and bottom panels) models following the THEMIS-AIS procedure. Closure phases are shown for three triangles at 5 UTC for April 5 as proxies for the $\chi^2$. Blue dots indicate the closure phases for the best-fit SSM associated with each simulation snapshot image. The red diamond shows the same for the average snapshot image. The green triangle shows the observed values on April 5, high-band, near 5 UTC. The manner in which the two models shown in the middle and bottom rows are excluded differs: in the middle case the reduced $\chi^2$ is too large, while in the bottom case the $\chi^2$ is too small to be consistent with that anticipated by the individual GRMHD model snapshots.

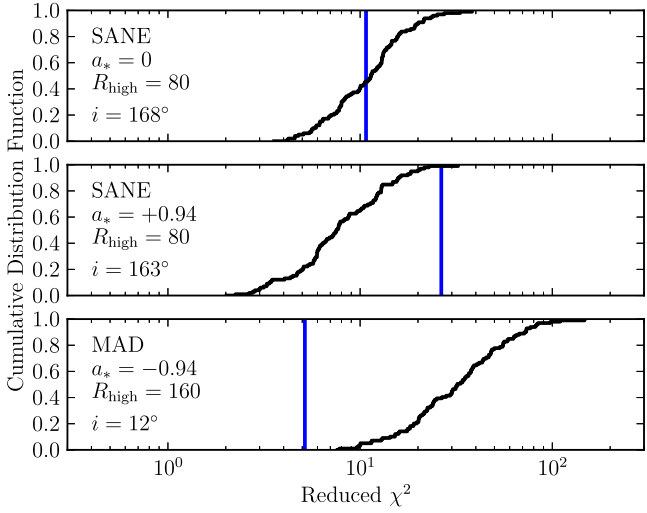

**Figure 28.** Anticipated cumulative distribution function of reduced $\chi^2$ for the SSM for the same three representative models shown in Figure 27 for the April 5 high-band data. For comparison, the measured reduced $\chi^2$ is indicated by the vertical blue line for each model.

## Appendix G
## GRMHD Model Parameter Estimation

The presence of a significant stochastic component in the snapshot images from GRMHD models complicates the interpretation of posterior distributions of the SSM parameters. This is exacerbated by the very incomplete coverage of the turbulent realizations by the simulation snapshots (relative to the observational noise) in a way similar to that found during model selection (Appendix F). Here we explicitly describe how we address this.

### G.1. Ensemble-based Posterior Construction

The GRMHD models effectively comprise an ensemble of snapshot images, which simultaneously identify a "typical" image and the statistical distribution of the stochastic features within the image. In principle, this stochasticity can be addressed by introducing hyper-parameters associated with this additional image structure, subject to priors obtained from the statistics of the fluctuations within the snapshot ensemble, and subsequently marginalizing over them. In practice, we approximate this procedure by instead fitting and marginalizing over each snapshot image independently. Thus, we formally define the joint model-parameter posterior distributions for the SSM model, following Bayes' theorem, by

$$P(\Theta, \mathcal{M}|\boldsymbol{D}) = \sum_{\mathcal{I}} \frac{P(\boldsymbol{D}|\Theta, \mathcal{I}) P(\mathcal{I}|\mathcal{M}) \pi(\Theta) \pi(\mathcal{M})}{P(\boldsymbol{D})}, \quad (62)$$

where $P(\boldsymbol{D}|\Theta, \mathcal{I})$ is formally the likelihood, $\mathcal{I}$ refers to a particular snapshot image from $\mathcal{M}$ obtained with probability $P(\mathcal{I}|\mathcal{M})$, and $\pi(\Theta)$ and $\pi(\mathcal{M})$ are the priors on parameters $\Theta$ and the model $\mathcal{M}$, respectively. We now discuss each of these terms separately.

The prior on a given model is obtained from the THEMIS-AIS procedure combined with theoretical priors based on X-ray luminosity, jet power, and radiative efficiency (see Table 2 of Paper V). Because the probability of finding a snapshot image within the ensemble generated for a particular GRMHD model that fits the EHT observations well when only observational errors are considered is small, we do not make significant distinctions between values of $p_{\mathrm{AIS}}(\mathcal{M}|\boldsymbol{D})$ above some threshold, $p_{\mathrm{AIS,0}}$ (see Appendix F). Thus, we set

$$\pi(\mathcal{M}) \approx \Theta[p_{\mathrm{AIS}}(\mathcal{M}|\boldsymbol{D}) - p_{\mathrm{AIS,0}}], \quad (63)$$

where $\Theta(x)$ is the Heaviside function. The priors on the SSM parameters, $\pi(\Theta)$, are described in Section 6.3.

The interpretation of the likelihood, $P(\boldsymbol{D}|\Theta, \mathcal{I})$, is again complicated by the substantial stochastic image components. In the presence of only data-based uncertainties, this likelihood is formally no different from that described in Section 4.1. In practice, we find that the spread in "best-fit" $\Theta_{p;\mathcal{I}}$ across different snapshots within a single GRMHD model is much larger than the distribution of the likelihood function for each individual snapshot (see Figure 29).

The value of $P(\boldsymbol{D}|\Theta_{p;\mathcal{I}}, \mathcal{I})$ is misleading due to the sparse sampling of the stochastic image components within the snapshot ensembles. That is, for models deemed acceptable by the AIS procedure, even large reduced $\chi^2$ (defined relative to the observational noise estimates) may be "high quality" in that they are well within the range anticipated by the stochastic variability. For the best of these, the expectation is that a realization of the turbulence close to that in the relevant snapshot image would constitute a "good" fit in the normal sense, i.e., a reduced $\chi^2$ of unity; that such a fit was not found is a consequence only of the necessarily sparse nature of the snapshot image ensemble. Therefore, these "high-quality" fits are indicative of parameters for which a realization of the stochastic component could adequately explain the EHT observations.

We address this in two steps. First, by specifying "high quality" within a particular model by ranking the fits relative to the likelihood (THEMIS) or $\chi^2$ (GENA) and placing a cut at some fractional level, e.g., for the THEMIS pipeline cutting on

$$q_{\mathcal{I},\mathcal{M}} = \frac{N_{>P(\boldsymbol{D}|\Theta, \mathcal{I})}}{N_{\mathcal{M}}}. \quad (64)$$

Fits with $q_{\mathcal{I},\mathcal{M}}$ above some cutoff $q_0$ are assumed to be sufficiently close to a formally acceptable fit in that a nearby realization of the stochastic image features could be found. Second, by making use of the comparatively narrow nature for the high-quality SSM posteriors

$$P(\boldsymbol{D}|\Theta, \mathcal{I}) \approx \delta^3(\Theta - \Theta_{p;\mathcal{I}}) \Theta(q_{\mathcal{I},\mathcal{M}} - q_0). \quad (65)$$

Combining these, the estimated posteriors on the SSM parameters after marginalizing over all models and images is then

$$\begin{aligned}
P(\Theta|\boldsymbol{D}) &= \sum_{\mathcal{M}} P(\Theta, \mathcal{M}|\boldsymbol{D}) \\
&= \frac{1}{P(\boldsymbol{D})} \sum_{\mathcal{M},\mathcal{I}} \delta^3(\Theta - \Theta_{p;\mathcal{I}}) \Theta(q_{\mathcal{I},\mathcal{M}} - q_0) \Theta \\
&\quad \times [p_{\mathrm{AIS}}(\mathcal{M}|\boldsymbol{D}) - p_{\mathrm{AIS,0}}] \pi(\Theta).
\end{aligned} \quad (66)$$

Key meta-parameters in this estimate are the quality cuts on models, $p_{\mathrm{AIS,0}}$, and on snapshot images within a model, $q_0$. In the following, we note that the posteriors are only weak functions of $p_{\mathrm{AIS,0}}$, and are converged for $q_0 < 0.5$, i.e., accepting snapshots with fit qualities above the median.

This procedure, while credible, does require a number of assumptions about the utility of what are ultimately low likelihood fits when only observational noise is considered. While practically these comport with the expectation from the underlying GRMHD simulations (see Section 6.4), it may result in intrinsic biases. Therefore, we now turn to validating the method with mock analyses.

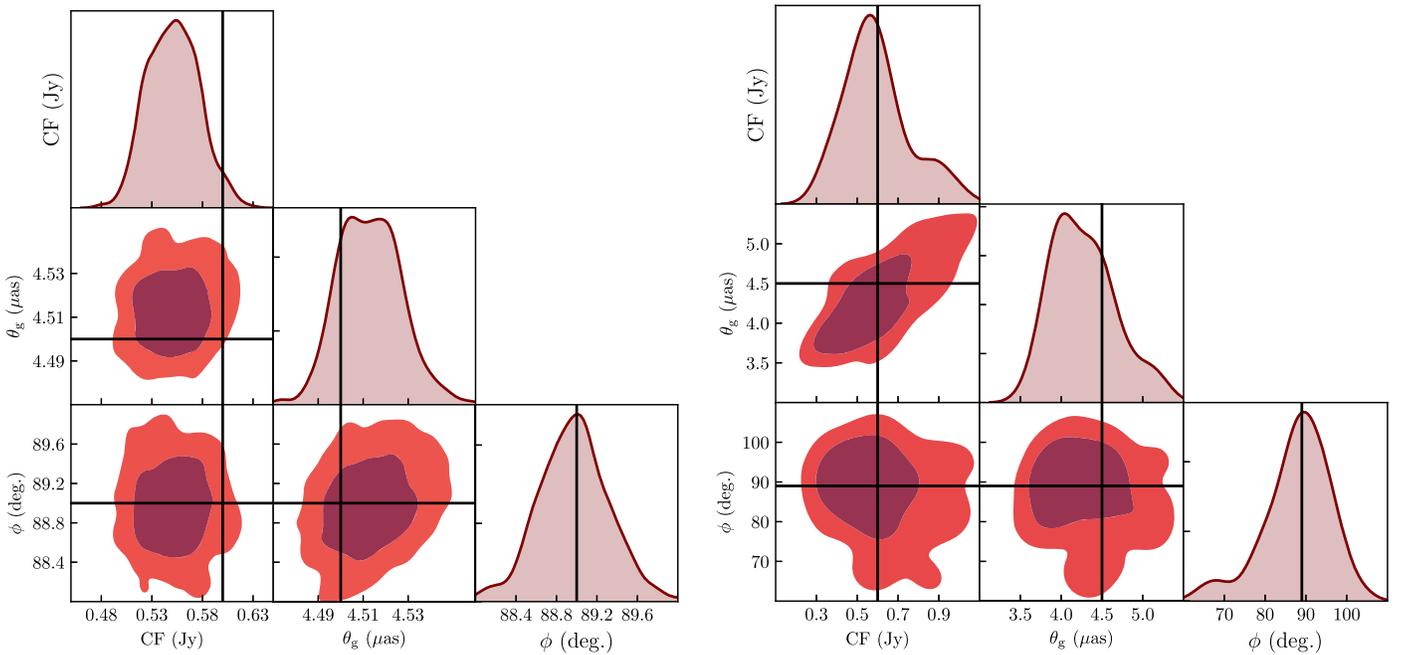

**Figure 29.** Example joint posterior plots for the SSM parameters from one of the many mock analyses performed as part of the ensemble-based posterior estimation validation. Shown are the distributions for compact flux density, $\theta_g$, and PA from SSM analyses of simulated data sets using individual frame snapshots. Left panel: the posteriors from fitting *only* the frame from which the data was generated. Right panel: the stacked posteriors of snapshots with a likelihood larger than the median for the GRMHD simulation from which the simulated data was generated (i.e., $q < 0.5$). Note that the ranges on the right are much larger than those on the left. The underlying model was a SANE-type accretion flow with $a_* = 0.94$, $i = 163°$, $R_{high} = 80$, and mass rescaled such that $\theta_g = 4.5$ μas. The data was generated for the April 5 $(u, v)$-coverage. Contours enclose 68% and 95% of the posterior.

### G.2. Validation with Mock Analyses

The first set of validation tests involves the construction of mock data from a GRMHD snapshot and subsequent analysis. In the creation of the mock data, the $(u,v)$-coverage from M87 observations was employed and realistic realizations of all known systematic errors were included (e.g., polarization leakage and gain fluctuations). In some cases, these tests have been repeated hundreds of times. See Appendix D.1 and Paper IV for descriptions of synthetic data generation.

Analyses using an SSM based on the snapshot from which the mock data was generated demonstrate the ability to reconstruct the SSM parameters, i.e., reconstruct $P(\boldsymbol{D}|\Theta, \mathcal{I})$. An example, taken from the calibration data sets employed in Appendix D, is shown in the left panel of Figure 29. This analysis has been performed for all 130 of the sets described in Appendix D, and for more than 1200 additional data sets as part of the THEMIS-AIS scoring procedure and internal validation. Generally, we find that the input SSM parameters are recovered within the very narrow posterior distributions.

Ensemble-based posterior reconstruction was performed for all 130 simulated data sets described in Appendix D and the more than 1200 sets constructed as part of the THEMIS-AIS procedure using the "truth" model, i.e., for the GRMHD model from which a snapshot was selected from which to generate the data. This directly tests the ability to reconstruct $P(\Theta|\boldsymbol{D})$ for a *single* GRMHD model, i.e., Equation (66). An example is shown in the right panel of Figure 29 for the same example data

set shown in the left panel. As expected, the true parameters lie within the ensemble-based posterior distribution. Small net biases, defined by the difference of mean parameter value and the "truth" in units of the posterior width (measured by standard deviation, $\sigma$), have been identified: less than $0.3\sigma$ in the compact flux, less than $0.07\sigma$ in $\theta_g$, and less than $0.02\sigma$ in $PA_{FJ}$.

For two mock data sets, the ensemble-based posterior estimate, $P(\Theta|\boldsymbol{D})$ in Equation (66), was reconstructed for the full GRMHD model library *without* model selection, i.e., $p_{AIS,0} = 0$. Again, as expected, the true parameters lie within the ensemble-based posterior distribution.

### G.3. Convergence Tests

There are a number of what might be deemed convergence tests, i.e., comparisons of the ensemble-based posteriors constructed in various ways that differ in the mechanics but not the underlying physical inputs.

Were the posteriors sensitive to outliers, we may have anticipated them to change with total number of snapshots used. A subset of GRMHD models had 500 snapshots generated from them, in comparison to the 100 snapshots that was standard. These do not show materially different ensemble-based posteriors when a subsample of 100 snapshots are used and when the full 500 snapshots are used.

The ensemble-based posteriors are typically generated from only a subset of the SSM analyses associated with a given GRMHD model, determined by a fit quality cut. This is characterized in terms of a likelihood or $\chi^2$ percentile, with the expectation that particular bad fits are likely dominated by the peculiarities of the realization of the stochastic components. The ensemble-based posteriors cease to evolve after reducing the percentile cut below 50%; i.e., once the likelihoods ($\chi^2$) that are included are above (below) the median value. This suggests that these are not contaminated by particularly poor fits.

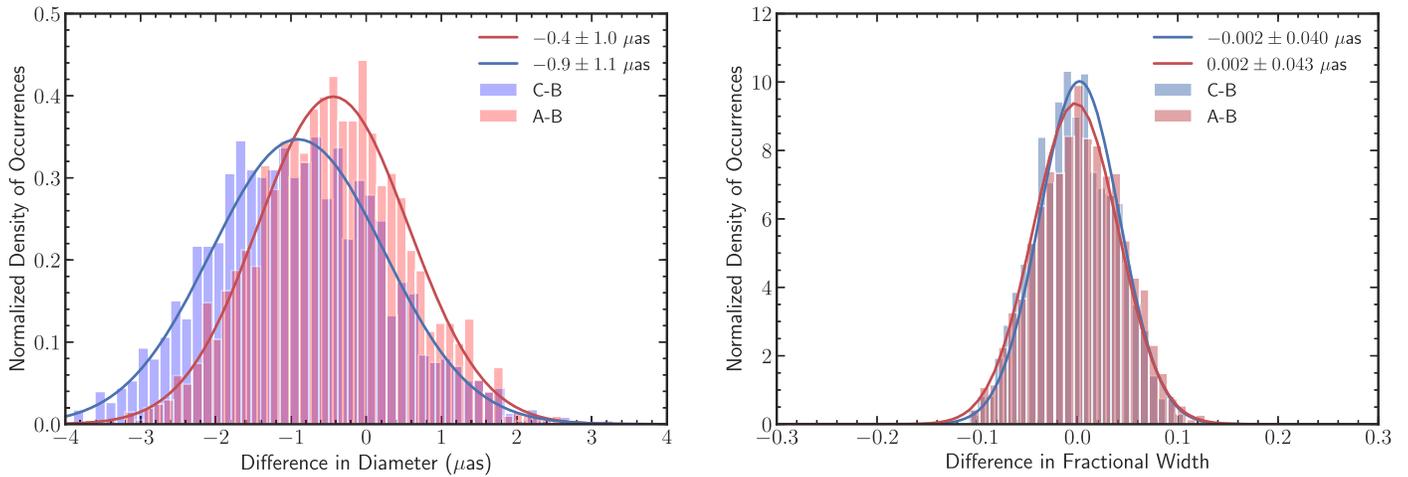

**Figure 30.** Differences in the measured (left panel) diameters and (right panel) fractional widths of the Top Set images reconstructed with eht-imaging using the April 5 M87 data when applying three different variants (A, B, and C; see the text for details) of the method for measuring these parameters. In both panels, the solid lines show Gaussian distributions with means and standard deviations equal to those of the histograms. The typical differences between the parameters measured with the three variants are smaller than the pixel size.

A subset of GRMHD models have been run with different GRMHD simulation codes and resolutions (and imaging resolutions); i.e., for the same black hole and initial accretion flow parameters, multiple GRMHD simulations have been performed. Individual SSM analyses do depend modestly on the GRMHD simulation resolution and the resolution of the snapshot image, though this is far smaller than the typical size of the ensemble-based posterior. Comparisons of the ensemble-based posteriors for identical simulation setups among GRMHD codes indicate that they are very consistent, implying that the number of frames is sufficient to adequately cover the posterior.

## Appendix H
## Image Domain Feature Extraction Method Tests

In order to explore and quantify the sensitivity of our measurement of the image parameters to the presence of noise in the images, the effect of pixelization, and the presence of outliers in the pixel brightness distribution, we performed a number of validation tests. We used the images reconstructed with eht-imaging data on April 5 and analyzed them using two different algorithmic implementations and three variants of the methods described above. Case A refers to the above definitions. Case B refers to using the same definitions but after having smoothed with a 2 μas Gaussian (Paper IV Section 9 uses this method to find the ring center before measuring the width and diameter from the unblurred image). Case C uses unsmoothed images but locates the centers by minimizing the difference between the 25th and the 75th percentile distances to the peak brightness over all azimuthal angles. It then calculates the image diameter using the median of the peak brightness distances from the center over all azimuthal angles. Case B tests the sensitivity of the measured parameters against the presence of noise in the image reconstruction. Case C tests against biases that might be introduced by outliers in the azimuthal cross sections.

Figure 30 shows the fractional differences in the measured diameters and fractional widths inferred using these three different variants. The differences in diameter between the three cases have means of 0.4–0.9 μas and comparable standard deviations, whereas the differences in fractional width have a mean of zero and a standard deviation of 0.04. These are smaller than the differences in the feature parameters inferred from the same data using different visibility- and image-domain techniques. In Section 7, we use Case B.

## Appendix I
## Prior Measurements of the Distance to M87 and the Dynamical Mass of Its Central Black Hole

### I.1. Prior Distance Measurements

We provide the details of the methods and the uncertainties included in each distance measurement here.

*TRGB method.* The TRGB method leads to a direct distance measurement by utilizing the bright-end distributions of the red giant stars as a primary standard candle. Bird et al. (2010) reported the absolute distance modulus of $\mu = 31.12 \pm 0.14$, using data sets of the *HST* Advanced Camera Survey (ACS) Virgo Cluster Survey (VCS) that resolved the brightest red giant stars in M87 for the first time.

*SBF method.* The SBF method also utilizes red giant stars. It is often categorized as a secondary standard-candle method because it relies for its calibration on nearby galaxies for which distances are independently measured with resolved-star methods. Therefore, SBF measurements in Virgo Cluster surveys provide *relative* distances of Virgo galaxies to the calibrated mean distance of the entire Virgo Cluster. To date, there are two distinct measurements that fall into this category. The first one by Blakeslee et al. (2009) uses the *HST* ACS-VCS data and finds $\mu = 31.11 \pm 0.08$ (denoted hereafter as $\mu_{SBF1}$; see also Mei et al. 2007). A second, more recent study by Cantiello et al. (2018a) reported $\mu = 31.15 \pm 0.04$ (denoted hereafter as $\mu_{SBF2}$) based on data from the Next Generation Virgo Cluster Survey (NGVCS), obtained using ground-based adaptive-optics with the Canada French Hawaii Telescope.

There are a couple of other sources of systematic errors in SBF measurements. The first one is a systematic error of 0.1 mag on the distance zero-point (Cantiello et al. 2018b),

**Table 9**
Relevant Literature and Derived Quantities

| Method | Measurement Type | Units | Value | References |
|---|---|---|---|---|
| TRGB | $D$ | Mpc | $16.75^{+1.11}_{-1.04}$ | Bird et al. (2010) |
| SBF | $D$ | Mpc | $16.67^{+1.02}_{-0.96}$ | Blakeslee et al. (2009) |
| SBF | $D$ | Mpc | $16.98^{+0.96}_{-0.91}$ | Cantiello et al. (2018a) |
| Gas dynamics | $\theta_{dyn}$ | $\mu$as | $2.05^{+0.48}_{-0.16}$ | Walsh et al. (2013) |
| Stellar dynamics | $\theta_{dyn}$ | $\mu$as | $3.62^{+0.60}_{-0.34}$ | Gebhardt et al. (2011) |
| Product of two SBF measurements | $D$ | Mpc | $16.82^{+1.02}_{-0.86}$ | This work |
| Product of three measurements | $D$ | Mpc | $16.76^{+0.79}_{-0.66}$ | This work |
| Gas dynamics | $M$ | $10^9\,M_\odot$ | $3.45^{+0.85}_{-0.26}$ | Walsh et al. (2013); this work |
| Stellar dynamics | $M$ | $10^9\,M_\odot$ | $6.14^{+1.07}_{-0.62}$ | Gebhardt et al. (2011); this work |

**Note.** We report median values and 68% confidence intervals.

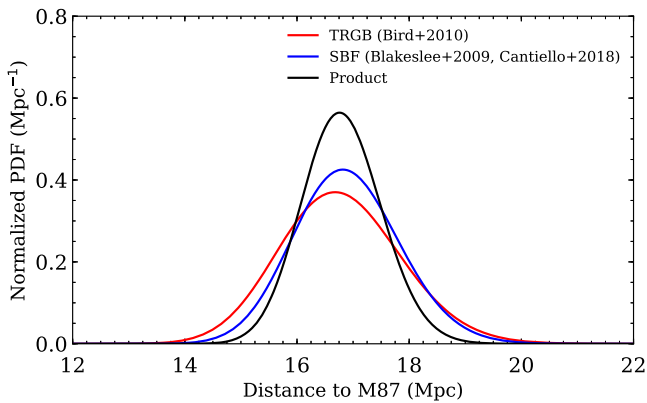

**Figure 31.** Normalized posteriors for the distance measurements to M87. Two colored lines show the posterior from the measurements using the TRGB method (Bird et al. 2010) and combined posterior from the SBF method (Blakeslee et al. 2009; Cantiello et al. 2018a). The black line shows the normalized product of three posteriors obtained by combining these measurements.

which was not accounted for in either of the two measurements. A second one, considered in Blakeslee et al. (2009) but not in Cantiello et al. (2018a), arise from the intrinsic scatter in the absolute fluctuation magnitude attributed to stellar populations (cosmic scatter; Tonry et al. 2000) and is estimated to be 0.06 mag (Blakeslee et al. 2009). Adding these in quadrature to the formal errors quoted above, we obtain for the absolute distant moduli $\mu_{SBF1} = 31.11 \pm 0.13$ (Blakeslee et al. 2009) and $\mu_{SBF2} = 31.15 \pm 0.12$ (Cantiello et al. 2018a). Finally, we average the two measurements weighted by their errors to get $\mu = 31.14 \pm 0.12$. We report these measurements in Table 9, together with the combined posterior as their product in Figure 31.

Assuming Gaussian posteriors $P(\mu)$ for the distance modulus measurements with the means and standard deviations given in Table 9, we calculate the posteriors for the distance $P(D)$ to M87 using the relation $\mu = 5\log_{10}(D) - 5$ as

$$P(D) = \frac{5}{\log 10}\frac{P(\mu)}{D}. \tag{67}$$

Finally, considering the three distance measurements to be independent, we calculate the combined posterior as their product. We show the posteriors in Figure 31 and quote their most likely values and uncertainties in Table 9.

### I.2. Mass Determination

Both the stellar dynamical and the gas dynamical mass measurement techniques that we described in Section 8.2 measure, in reality, a mass-to-distance ratio rather than the mass of the central object directly. To convert these into posteriors over a mass, we first convert the measurements into posteriors over the mass-to-distance ratio $\mathcal{R}$ and then rescale them based on the posterior over distance that we described in the previous section.

We use the marginal $\chi^2$ distributions for mass $M_{BH}$ shown in Figure 6 of Gebhardt et al. (2011) and Figure 6 of Walsh et al. (2013) and set $M_{BH} = \mathcal{R} \times (17.9\ \text{Mpc})$. We denote the posteriors for the stellar and gas dynamical measurements as $P(\mathcal{R}_{stars})$ and $P(\mathcal{R}_{gas})$, respectively, and write

$$P(\mathcal{R}_i) \equiv \frac{\exp[-\chi^2(\mathcal{R}_i \times 17.9\ \text{Mpc})/2]}{\int \exp[-\chi^2(\mathcal{R}_i \times 17.9\ \text{Mpc})/2]\, d\mathcal{R}_i} \tag{68}$$

for each model ($i$ = gas, stars). Table 9 summarizes the credible regions of these distributions.

We can also convert the mass-to-distance ratio from dynamical measurements directly to an angular size of one gravitational radius at the distance of M87, $\theta_{dyn} \equiv GR/c^2$. This posterior is given by

$$P(\theta_{dyn}) = \frac{c^2}{G}P(\mathcal{R}). \tag{69}$$

We list the credible intervals for this quantity in Table 9.

We can now use the posterior over the mass-to-distance ratio and that over distance to calculate the posterior over the mass for the stellar and gas dynamics measurements as

$$P(M_i) = \int \frac{1}{D}P(D)P(\mathcal{R}_i)\,dD, \tag{70}$$

where $\mathcal{R}_i = M_i/D$.

### ORCID iDs

Kazunori Akiyama  https://orcid.org/0000-0002-9475-4254
Antxon Alberdi  https://orcid.org/0000-0002-9371-1033
Rebecca Azulay  https://orcid.org/0000-0002-2200-5393
Anne-Kathrin Baczko  https://orcid.org/0000-0003-3090-3975

Mislav Baloković ◎ https://orcid.org/0000-0003-0476-6647

John Barrett ◎ https://orcid.org/0000-0002-9290-0764

Lindy Blackburn ◎ https://orcid.org/0000-0002-9030-642X

Katherine L. Bouman ◎ https://orcid.org/0000-0003-0077-4367

Geoffrey C. Bower ◎ https://orcid.org/0000-0003-4056-9982

Christiaan D. Brinkerink ◎ https://orcid.org/0000-0002-2322-0749

Roger Brissenden ◎ https://orcid.org/0000-0002-2556-0894

Silke Britzen ◎ https://orcid.org/0000-0001-9240-6734

Avery E. Broderick ◎ https://orcid.org/0000-0002-3351-760X

Do-Young Byun ◎ https://orcid.org/0000-0003-1157-4109

Andrew Chael ◎ https://orcid.org/0000-0003-2966-6220

Chi-kwan Chan ◎ https://orcid.org/0000-0001-6337-6126

Shami Chatterjee ◎ https://orcid.org/0000-0002-2878-1502

Ilje Cho ◎ https://orcid.org/0000-0001-6083-7521

Pierre Christian ◎ https://orcid.org/0000-0001-6820-9941

John E. Conway ◎ https://orcid.org/0000-0003-2448-9181

Geoffrey B. Crew ◎ https://orcid.org/0000-0002-2079-3189

Yuzhu Cui ◎ https://orcid.org/0000-0001-6311-4345

Jordy Davelaar ◎ https://orcid.org/0000-0002-2685-2434

Mariafelicia De Laurentis ◎ https://orcid.org/0000-0002-9945-682X

Roger Deane ◎ https://orcid.org/0000-0003-1027-5043

Jessica Dempsey ◎ https://orcid.org/0000-0003-1269-9667

Gregory Desvignes ◎ https://orcid.org/0000-0003-3922-4055

Jason Dexter ◎ https://orcid.org/0000-0003-3903-0373

Sheperd S. Doeleman ◎ https://orcid.org/0000-0002-9031-0904

Ralph P. Eatough ◎ https://orcid.org/0000-0001-6196-4135

Heino Falcke ◎ https://orcid.org/0000-0002-2526-6724

Vincent L. Fish ◎ https://orcid.org/0000-0002-7128-9345

Raquel Fraga-Encinas ◎ https://orcid.org/0000-0002-5222-1361

José L. Gómez ◎ https://orcid.org/0000-0003-4190-7613

Peter Galison ◎ https://orcid.org/0000-0002-6429-3872

Charles F. Gammie ◎ https://orcid.org/0000-0001-7451-8935

Boris Georgiev ◎ https://orcid.org/0000-0002-3586-6424

Roman Gold ◎ https://orcid.org/0000-0003-2492-1966

Minfeng Gu (顾敏峰) ◎ https://orcid.org/0000-0002-4455-6946

Mark Gurwell ◎ https://orcid.org/0000-0003-0685-3621

Kazuhiro Hada ◎ https://orcid.org/0000-0001-6906-772X

Ronald Hesper ◎ https://orcid.org/0000-0003-1918-6098

Luis C. Ho (何子山) ◎ https://orcid.org/0000-0001-6947-5846

Mareki Honma ◎ https://orcid.org/0000-0003-4058-9000

Chih-Wei L. Huang ◎ https://orcid.org/0000-0001-5641-3953

Shiro Ikeda ◎ https://orcid.org/0000-0002-2462-1448

Sara Issaoun ◎ https://orcid.org/0000-0002-5297-921X

David J. James ◎ https://orcid.org/0000-0001-5160-4486

Michael Janssen ◎ https://orcid.org/0000-0001-8685-6544

Britton Jeter ◎ https://orcid.org/0000-0003-2847-1712

Wu Jiang (江悟) ◎ https://orcid.org/0000-0001-7369-3539

Michael D. Johnson ◎ https://orcid.org/0000-0002-4120-3029

Svetlana Jorstad ◎ https://orcid.org/0000-0001-6158-1708

Taehyun Jung ◎ https://orcid.org/0000-0001-7003-8643

Mansour Karami ◎ https://orcid.org/0000-0001-7387-9333

Ramesh Karuppusamy ◎ https://orcid.org/0000-0002-5307-2919

Tomohisa Kawashima ◎ https://orcid.org/0000-0001-8527-0496

Garrett K. Keating ◎ https://orcid.org/0000-0002-3490-146X

Mark Kettenis ◎ https://orcid.org/0000-0002-6156-5617

Jae-Young Kim ◎ https://orcid.org/0000-0001-8229-7183

Junhan Kim ◎ https://orcid.org/0000-0002-4274-9373

Motoki Kino ◎ https://orcid.org/0000-0002-2709-7338

Jun Yi Koay ◎ https://orcid.org/0000-0002-7029-6658

Patrick M. Koch ◎ https://orcid.org/0000-0003-2777-5861

Shoko Koyama ◎ https://orcid.org/0000-0002-3723-3372

Michael Kramer ◎ https://orcid.org/0000-0002-4175-2271

Carsten Kramer ◎ https://orcid.org/0000-0002-4908-4925

Thomas P. Krichbaum ◎ https://orcid.org/0000-0002-4892-9586

Tod R. Lauer ◎ https://orcid.org/0000-0003-3234-7247

Sang-Sung Lee ◎ https://orcid.org/0000-0002-6269-594X

Yan-Rong Li (李彦荣) ◎ https://orcid.org/0000-0001-5841-9179

Zhiyuan Li (李志远) ◎ https://orcid.org/0000-0003-0355-6437

Michael Lindqvist ◎ https://orcid.org/0000-0002-3669-0715

Kuo Liu ◎ https://orcid.org/0000-0002-2953-7376

Elisabetta Liuzzo ◎ https://orcid.org/0000-0003-0995-5201

Laurent Loinard ◎ https://orcid.org/0000-0002-5635-3345

Ru-Sen Lu (路如森) ◎ https://orcid.org/0000-0002-7692-7967

Nicholas R. MacDonald ◎ https://orcid.org/0000-0002-6684-8691

Jirong Mao (毛基荣) ◎ https://orcid.org/0000-0002-7077-7195

Sera Markoff ◎ https://orcid.org/0000-0001-9564-0876

Daniel P. Marrone ◎ https://orcid.org/0000-0002-2367-1080

Alan P. Marscher ◎ https://orcid.org/0000-0001-7396-3332

Iván Martí-Vidal ◎ https://orcid.org/0000-0003-3708-9611

Lynn D. Matthews ◎ https://orcid.org/0000-0002-3728-8082

Lia Medeiros ◎ https://orcid.org/0000-0003-2342-6728

Karl M. Menten ◎ https://orcid.org/0000-0001-6459-0669

Yosuke Mizuno ◎ https://orcid.org/0000-0002-8131-6730

Izumi Mizuno ◎ https://orcid.org/0000-0002-7210-6264

James M. Moran ◎ https://orcid.org/0000-0002-3882-4414

Kotaro Moriyama ◎ https://orcid.org/0000-0003-1364-3761

Monika Moscibrodzka ◎ https://orcid.org/0000-0002-4661-6332

Cornelia Müller ◎ https://orcid.org/0000-0002-2739-2994

Hiroshi Nagai ◎ https://orcid.org/0000-0003-0292-3645

Neil M. Nagar ◎ https://orcid.org/0000-0001-6920-662X

Masanori Nakamura ◎ https://orcid.org/0000-0001-6081-2420

Ramesh Narayan ◎ https://orcid.org/0000-0002-1919-2730

Iniyan Natarajan ◎ https://orcid.org/0000-0001-8242-4373

Chunchong Ni ◎ https://orcid.org/0000-0003-1361-5699

Aristeidis Noutsos ◎ https://orcid.org/0000-0002-4151-3860

Héctor Olivares ◎ https://orcid.org/0000-0001-6833-7580

Daniel C. M. Palumbo ◎ https://orcid.org/0000-0002-7179-3816

Ue-Li Pen ◎ https://orcid.org/0000-0003-2155-9578

Dominic W. Pesce ◎ https://orcid.org/0000-0002-5278-9221

Oliver Porth ◎ https://orcid.org/0000-0002-4584-2557

Ben Prather ◎ https://orcid.org/0000-0002-0393-7734

Jorge A. Preciado-López ◎ https://orcid.org/0000-0002-4146-0113

Hung-Yi Pu ◎ https://orcid.org/0000-0001-9270-8812

Venkatessh Ramakrishnan ◎ https://orcid.org/0000-0002-9248-086X

Ramprasad Rao ◎ https://orcid.org/0000-0002-1407-7944

Alexander W. Raymond ◎ https://orcid.org/0000-0002-5779-4767

Luciano Rezzolla ◎ https://orcid.org/0000-0002-1330-7103

Bart Ripperda ◎ https://orcid.org/0000-0002-7301-3908

Freek Roelofs ◎ https://orcid.org/0000-0001-5461-3687

Eduardo Ros ◎ https://orcid.org/0000-0001-9503-4892

Mel Rose ◎ https://orcid.org/0000-0002-2016-8746

Alan L. Roy ◎ https://orcid.org/0000-0002-1931-0135

Chet Ruszczyk ◎ https://orcid.org/0000-0001-7278-9707

Benjamin R. Ryan ◎ https://orcid.org/0000-0001-8939-4461

Kazi L. J. Rygl ◎ https://orcid.org/0000-0003-4146-9043

David Sánchez-Arguelles ◎ https://orcid.org/0000-0002-7344-9920

Mahito Sasada ◎ https://orcid.org/0000-0001-5946-9960

Tuomas Savolainen  https://orcid.org/0000-0001-6214-1085
Lijing Shao  https://orcid.org/0000-0002-1334-8853
Zhiqiang Shen (沈志强)  https://orcid.org/0000-0003-3540-8746
Des Small  https://orcid.org/0000-0003-3723-5404
Bong Won Sohn  https://orcid.org/0000-0002-4148-8378
Jason SooHoo  https://orcid.org/0000-0003-1938-0720
Fumie Tazaki  https://orcid.org/0000-0003-0236-0600
Paul Tiede  https://orcid.org/0000-0003-3826-5648
Remo P. J. Tilanus  https://orcid.org/0000-0002-6514-553X
Michael Titus  https://orcid.org/0000-0002-3423-4505
Kenji Toma  https://orcid.org/0000-0002-7114-6010
Pablo Torne  https://orcid.org/0000-0001-8700-6058
Sascha Trippe  https://orcid.org/0000-0003-0465-1559
Ilse van Bemmel  https://orcid.org/0000-0001-5473-2950
Huib Jan van Langevelde  https://orcid.org/0000-0002-0230-5946
Daniel R. van Rossum  https://orcid.org/0000-0001-7772-6131
John Wardle  https://orcid.org/0000-0002-8960-2942
Jonathan Weintroub  https://orcid.org/0000-0002-4603-5204
Norbert Wex  https://orcid.org/0000-0003-4058-2837
Robert Wharton  https://orcid.org/0000-0002-7416-5209
Maciek Wielgus  https://orcid.org/0000-0002-8635-4242
George N. Wong  https://orcid.org/0000-0001-6952-2147
Qingwen Wu (吴庆文)  https://orcid.org/0000-0003-4773-4987
André Young  https://orcid.org/0000-0003-0000-2682
Ken Young  https://orcid.org/0000-0002-3666-4920
Ziri Younsi  https://orcid.org/0000-0001-9283-1191
Feng Yuan (袁峰)  https://orcid.org/0000-0003-3564-6437
J. Anton Zensus  https://orcid.org/0000-0001-7470-3321
Guangyao Zhao  https://orcid.org/0000-0002-4417-1659
Shan-Shan Zhao  https://orcid.org/0000-0002-9774-3606
Joseph R. Farah  https://orcid.org/0000-0003-4914-5625
Daniel Michalik  https://orcid.org/0000-0002-7618-6556
Andrew Nadolski  https://orcid.org/0000-0001-9479-9957
Rurik A. Primiani  https://orcid.org/0000-0003-3910-7529
Paul Yamaguchi  https://orcid.org/0000-0002-6017-8199

## The Event Horizon Telescope Collaboration

Kazunori Akiyama[1,2,3,4] , Antxon Alberdi[5] , Walter Alef[6], Keiichi Asada[7], Rebecca Azulay[8,9,6] , Anne-Kathrin Baczko[6] ,
David Ball[10], Mislav Baloković[4,11] , John Barrett[2] , Dan Bintley[12], Lindy Blackburn[4,11] , Wilfred Boland[13],
Katherine L. Bouman[4,11,14] , Geoffrey C. Bower[15] , Michael Bremer[16], Christiaan D. Brinkerink[17] , Roger Brissenden[4,11] ,
Silke Britzen[6] , Avery E. Broderick[18,19,20] , Dominique Broguiere[16], Thomas Bronzwaer[17], Do-Young Byun[21,22] ,
John E. Carlstrom[23,24,25,26], Andrew Chael[4,11] , Chi-kwan Chan[10,27] , Shami Chatterjee[28] , Koushik Chatterjee[29],
Ming-Tang Chen[15], Yongjun Chen (陈永军)[30,31], Ilje Cho[21,22] , Pierre Christian[10,11] , John E. Conway[32] , James M. Cordes[28],
Geoffrey B. Crew[2] , Yuzhu Cui[33,34] , Jordy Davelaar[17] , Mariafelicia De Laurentis[35,36,37] , Roger Deane[38,39] ,
Jessica Dempsey[12] , Gregory Desvignes[6] , Jason Dexter[40] , Sheperd S. Doeleman[4,11] , Ralph P. Eatough[6] ,
Heino Falcke[17] , Vincent L. Fish[2] , Ed Fomalont[41], Raquel Fraga-Encinas[17] , Per Friberg[12], Christian M. Fromm[36],
José L. Gómez[5] , Peter Galison[4,11,42] , Charles F. Gammie[43,44] , Roberto García[16], Olivier Gentaz[16], Boris Georgiev[19,20] ,
Ciriaco Goddi[17,45], Roman Gold[36] , Minfeng Gu (顾敏峰)[30,46] , Mark Gurwell[11] , Kazuhiro Hada[33,34] , Michael H. Hecht[2],
Ronald Hesper[47] , Luis C. Ho (何子山)[48,49] , Paul Ho[7], Mareki Honma[33,34] , Chih-Wei L. Huang[7] , Lei Huang (黄磊)[30,46] ,
David H. Hughes[50], Shiro Ikeda[3,51,52,53] , Makoto Inoue[7], Sara Issaoun[17] , David J. James[4,11] , Buell T. Jannuzi[10],
Michael Janssen[17] , Britton Jeter[19,20] , Wu Jiang (江悟)[30] , Michael D. Johnson[4,11] , Svetlana Jorstad[54,55] ,
Taehyun Jung[21,22] , Mansour Karami[56,19] , Ramesh Karuppusamy[6] , Tomohisa Kawashima[3] , Garrett K. Keating[11] ,
Mark Kettenis[57] , Jae-Young Kim[6] , Junhan Kim[10] , Jongsoo Kim[21], Motoki Kino[3,58] , Jun Yi Koay[7] ,
Patrick M. Koch[7] , Shoko Koyama[7] , Michael Kramer[6] , Carsten Kramer[16], Thomas P. Krichbaum[6] , Cheng-Yu Kuo[59],
Tod R. Lauer[60] , Sang-Sung Lee[21] , Yan-Rong Li (李彦荣)[61] , Zhiyuan Li (李志远)[62,63] , Michael Lindqvist[32] ,
Kuo Liu[6] , Elisabetta Liuzzo[64] , Wen-Ping Lo[7,65], Andrei P. Lobanov[6], Laurent Loinard[66,67] , Colin Lonsdale[2], Ru-Sen Lu
(路如森)[30,6] , Nicholas R. MacDonald[6], Jirong Mao (毛基荣)[68,69,70] , Sera Markoff[29,71] , Daniel P. Marrone[10] ,
Alan P. Marscher[54] , Iván Martí-Vidal[72,73] , Satoki Matsushita[7], Lynn D. Matthews[2] , Lia Medeiros[10,74] ,
Karl M. Menten[6] , Yosuke Mizuno[36] , Izumi Mizuno[12] , James M. Moran[4,11] , Kotaro Moriyama[33,2] ,
Monika Moscibrodzka[17] , Cornelia Müller[6,17] , Hiroshi Nagai[3,35] , Neil M. Nagar[75] , Masanori Nakamura[7] ,
Ramesh Narayan[4,11] , Gopal Narayanan[76], Iniyan Natarajan[39] , Roberto Neri[16], Chunchong Ni[19,20] , Aristeidis Noutsos[6] ,
Hiroki Okino[33,77], Héctor Olivares[36] , Tomoaki Oyama[33], Feryal Özel[10], Daniel C. M. Palumbo[4,11] , Nimesh Patel[11],
Ue-Li Pen[78,79,80,81] , Dominic W. Pesce[4,11] , Vincent Piétu[16], Richard Plambeck[82], Aleksandar PopStefanija[76],
Oliver Porth[29,36] , Ben Prather[43] , Jorge A. Preciado-López[18] , Dimitrios Psaltis[10], Hung-Yi Pu[18] ,
Venkatessh Ramakrishnan[75] , Ramprasad Rao[15] , Mark G. Rawlings[12], Alexander W. Raymond[4,11] , Luciano Rezzolla[36] ,
Bart Ripperda[36] , Freek Roelofs[17] , Alan Rogers[2], Eduardo Ros[6] , Mel Rose[10] , Arash Roshanineshat[10], Helge Rottmann[6],
Alan L. Roy[6] , Chet Ruszczyk[2] , Benjamin R. Ryan[83,84] , Kazi L. J. Rygl[64] , Salvador Sánchez[85],
David Sánchez-Arguelles[50,86] , Mahito Sasada[33,87] , Tuomas Savolainen[6,88,89] , F. Peter Schloerb[76], Karl-Friedrich Schuster[16],
Lijing Shao[6,49] , Zhiqiang Shen (沈志强)[30,31] , Des Small[57] , Bong Won Sohn[21,22,90] , Jason SooHoo[2] , Fumie Tazaki[33] ,
Paul Tiede[19,20] , Remo P. J. Tilanus[17,45,91] , Michael Titus[2] , Kenji Toma[92,93] , Pablo Torne[6,85] , Tyler Trent[10],
Sascha Trippe[94] , Shuichiro Tsuda[95], Ilse van Bemmel[57] , Huib Jan van Langevelde[57,96] , Daniel R. van Rossum[17] ,
Jan Wagner[6], John Wardle[97] , Jonathan Weintroub[4,11] , Norbert Wex[6] , Robert Wharton[6] , Maciek Wielgus[4,11] ,
George N. Wong[43] , Qingwen Wu (吴庆文)[98] , André Young[17] , Ken Young[11] , Ziri Younsi[99,36] , Feng Yuan
(袁峰)[30,46,100] , Ye-Fei Yuan (袁业飞)[101] , J. Anton Zensus[6] , Guangyao Zhao[21], Shan-Shan Zhao[17,62] , Ziyan Zhu[42],
Joseph R. Farah[11,102,4] , Zheng Meyer-Zhao[7,103] , Daniel Michalik[104,105] , Andrew Nadolski[44] , Hiroaki Nishioka[7],
Nicolas Pradel[7], Rurik A. Primiani[106] , Kamal Souccar[76] , Laura Vertatschitsch[11,106] , and Paul Yamaguchi[11]

[1] National Radio Astronomy Observatory, 520 Edgemont Rd, Charlottesville, VA 22903, USA
[2] Massachusetts Institute of Technology Haystack Observatory, 99 Millstone Road, Westford, MA 01886, USA
[3] National Astronomical Observatory of Japan, 2-21-1 Osawa, Mitaka, Tokyo 181-8588, Japan
[4] Black Hole Initiative at Harvard University, 20 Garden Street, Cambridge, MA 02138, USA
[5] Instituto de Astrofísica de Andalucía-CSIC, Glorieta de la Astronomía s/n, E-18008 Granada, Spain
[6] Max-Planck-Institut für Radioastronomie, Auf dem Hügel 69, D-53121 Bonn, Germany

[7] Institute of Astronomy and Astrophysics, Academia Sinica, 11F of Astronomy-Mathematics Building, AS/NTU No. 1, Section 4, Roosevelt Rd, Taipei 10617, Taiwan, R.O.C.

[8] Departament d'Astronomia i Astrofísica, Universitat de València, C. Dr. Moliner 50, E-46100 Burjassot, València, Spain

[9] Observatori Astronòmic, Universitat de València, C. Catedrático José Beltrán 2, E-46980 Paterna, València, Spain

[10] Steward Observatory and Department of Astronomy, University of Arizona, 933 N. Cherry Ave., Tucson, AZ 85721, USA

[11] Center for Astrophysics | Harvard & Smithsonian, 60 Garden Street, Cambridge, MA 02138, USA

[12] East Asian Observatory, 660 N. A'ohoku Pl., Hilo, HI 96720, USA

[13] Nederlandse Onderzoekschool voor Astronomie (NOVA), P.O. Box 9513, 2300 RA Leiden, The Netherlands

[14] California Institute of Technology, 1200 East California Boulevard, Pasadena, CA 91125, USA

[15] Institute of Astronomy and Astrophysics, Academia Sinica, 645 N. A'ohoku Place, Hilo, HI 96720, USA

[16] Institut de Radioastronomie Millimétrique, 300 rue de la Piscine, F-38406 Saint Martin d'Hères, France

[17] Department of Astrophysics, Institute for Mathematics, Astrophysics and Particle Physics (IMAPP), Radboud University, P.O. Box 9010, 6500 GL Nijmegen, The Netherlands

[18] Perimeter Institute for Theoretical Physics, 31 Caroline Street North, Waterloo, ON, N2L 2Y5, Canada

[19] Department of Physics and Astronomy, University of Waterloo, 200 University Avenue West, Waterloo, ON, N2L 3G1, Canada

[20] Waterloo Centre for Astrophysics, University of Waterloo, Waterloo, ON N2L 3G1 Canada

[21] Korea Astronomy and Space Science Institute, Daedeok-daero 776, Yuseong-gu, Daejeon 34055, Republic of Korea

[22] University of Science and Technology, Gajeong-ro 217, Yuseong-gu, Daejeon 34113, Republic of Korea

[23] Kavli Institute for Cosmological Physics, University of Chicago, 5640 South Ellis Avenue, Chicago, IL 60637, USA

[24] Department of Astronomy and Astrophysics, University of Chicago, 5640 South Ellis Avenue, Chicago, IL 60637, USA

[25] Department of Physics, University of Chicago, 5720 South Ellis Avenue, Chicago, IL 60637, USA

[26] Enrico Fermi Institute, University of Chicago, 5640 South Ellis Avenue, Chicago, IL 60637, USA

[27] Data Science Institute, University of Arizona, 1230 N. Cherry Ave., Tucson, AZ 85721, USA

[28] Cornell Center for Astrophysics and Planetary Science, Cornell University, Ithaca, NY 14853, USA

[29] Anton Pannekoek Institute for Astronomy, University of Amsterdam, Science Park 904, 1098 XH, Amsterdam, The Netherlands

[30] Shanghai Astronomical Observatory, Chinese Academy of Sciences, 80 Nandan Road, Shanghai 200030, People's Republic of China

[31] Key Laboratory of Radio Astronomy, Chinese Academy of Sciences, Nanjing 210008, People's Republic of China

[32] Department of Space, Earth and Environment, Chalmers University of Technology, Onsala Space Observatory, SE-439 92 Onsala, Sweden

[33] Mizusawa VLBI Observatory, National Astronomical Observatory of Japan, 2-12 Hoshigaoka, Mizusawa, Oshu, Iwate 023-0861, Japan

[34] Department of Astronomical Science, The Graduate University for Advanced Studies (SOKENDAI), 2-21-1 Osawa, Mitaka, Tokyo 181-8588, Japan

[35] Dipartimento di Fisica "E. Pancini," Universitá di Napoli "Federico II," Compl. Univ. di Monte S. Angelo, Edificio G, Via Cinthia, I-80126, Napoli, Italy

[36] INFN Sez. di Napoli, Compl. Univ. di Monte S. Angelo, Edificio G, Via Cinthia, I-80126, Napoli, Italy

[37] Institut für Theoretische Physik, Goethe-Universität Frankfurt, Max-von-Laue-Straße 1, D-60438 Frankfurt am Main, Germany

[38] Department of Physics, University of Pretoria, Lynnwood Road, Hatfield, Pretoria 0083, South Africa

[39] Centre for Radio Astronomy Techniques and Technologies, Department of Physics and Electronics, Rhodes University, Grahamstown 6140, South Africa

[40] Max-Planck-Institut für Extraterrestrische Physik, Giessenbachstr. 1, D-85748 Garching, Germany

[41] Department of History of Science, Harvard University, Cambridge, MA 02138, USA

[42] Department of Physics, Harvard University, Cambridge, MA 02138, USA

[43] Department of Physics, University of Illinois, 1110 West Green Street, Urbana, IL 61801, USA

[44] Department of Astronomy, University of Illinois at Urbana-Champaign, 1002 West Green Street, Urbana, IL 61801, USA

[45] Leiden Observatory—Allegro, Leiden University, P.O. Box 9513, 2300 RA Leiden, The Netherlands

[46] Key Laboratory for Research in Galaxies and Cosmology, Chinese Academy of Sciences, Shanghai 200030, People's Republic of China

[47] NOVA Sub-mm Instrumentation Group, Kapteyn Astronomical Institute, University of Groningen, Landleven 12, 9747 AD Groningen, The Netherlands

[48] Department of Astronomy, School of Physics, Peking University, Beijing 100871, People's Republic of China

[49] Kavli Institute for Astronomy and Astrophysics, Peking University, Beijing 100871, People's Republic of China

[50] Instituto Nacional de Astrofísica, Óptica y Electrónica. Apartado Postal 51 y 216, 72000. Puebla Pue., México

[51] The Institute of Statistical Mathematics, 10-3 Midori-cho, Tachikawa, Tokyo, 190-8562, Japan

[52] Department of Statistical Science, The Graduate University for Advanced Studies (SOKENDAI), 10-3 Midori-cho, Tachikawa, Tokyo 190-8562, Japan

[53] Kavli Institute for the Physics and Mathematics of the Universe, The University of Tokyo, 5-1-5 Kashiwanoha, Kashiwa, 277-8583, Japan

[54] Institute for Astrophysical Research, Boston University, 725 Commonwealth Ave., Boston, MA 02215, USA

[55] Astronomical Institute, St. Petersburg University, Universitetskij pr., 28, Petrodvorets, 198504 St. Petersburg, Russia

[56] Perimeter Institute for Theoretical Physics, 31 Caroline Street North, Waterloo, Ontario N2L 2Y5, Canada

[57] Joint Institute for VLBI ERIC (JIVE), Oude Hoogeveensedijk 4, 7991 PD Dwingeloo, The Netherlands

[58] Kogakuin University of Technology & Engineering, Academic Support Center, 2665-1 Nakano, Hachioji, Tokyo 192-0015, Japan

[59] Physics Department, National Sun Yat-Sen University, No. 70, Lien-Hai Rd, Kaosiung City 80424, Taiwan, R.O.C

[60] National Optical Astronomy Observatory, 950 North Cherry Ave., Tucson, AZ 85719, USA

[61] Key Laboratory for Particle Astrophysics, Institute of High Energy Physics, Chinese Academy of Sciences, 19B Yuquan Road, Shijingshan District, Beijing, People's Republic of China

[62] School of Astronomy and Space Science, Nanjing University, Nanjing 210023, People's Republic of China

[63] Key Laboratory of Modern Astronomy and Astrophysics, Nanjing University, Nanjing 210023, People's Republic of China

[64] Italian ALMA Regional Centre, INAF-Istituto di Radioastronomia, Via P. Gobetti 101, I-40129 Bologna, Italy

[65] Department of Physics, National Taiwan University, No.1, Sect.4, Roosevelt Rd., Taipei 10617, Taiwan, R.O.C

[66] Instituto de Radioastronomía y Astrofísica, Universidad Nacional Autónoma de México, Morelia 58089, México

[67] Instituto de Astronomía, Universidad Nacional Autónoma de México, CdMx 04510, México

[68] Yunnan Observatories, Chinese Academy of Sciences, 650011 Kunming, Yunnan Province, People's Republic of China

[69] Center for Astronomical Mega-Science, Chinese Academy of Sciences, 20A Datun Road, Chaoyang District, Beijing, 100012, People's Republic of China

[70] Key Laboratory for the Structure and Evolution of Celestial Objects, Chinese Academy of Sciences, 650011 Kunming, People's Republic of China

[71] Gravitation Astroparticle Physics Amsterdam (GRAPPA) Institute, University of Amsterdam, Science Park 904, 1098 XH Amsterdam, The Netherlands

[72] Department of Space, Earth and Environment, Chalmers University of Technology, Onsala Space Observatory, SE-43992 Onsala, Sweden

[73] Centro Astronómico de Yebes (IGN), Apartado 148, E-19180 Yebes, Spain

[74] Department of Physics, Broida Hall, University of California Santa Barbara, Santa Barbara, CA 93106, USA

[75] Astronomy Department, Universidad de Concepción, Casilla 160-C, Concepción, Chile

[76] Department of Astronomy, University of Massachusetts, 01003, Amherst, MA, USA

[77] Department of Astronomy, Graduate School of Science, The University of Tokyo, 7-3-1 Hongo, Bunkyo-ku, Tokyo 113-0033, Japan

[78] Canadian Institute for Theoretical Astrophysics, University of Toronto, 60 St. George Street, Toronto, ON M5S 3H8, Canada

[79] Dunlap Institute for Astronomy and Astrophysics, University of Toronto, 50 St. George Street, Toronto, ON M5S 3H4, Canada
[80] Canadian Institute for Advanced Research, 180 Dundas St West, Toronto, ON M5G 1Z8, Canada
[81] Perimeter Institute for Theoretical Physics, 31 Caroline Street North, Waterloo, ON N2L 2Y5, Canada
[82] Radio Astronomy Laboratory, University of California, Berkeley, CA 94720, USA
[83] CCS-2, Los Alamos National Laboratory, P.O. Box 1663, Los Alamos, NM 87545, USA
[84] Center for Theoretical Astrophysics, Los Alamos National Laboratory, Los Alamos, NM 87545, USA
[85] Instituto de Radioastronomía Milimétrica, IRAM, Avenida Divina Pastora 7, Local 20, E-18012, Granada, Spain
[86] Consejo Nacional de Ciencia y Tecnología, Av. Insurgentes Sur 1582, 03940, Ciudad de México, México
[87] Hiroshima Astrophysical Science Center, Hiroshima University, 1-3-1 Kagamiyama, Higashi-Hiroshima, Hiroshima 739-8526, Japan
[88] Aalto University Department of Electronics and Nanoengineering, PL 15500, FI-00076 Aalto, Finland
[89] Aalto University Metsähovi Radio Observatory, Metsähovintie 114, FI-02540 Kylmälä, Finland
[90] Department of Astronomy, Yonsei University, Yonsei-ro 50, Seodaemun-gu, 03722 Seoul, Republic of Korea
[91] Netherlands Organisation for Scientific Research (NWO), Postbus 93138, 2509 AC Den Haag , The Netherlands
[92] Frontier Research Institute for Interdisciplinary Sciences, Tohoku University, Sendai 980-8578, Japan
[93] Astronomical Institute, Tohoku University, Sendai 980-8578, Japan
[94] Department of Physics and Astronomy, Seoul National University, Gwanak-gu, Seoul 08826, Republic of Korea
[95] Mizusawa VLBI Observatory, National Astronomical Observatory of Japan, Hoshigaoka 2-12, Mizusawa-ku, Oshu-shi, Iwate 023-0861, Japan
[96] Leiden Observatory, Leiden University, Postbus 2300, 9513 RA Leiden, The Netherlands
[97] Physics Department, Brandeis University, 415 South Street, Waltham, MA 02453, USA
[98] School of Physics, Huazhong University of Science and Technology, Wuhan, Hubei, 430074, People's Republic of China
[99] Mullard Space Science Laboratory, University College London, Holmbury St. Mary, Dorking, Surrey, RH5 6NT, UK
[100] School of Astronomy and Space Sciences, University of Chinese Academy of Sciences, No. 19A Yuquan Road, Beijing 100049, People's Republic of China
[101] Astronomy Department, University of Science and Technology of China, Hefei 230026, People's Republic of China
[102] University of Massachusetts Boston, 100 William T, Morrissey Blvd, Boston, MA 02125, USA
[103] ASTRON, Oude Hoogeveensedijk 4, 7991 PD Dwingeloo, The Netherlands
[104] Science Support Office, Directorate of Science, European Space Research and Technology Centre (ESA/ESTEC),
Keplerlaan 1, 2201 AZ Noordwijk, The Netherlands
[105] University of Chicago, 5640 South Ellis Avenue, Chicago, IL 60637, USA
[106] Systems & Technology Research, 600 West Cummings Park, Woburn, MA 01801, USA